\documentclass[pr,twocolumn,notitlepage]{revtex4}
\usepackage{amsmath}
\usepackage{amssymb}
\usepackage{bm}
\usepackage{epsfig}
\usepackage{graphicx}
\usepackage{color}
\usepackage[T2A]{fontenc}
\usepackage[cp1251]{inputenc}
\usepackage[english]{babel}
\usepackage{enumitem}

\renewcommand{\Re}{\mathop{\rm Re}}
\renewcommand{\Im}{\mathop{\rm Im}}
\renewcommand{\cosh}{\mathop{\rm ch}}
\renewcommand{\tanh}{\mathop{\rm th}}
\newcommand{\hF}{\mathop{{}_2\rm{F}_1}}

\newcommand{\Arch}{\mathop{\rm Arch}}

\newcount\timehh  \newcount\timemm
\timehh=\time \divide\timehh by 60
\timemm=\time
\begin{document}

\title{Coherent spin dynamics of electrons and excitons in
  nanostructures (Review)}
\author{M.M. Glazov}
\affiliation{Ioffe Physical-Technical Institute of the RAS, 
194021 St.-Petersburg, Russia}

\date{\today, file = \jobname.tex, printing time = \number\timehh\,:\,\ifnum\timemm<10 0\fi \number\timemm}

\begin{abstract}
The studies of spin phenomena in semiconductor low dimensional systems
have grown into the rapidly developing area of the condensed matter
physics: spintronics. The most urgent problems in this area, both
fundamental and applied, are the creation of charge carrier spin
polarization and its detection as well as electron spin control by
nonmagnetic methods. Here we present a review of recent
achievements in the studies of spin dynamics of electrons, holes and
their complexes in the pump-probe method. The microscopic mechanisms
of spin orientation of charge carriers and their complexes by short
circularly polarized optical pulses and the formation processes of the
spin signals of 
Faraday and Kerr rotation of the probe pulse polarization plane as
well as induced ellipticity are discussed. A special attention is paid
to the comparison of theoretical concepts with experimental data
obtained on the $n$-type quantum well and quantum dot array samples.  
\end{abstract}

\maketitle
%\end{document}

\tableofcontents

\section{Introduction}\label{sec:intro}

The studies of spin effects in semiconductors have started in the end
of 1960s after the discovery of the optical orientation of electron
spins in Silicon~\cite{lampel68}. The investigation of circular
polarization of luminescence at a constant wave pumping as a function
of magnetic field made it possible to establish the main mechanisms of 
the nonequilibrium spin generation of free and localized charge
carriers in semiconductors and to study the processes of electron and
hole spin relaxation, as well as to explore the interaction of
electron and nuclear spin systems~\cite{optor:eng}.

An interest in the studies of spin dynamics in bulk semiconductors and
semiconductor low dimensional systems has revived in the end of 90s of
the last century. The pump-probe method~\cite{PhysRevLett.55.1128,Zheludev1994823} has played a role of no small
importance in it, this method has enabled scientists to study spin
coherence with temporal resolution. It is certain, that the precision
measurements of ultralong spin relaxation times in bulk
materials~\cite{Kikkawa98}, quantum wells~\cite{Kikkawa29081997} and
quantum dots~\cite{A.Greilich07212006}, a visualization of spin
precession, relaxation~\cite{brand02,leyland06,PhysRevB.80.241314}
and the spin transport in bulk materials and
nanostructures~\cite{Kikkawa1999,kato04,PhysRevLett.94.236601} carried
out in the framework of this two beam technique, have laid the foundation
of spintronics: a new area of science and technology, where the
electron spin along with its charge finds application for the
information transfer and processing,
see~\cite{awschalom_book,dyakonov_book,ssc:optor} and references therein.  

\begin{figure*}[hptb]
\includegraphics[width=0.75\textwidth]{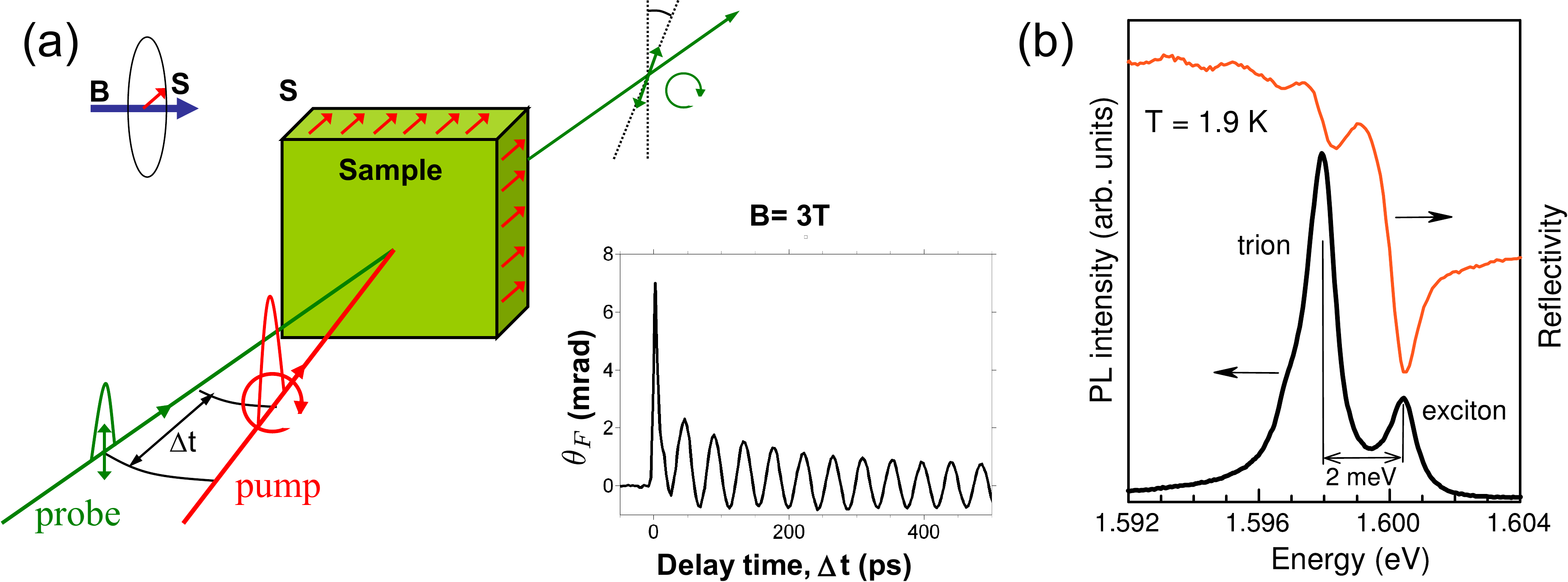}
\caption{(a) Illustration of the pump-probe technique. Pump and probe
  denote circularly polarized pump and linearly polarized probe
  pulses, respectively.
Inset shows a typical Faraday rotation signal as a function of the delay
between pump and probe pulses, $\Delta t$. 
(b) Photoluminescence and reflection spectra of the five
CdTe/Cd$_{0.78}$Mg$_{0.22}$Te quantum well sample, each well of the
width of 20 nm 
    contains an electron gas with the density $N\approx
    10^{10}$~cm$^{-2}$. Photoluminescence was measured at a
    nonresonant constant wave pumping with the photon energy being
    2.33~eV. Data are reproduced from Ref.~\cite{zhukov10}.}\label{fig:scheme} 
\end{figure*}

\begin{figure}[hptb]
\includegraphics[width=\linewidth]{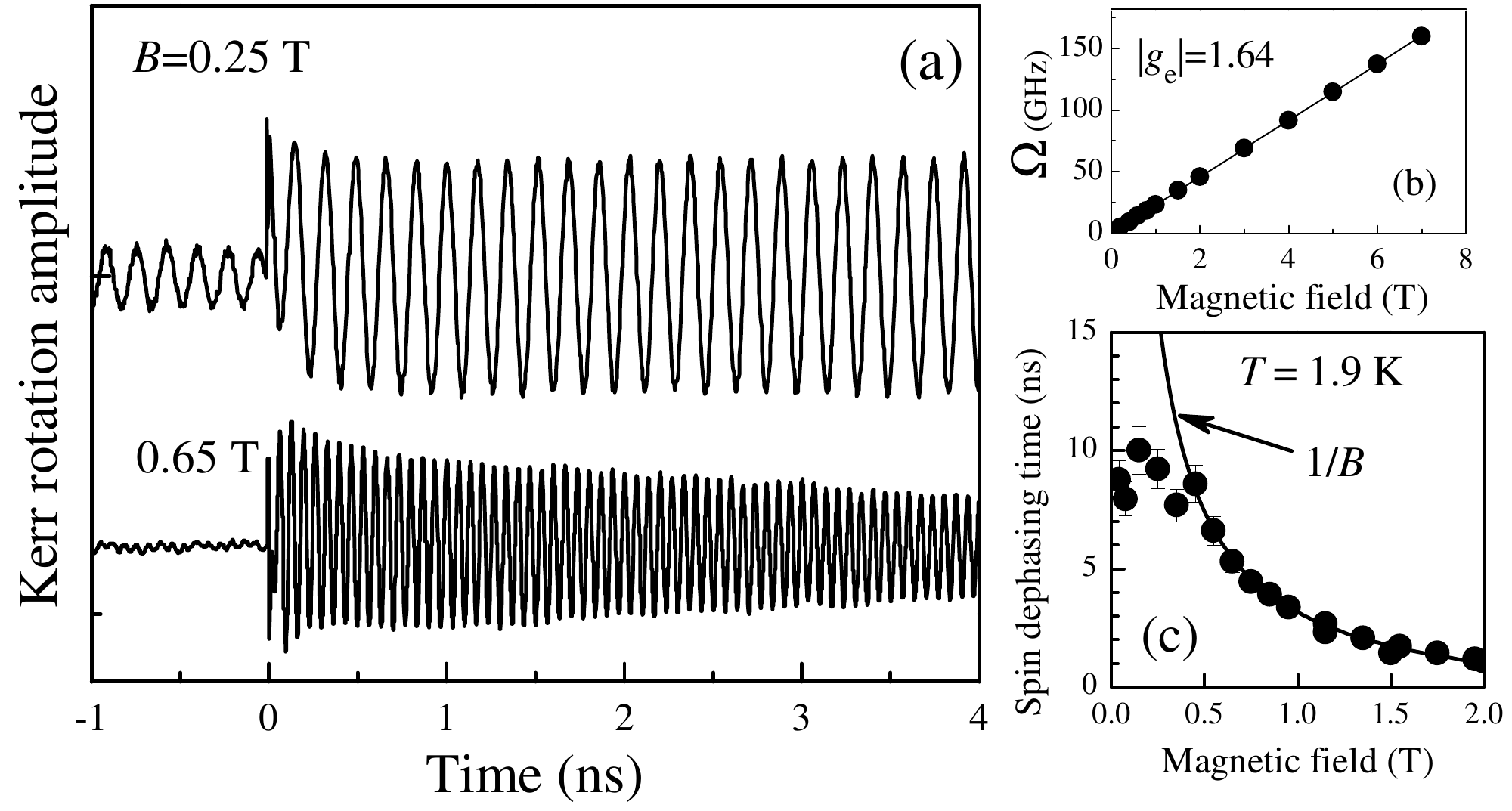}
\caption{(a) Typical Kerr rotation signals as functions of delay
  between the pump and probe pulses measured in the five
  CdTe/Cd$_{0.78}$Mg$_{0.22}$Te quantum well sample, each well has a
  width of 20 nm and contains the electron gas with density $N\approx
  10^{10}$~cm$^{-2}$. (b) Dependence of the spin beats frequency on the
  external magnetic field. Points represent experimental data,
  straight line is a fit according to Eq.~\eqref{larmor}. (c)
  Dependence of spin beats decay time on magnetic field. Points represent
  the experimental data, solid curve is a theory result, assuming the
  spread of electron $g$-factor values. Data are reproduced from
  Ref.~\cite{yakovlev_bayer}.}\label{fig:demo}   
\end{figure}

The essence of the pump-probe method is schematically shown in
Fig.~\ref{fig:scheme}(a). The core of this technique is as follows:
a sample is subject to a sufficiently strong circularly polarized
pump pulse, whose absorption causes the spin orientation of charge
carriers and their complexes: excitons, trions. After a certain delay, a
much weaker linearly polarized probe pulse arrives at the sample. The
presence of the nonequilibrium spin polarization in the sample makes
the system optically active: the polarization plane of the probe pulse
rotates in the transmission geometry (magnetooptical or spin Faraday
effect) and in the reflection geometry (spin Kerr
effect)~\cite{aronovivchenko:eng}. Moreover, the probe pulse passed and
reflected from the sample acquires partially a circular polarization,
that is the ellipticity. The polarization plane rotation angle as well as
induced ellipticity are proportional to the spin polarization in the
system. If the sample is subject to a magnetic field in Voigt geometry
(field is applied in the plane perpendicular to the pump and probe
pulses propagation direction), the
spins of electrons, holes and their complexes precess around the
external field. Therefore, the spin ellipticity, Faraday and Kerr rotation
angles oscillate as functions of the delay between the pump and probe
pulses, reflecting spin precession, see inset in
Fig.~\ref{fig:scheme}(a) and Fig.~\ref{fig:demo}(a). 

Pump-probe method is very sensitive to the momentary values of
electron and hole spin polarization in semiconductors. The dependence
of the signals on the temporal delay between the pulses allows one to
study straightforwardly the spin dynamics of electron system in solids
and extract 
 spin precession frequencies, spin relaxation and
decoherence times directly from the experimental data. It is shown in
Fig.~\ref{fig:demo}, where two typical curves of Kerr rotation signal
as function of delay measured in the CdTe/CdMgTe quantum well structure
are presented [Fig.~\ref{fig:demo}(a)] along with the results of its treatment
[Fig.~\ref{fig:demo}(b) and (c)]. In Fig.~\ref{fig:demo}(b),(c) the
dependence of 
spin precession frequency and electron spin dephasing time on magnetic
field is shown. The detailed analysis of the spin signals temporal
dependence measured in the pump-probe technique is presented below in
Sec.~\ref{sec:time}. We just mention here, that as distinct from the
methods based on polarization of luminescence studies, in the
pump-probe technique the charge
carriers spin dynamics can be studied  on
the time scales which exceed by far the luminescence decay
time. Moreover, application of an additional circularly or linearly
polarized ``control''
pulse~\cite{PhysRevB.77.165309,phelps:237402,Greilich2009,zhukov10}
opens up possibilities to control spin dynamics by nonmagnetic means.

Among the systems of every sort and kind, where the pump-probe
technique is successfully applied, the special place is taken by the
structures with singly charged quantum dot arrays or quantum well
structures with low density electron gas, where the condition $Na_{\rm
  B}^2 \lesssim 1$ is fulfilled~\cite{Chen20101803}. Here $N$ is
two-dimensional electron density,  $a_{\rm B}$ is the Bohr
radius. Mentioned systems possess important features: firstly, due to
electron localization in quantum dots or at quantum well potential
fluctuations, the spin relaxation times of resident carriers are
strongly increased, secondly, in these very systems, the role of
Coulomb interaction is high, and Coulomb complexes: excitons and
trions manifest themselves most brightly
[Fig.~\ref{fig:scheme}(b)]. It makes possible to study the microscopic 
processes responsible for the excitation, control and detection of
spin polarization with spectral sensitivity, while long spin
relaxation times are important for device applications in the field of
spintronics.

In these structures one deals with an electron ensemble. On one hand,
its inevitable inhomogeneity results in an effective dephasing of
electron spins, e.g., in magnetic field due to the spread of electron
$g$-factor values. The influence of inhomogeneity can be avoided by
studying single quantum dots~\cite{mikkelsen07,atature07}, however,
a wide application of the pump-probe method for the single dots is
quite hampered owing to weak signals and small ``signal-to-noise''
ratio. On the other hand, in quantum dot ensembles the electron spin
precession mode-locking is observed, in which case about  $10^6$ spins
precess with commensurable frequencies~\cite{A.Greilich07212006},
which allows in a certain degree to overcome the inhomogeneity
effects. In the review we focus on these particular systems: singly
charged quantum dots and quantum wells with low density electron gas.

\section{Macroscopic description of spin coherence generation and detection}\label{sec:macro}

The semiphenomenological theory of resident charge carriers spin
coherence generation and 
detection processes in quantum wells and quantum dot ensembles is
given in this Section. The consistent quantum-mechanical description
of the interaction of short optical pulses with localized carriers
is presented below in Sec.~\ref{sec:micro}. Here we focus on 
simple physical models, describing spin generation at trion and
exciton excitation, as well as on the macroscopic description of the
spin Faraday, Kerr and ellipticity effects in quantum well and quantum
dot structures.

\subsection{Resident carriers spin orientation mechanisms}\label{subsec:macro:orient}

\subsubsection{Resonant excitation of trions}\label{subsec:macro:orient:tr}

In quantum wells with a low density electron gas and in singly charged
quantum dots the optical absorption results in the formation of
 $X^-$ trions: the three particle complexes, consisting of an
 electron pair and a hole. In the absence of an external magnetic
 field or in moderate magnetic fields (up to several Tesla) the total
 spin of the electron pair in the trion ground state equals to zero,
 therefore \emph{prima facie} it is not clear whether the resident
 electrons spin
 coherence can arise in the system.

\begin{figure}[hptb]
\includegraphics[width=0.45\textwidth]{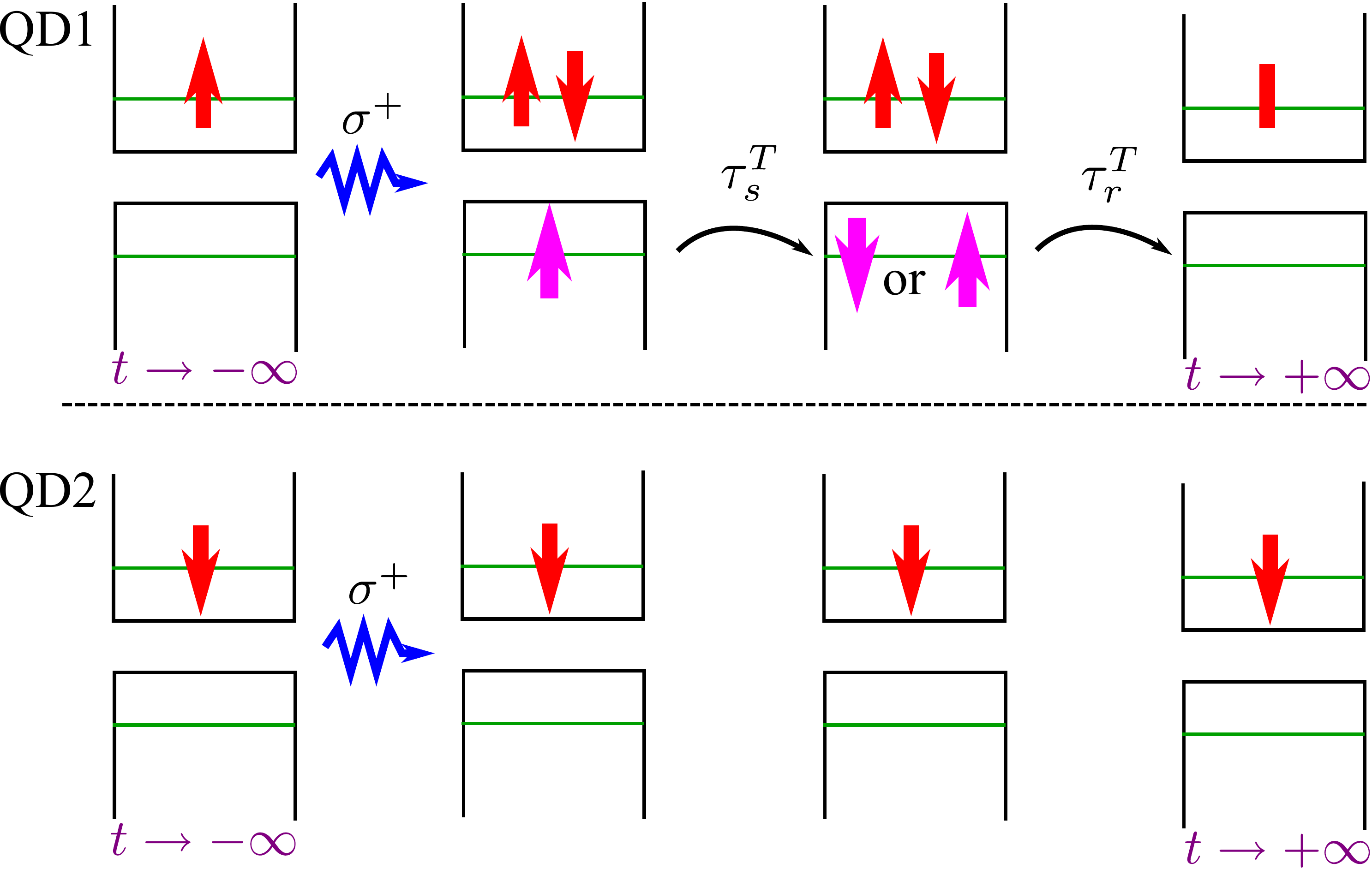}
\caption{Scheme of resident electron spin orientation at the resonant
  trion excitation.  QD1 and QD2 are two quantum dots where the
  electron spins before the pump pulse arrival are opposite. As a
  result of the
  $\sigma^+$ photon absorption the trion is formed in the dot  QD1
  only.}\label{fig:excitation} 
\end{figure}

One can be convinced, however, that the trion formation process is
spin-dependent. Indeed, in accordance with the selection rules for the
quantum well and self-organized quantum dot structures the absorption
of the circularly polarized photon is accompanied by the formation of
electron ($e$) and heavy-hole ($hh$) pair: $(e,s_z=-1/2;hh,j_z=+3/2)$
for $\sigma^+$ polarized light propagating in the positive direction
of $z$ axis and $(e,s_z=+1/2;hh,j_z=-3/2)$ for $\sigma^-$ polarized
quantum. Here $s_z$, $j_z$ are the projections of electron and hole
spins onto  $z$ axis. Therefore, e.g., only electrons with spin
component being equal to $s_z=+1/2$ participate in the trion formation
for the
$\sigma^+$ polarized pump pulse, as it is schematically shown in
Fig.~\ref{fig:excitation}.  

Let us assume now that the hole-in-trion spin relaxation time
$\tau_s^T$ is small as compared with the trion radiative lifetime
$\tau_r^T$. Hence, at trion recombination the depolarized electrons
return to the system. Thereby, due to the trion formation, $\sigma^+$
pump pulse depolarizes electrons with the spin component $1/2$ and
does not affect electrons with the spin component
$-1/2$. Since at low temperatures resident electron spin relaxation
time  $\tau_s>10$~ns and exceeds by far the trion lifetime $\tau_r^T\sim 10 \div
100$~ps~\cite{dzhioev97:eng,Kikkawa98,A.Greilich07212006,fokina-2010},
after the trions recombination, the imbalance of electrons with
opposite spin projection arises, that is spin
polarization~\cite{shabaev:201305,PhysRevB.68.235316}. Note, that the
spin accumulated in the system is directed in the same way as the spin
of photocreated electrons: in opposite to $z$ axis direction at the
$\sigma^+$ pumping and along  $z$ axis direction in the case of
$\sigma^-$ pumping.

Specified mechanism of the long living electron spin polarization
generation at a trion resonant excitation is quite analogous to the
classical spin pumping of majority carriers in
semiconductors~\cite{dp_orientation:eng}. If hole spin relaxation is
suppressed, $\tau_s^T\gg \tau_r^T$, then in the absence of magnetic
field the electron spin orientation in uneffective: spin of electrons
returning after trions recombination exactly compensates the
polarization of carriers which did not pariticipate in the trions
formation. The magnetic field presence results in the electron and
hole spin precession. For instance, in quantum well structures where
the in-plane heavy hole $g$-factor is small~\cite{Mar99}, the trion
state remains unchanged, while electron spin rotates. Hence, after the
trion recombination the spin compensation is broken in the magnetic
field which leads to the appearance of resident electron spin
polarization~\cite{kennedy:045307,PhysRevLett.94.227403}. 

We introduce the electron spin pseudovector $\bm S = (S_x,S_y,S_z)$,
describing mean values of spin components of the charge carriers
ensemble in the quantum well or in the quantum dot array.  Trion spin
states can also be characterized by an effective spin $J_z =
(T_+-T_-)/2$, where $T_\pm$ is the number of trions with the spin
projections onto  $z$ axis equal to  $\pm 3/2$, respectively. Kinetic
equations describing spin dynamics of excitons and trions after the
system 
photoexcitation have form~\cite{zhu07,PhysRevB.75.115320}
\begin{subequations}
\label{system:kin}
\begin{equation}
\label{system:ez}
\frac{\mathrm d S_z}{\mathrm dt} = S_y \Omega - \frac{S_z}{\tau_s} + \frac{J_z}{\tau_r^T},
\end{equation}
\begin{equation}
\label{system:ey}
\frac{\mathrm d S_y}{\mathrm dt} = -S_z \Omega - \frac{S_y}{\tau_s},
\end{equation}
\begin{equation}
\label{system:tz}
\frac{\mathrm d J_z}{\mathrm dt} = - \frac{J_z}{\tau_s^T} - \frac{J_z}{\tau_r^T}.
\end{equation}
\end{subequations}
It is assumed here that the magnetic field  $\bm B$ is applied along
$x$ axis in the structure plane, therefore $x$ component of electron
spin 
$S_x(t)\equiv 0$, 
\begin{equation}
\label{larmor}
\Omega = g\mu_B B/\hbar
\end{equation} 
is the electron spin precession frequency in the external field, $g$
is the electron $g$-factor. A simple form of Eq.~\eqref{system:tz} for
the trion spin follows from the fact that the hole spin rotation angle during
its lifetime is small as compared with that of an electron, inasmuch
as hole $g$-factor in the quantum well or dot plane is small as
compared with electron $g$-factor (general case is considered in
Ref.~\cite{sokolova09}). Following relations serve as initial
conditions for the system \eqref{system:kin}
\begin{equation}
\label{init:class}
S_y(0)=0, \quad S_z(0) = -J_z(0) = - \frac{N_0^T}{2},
\end{equation}
where $N_0^T$ is the number of photoexcited trions. The system of
kinetic equations~\eqref{system:kin} describes well the spin beats in the
quantum well structures containing low density electron gas~\cite{zhu07,ast08}.

Here and in what follows we focus on $n$-type structures. In $p$-type
quantum
wells and quantum dots the physical principles of hole spin coherence
excitation are analogous to those considered above. The specifics of
these structures lies in the fact, that the hole $g$-factor in the
structure plane is much smaller than the $g$-factor of the electron,
which leads to some features of spin
dynamics~\cite{Korn2010415,korn_njp,Machnikowski10,PhysRevB.81.045322}.  

To conclude this Subsection we note, that the physical origin of the
resident electron spin orientation at the
\emph{resonant} excitation of trion:  spin-dependent trion
formation at the circularly polarized light action and the imbalance
formation between the electron spins returning after the trion
recombination and those, which did not participate in the trion formation, is
the same for quantum well and quantum dot
structures~\cite{shabaev:201305,kennedy:045307}. The specifics of
electron and hole spin orientation at the 
\emph{nonresonant} excitation of singly charged quantum dots,
including negative circular polarization of luminescence phenomenon,
was studied in detain in number of
works~\cite{dzhioev98:eng,cortez02,Ignatiev05,laurent06,springerlink:10.1134/S0030400X09030114}. 

\subsubsection{Resonant excitation of excitons}\label{subsec:macro:orient:exc}

The line responsible for exciton resonance is present in the optical
spectra of quantum well structures containing low density electron
gas, see Fig.~\ref{fig:scheme}(b). Let us briefly analyze the
mechanisms of resident electron spin orientation at the exciton
excitation in these systems.

If the system temperature expressed in the units of energy is small
compared to the trion binding energy, the photocreated excitons form
trions capturing those electrons from the resident ensemble, whose spin
orientation is opposite to that of electron-in-exciton. The
following scenario of resident electron spin coherence generation is
quite analogous to the described above for resonant trion excitation.

It was shown in Ref.~\cite{zhu07}
(see also~\cite{fokina-2010}) that the effective resident electron
spin orientation is even possible in situations where the trion
formation is impossible, but excitons are still stable, e.g. at 
temperatures exceeding trion binding energy or in a relatively dense
electron gas. In these cases, the exchange scattering processes of resident
electrons by excitons are important~\cite{tarasenko98}.  

The resident charge carriers spin coherence excitation scenario
consists of two stages: first, polarized pump pulse forms excitons
with definite spin projections of electron and hole
(for example, $s_z = -1/2$, $j_z=3/2$ for $\sigma^+$ polarized
pulse). At a second stage, the spin transfer from
electrons-in-excitons to resident electrons takes place due to
exchange flip-flop scattering. At that, the resident electrons turn out
to be partially spin polarized, and for relatively fast hole spin
relaxation excitons recombine regardless the electron spin orientation
in excitons. Mathematical description of this scenario is presented in
Ref.~\cite{zhu07}.

We note that the long living spin coherence may arise also at the
nonresonant excitation of the quantum well structures by circularly
polarized light. Microscopic mechanisms of these processes are related
both the excitons and trions formation during the relaxation of
photocreated carriers, and with the classical optical pumping of
electron spins~\cite{zhu07,ast08}.

\subsection{Detection of charge carriers spin
  coherence}\label{subsec:macro:detect} 

The spin polarization of electrons and electron-hole complexes leads
to an optical activity of a medium: interaction efficiencies for right-
and left- circularly polarized electromagnetic waves with such a
system turn out to be different. The response of the quantum well
structures and
planar quantum dot arrays to the electromagnetic radiation
can be conveniently characterized by frequency and polarization
dependent light reflection coefficient, $r_{\pm}(\omega)$, which in the
vicinity of the exciton or trion resonance has the form
\begin{equation}
\label{rpm}
r_{\pm}(\omega) = \frac{\mathrm i \Gamma_{0,\pm}}{\omega_{0,\pm}-\omega-\mathrm i
  (\Gamma_{0,\pm} + \Gamma_{\pm})}.
\end{equation}
Here $\omega$ is the probe pulse frequency, subscripts $+$ and 
$-$ refer to  $\sigma^+$ and  $\sigma^-$ components of the pulse,
respectively,  $\omega_0$ is the exciton or trion resonant frequency,
$\Gamma_0$ is its radiative and  $\Gamma$ is its nonradiative damping.
The difference of the resonance parameters for right- and left-
circularly polarized radiation is related with the spin polarization
of charge carriers:
\[
(\Gamma_{0,+}-\Gamma_{0,-}) , (\Gamma_{+}-\Gamma_{-}), (\omega_{0,+}  -
\omega_{0,-}) \propto S_z,
\]
and the spin Faraday and Kerr rotation
as well as induced ellipticity signals are formed  owing to this very
difference.  It is important to
note that in contrast to the classical magnetooptical
effects~\cite{sizov_book},  the spin Faraday and Kerr rotation signals
as well as spin 
ellipticity are determined just by the components of nonequilibrium
spin polarization of electrons, rather than by an external magnetic
field. The signals sensitivity to the electron spin $z$ component is
determined by the selection rules related with the heavy hole
excitation under normal incidence of light.  One can investigate the
dynamics of all spin pseudovector components making use of the light
hole related resonances~\cite{Kosaka2009}. 

The connection between the trion and exciton resonance parameters with
the resident electron spin polarization is analyzed in the following
Subsections. The structure with high density electron gas where 
electron-hole complexes are not stable and the response character
differs from the resonant one, described by expression~\eqref{rpm}, is
also considered below.

Let us establish the link between the light reflection coefficients
and the spin Faraday, Kerr and induced ellipticity signals. Assume that the
probe pulse propagates along the structure normal, i.e.  $z$ axis, and
let its electric field be polarized along $x$. In Faraday effect
studies the probe pulse is split into two ones being linearly
polarized at $\pm 45^{\rm o}$ angles with respect to the initial
polarization. The time integrated difference of intensities of these
pulses as function of the pump-probe delay is measured. Hence, the
spin Faraday rotation signal equals to~\cite{yugova09}
\begin{equation}
\label{spinF}
\mathcal F = \lim_{z\to + \infty} \int_0^{T_{\rm exp}}
\left[\left|E^{(t)}_{x'}(z,t)\right|^2 -
  \left|E^{(t)}_{y'}(z,t)\right|^2 \right]
\mathrm dt.
\end{equation}
Here $x'$, $y'$ axes are oriented at $45^{\rm o}$ with respect to the
initial frame $x$, $y$; $E^{(t)}_{x'}(z,t)$ and $E^{(t)}_{y'}(z,t)$
are the components of the field transmitted through the sample. The
integration in Eq.~\eqref{spinF} is carried out over the measurement
time,  $T_{\rm  exp}$, which exceeds by far all other time constants in the
system. The Kerr effect is studied in the reflection geometry and
its magnitude is defined as
\begin{equation}
\label{spinK}
\mathcal K = \lim_{z\to - \infty} \int_0^{T_{\rm exp}}
\left[\left|E^{(r)}_{x'}(z,t)\right|^2 -
  \left|E^{(r)}_{y'}(z,t)\right|^2 \right]
\mathrm dt,
\end{equation}
where the superscript $r$ indicates, that the fields of reflected wave
enter Eq.~\eqref{spinK}. In pump-probe experiments the induced
ellipticity effect is also measured, which in the transmission
geometry, is described by the following expression
\begin{equation}
\label{spinE}
\mathcal E = \lim_{z\to + \infty} \int_0^{T_{\rm exp}}
\left[\left|E^{(t)}_{\sigma^+}(z,t)\right|^2 -
  \left|E^{(t)}_{\sigma^-}(z,t)\right|^2 \right]
\mathrm dt.
\end{equation}
In this case the difference of transmitted wave circularly polarized
components,
$E_{\sigma^\pm}^{(t)} = (E_x^{(t)}\mp \mathrm i E_y^{(t)})/\sqrt{2}$, is analyzed.

In the single quantum well structure and single layers of quantum dots
the transmission coefficients through the layer, $t_{\pm}(\omega)$,
are related with the reflection coefficients, $r_{\pm}(\omega)$, by
the simple expression
\[
t_{\pm}(\omega) = 1 + r_{\pm}(\omega).
\]
Since in real systems $|r_{\pm}(\omega)| \ll 1$ and
$|r_+(\omega) - r_-(\omega)| \ll |r_{\pm}(\omega)|$ as  a rule, the
spin Faraday and ellipticity signals are described by a simplified formula~\cite{zhu07,fokina-2010}:
\begin{equation}
\label{EF}
\mathcal E + \mathrm i \mathcal F \propto r_{+}(\omega) - r_{-}(\omega).
\end{equation}
The Kerr effect is associated with the light reflection from the sample,
therefore, it is determined by the interference of the beams,
reflected from the sample surface (vacuum-sample boundary) and from
the well or dot layer. The interference brings about an additional
phase equal to  $2qL$, where $L$ is the cap layer thickness (distance
from the boundary with vacuum and the well or dot array) and  $q$ is
the light wave vector in the cap
layer~\cite{springerlink:10.1134/1.1130188}. As a result, 
Kerr signal is related with the reflection coefficients from the
system as~\cite{zhu07}
\begin{equation}
\label{K}
\mathcal K \propto \Im\{\mathrm e^{2\mathrm i qL} [r_{+}(\omega) - r_{-}(\omega)]\}.
\end{equation}
We stress that the description of the spin Kerr and Faraday effects,
as well as the induced ellipticity in the macroscopic approach can be
applied not only for the analysis of experimental data obtained in the
pump-probe technique, but also to study the spin dynamics and
magnetization at the constant wave pumping~\cite{PhysRevB.74.073407},
as well as in quasi equilibrium conditions for magnetic and
superconducting structures~\cite{gourdon:230}.

\subsubsection{Detection at trion resonance}\label{subsec:macro:detect:tr}

\begin{figure}[hptb]
\includegraphics[width=0.45\textwidth]{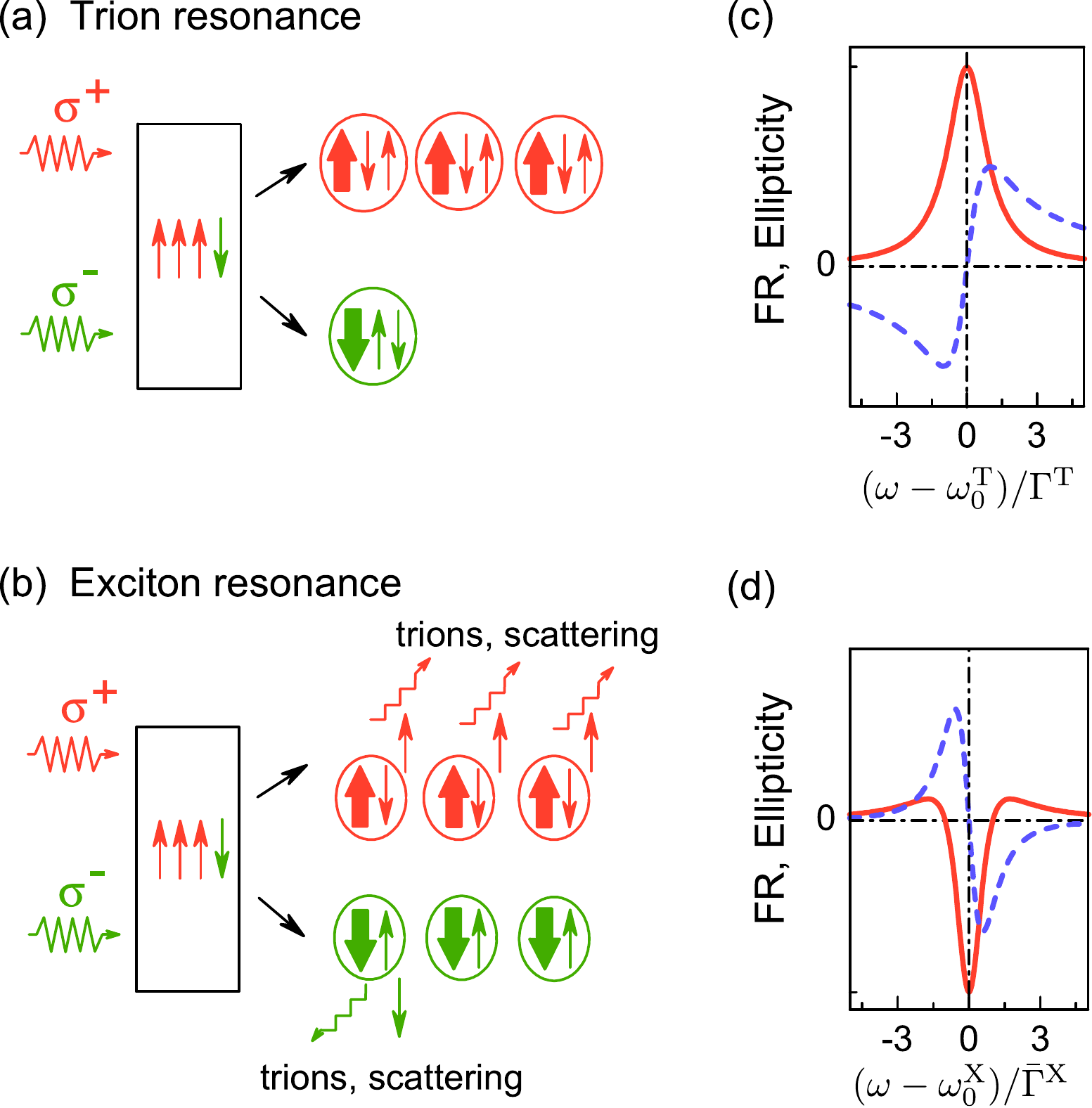}
\caption{(a) Schematic illustration of the trion oscillator strength
  polarization dependence in the spin-polarized electron gas. Right
  circularly polarized component of the probe pulse
  ($\sigma^+$) is absorbed stronger as compared with left circularly
  polarized one
  ($\sigma^-$), since the number of resident electrons with the spin
  projection $+1/2$ onto  $z$ axis is larger than that of electrons
  with spin projection $-1/2$. (b) Illustration of nonradiative
  exciton decay polarization dependence in the spin-polarized electron
  gas. Excitons formed by absorption of  $\sigma^+$ component of the
  probe pulse decay faster than those formed by 
  $\sigma^-$ component absorption, due to more efficient trion
  formation and exchange scattering by electrons. (c) Spectral
  dependence of the Faraday rotation (dashed line) and ellipticity
  (solid line) at trion resonance detection. (d) Same as in panel (c)
  but for exciton resonance detection. Data are reproduced from
  Ref.~\cite{fokina-2010}. }\label{fig:freq:theor} 
\end{figure}

The trion oscillator strength in  $\sigma^+$ and
$\sigma^-$  circular polarizations is directly proportional to the
number of electrons with a given spin projection onto 
$z$ axis,  $N_{\pm 1/2}$~\cite{astakhov00,astakhov02}. Indeed, as
discussed in Sec.~\ref{subsec:macro:orient:tr} the resident electrons
with $s_z=1/2$ take part in the singlet trion formation by $\sigma^+$
polarized light. It is illustrated in
Fig.~\ref{fig:freq:theor}(a). Therefore, the trion radiative damping
being proportional to its oscillator strength can be represented in
the form:
\begin{equation}
\label{Gamma0T}
\Gamma_{0,\pm}^{\rm T} = \alpha_{\rm T} \Gamma_{0}^{\rm X} N_{\pm 1/2}.\nonumber
\end{equation}
Here $\alpha_{\rm T}$ is a constant, $\Gamma_{0}^{\rm X}$ is the
exciton radiative damping. Trion resonance frequency renormalization
$\omega^{\rm T}_0$ due to electron spin polarization and the
modification of its nonradiative damping  $\Gamma^{\rm T}$ are
negligibly small because they are determined by an exchange
interaction of an electron and hole. The trion resonance contribution
to the spin Faraday and ellipticity signals can be written as~\cite{fokina-2010}
\begin{equation}
\label{EFT}
\mathcal E + \mathrm i \mathcal F \propto \frac{\mathrm i \alpha_{\rm
    T} \Gamma_0^{\rm X}(N_{+1/2}-N_{-1/2})}{\omega_0^{\rm T} -
  \omega-\mathrm i \Gamma^{\rm T}}.
\end{equation}
We took into account that $\Gamma_0^{\rm T} \ll \Gamma^{\rm
  T}$ in derivation of Eq.~\eqref{EFT}. Hence, spin ellipticity and
Faraday rotation signals detected 
at the trion resonance are proportional to the total electron spin $z$
component. Frequency dependence of these signals calculated according
to Eq.~\eqref{EFT} is shown in Fig.~\ref{fig:freq:theor}(c). It is
seen, that the Faraday rotation signal is an odd function of the detuning
between the probe pulse and the trion resonance frequencies, while the
ellipticity signal as an even function. Physically, it is related with
the fact that the spin ellipticity is related with the absorption
dichroism which takes its maximum value at the resonance, while
Faraday rotation is related with the refraction of electromagnetic waves
in the medium.

\subsubsection{Detection at exciton resonance}\label{subsec:macro:detect:exciton}

Mechanism of the Faraday, Kerr and ellipticity effects formation at the
exciton resonance detection is different from that at trion resonance
detection. As we noted above, at low temperatures in low density
electron gas the exciton lifetime is determined by the electron
capture and trion formation process. Therefore, the nonradiative
damping of the exciton excited by the light with a given circular
polarization  can be presented as [Fig.~\ref{fig:freq:theor}(b)]
\begin{equation}
\label{GammaX}
\Gamma_{\pm}^{\rm X} = \bar{\Gamma}^{\rm X}
+ \beta_{\rm X} N_{\pm 1/2},\nonumber
\end{equation}
where $\bar{\Gamma}^{\rm X}$ is the spin polarization independent
exciton damping, $\beta_{\rm X}$ is a coefficient determined  by the
trion formation rate. An additional contribution to  $\beta_{\rm X}$
is given by the spin-dependent electron-exciton scattering
processes~\cite{tarasenko98}. An exchange (Hartree-Fock) interaction
of electron-in-exciton and of resident electron may give rise to
the exciton resonance frequencies renormalization, and at high pumping
and probing intensities, the exciton-exciton scattering may play a
role, see, e.g., Refs.~\cite{PhysRevLett.75.2554, wang04,
  shen05, PhysRevB.72.245202, PhysRevB.74.125316,
  PhysRevLett.103.056405}. Restricting ourselves with the low
temperatures and low pulse powers limit we obtain from
Eq.~\eqref{rpm}
\begin{equation}
\label{EFX}
\mathcal E + \mathrm i \mathcal F \propto -\frac{\mathrm i \beta_{\rm
    X} \Gamma_0^{\rm X}(N_{+1/2}-N_{-1/2})}{(\omega_0^{\rm X} -
  \omega-\mathrm i \bar{\Gamma}^{\rm X})^2}.
\end{equation}
Here, similarly to the derivation of Eq.~\eqref{EFT}, we took into
account that $\Gamma_0^{\rm X}
\ll \bar{\Gamma}^{\rm X}$. The frequency dependence of the Faraday
rotation and
ellipticity spin signals is shown in Fig.~\ref{fig:freq:theor}(d).

\begin{figure}[hptb]
\includegraphics[width=0.45\textwidth]{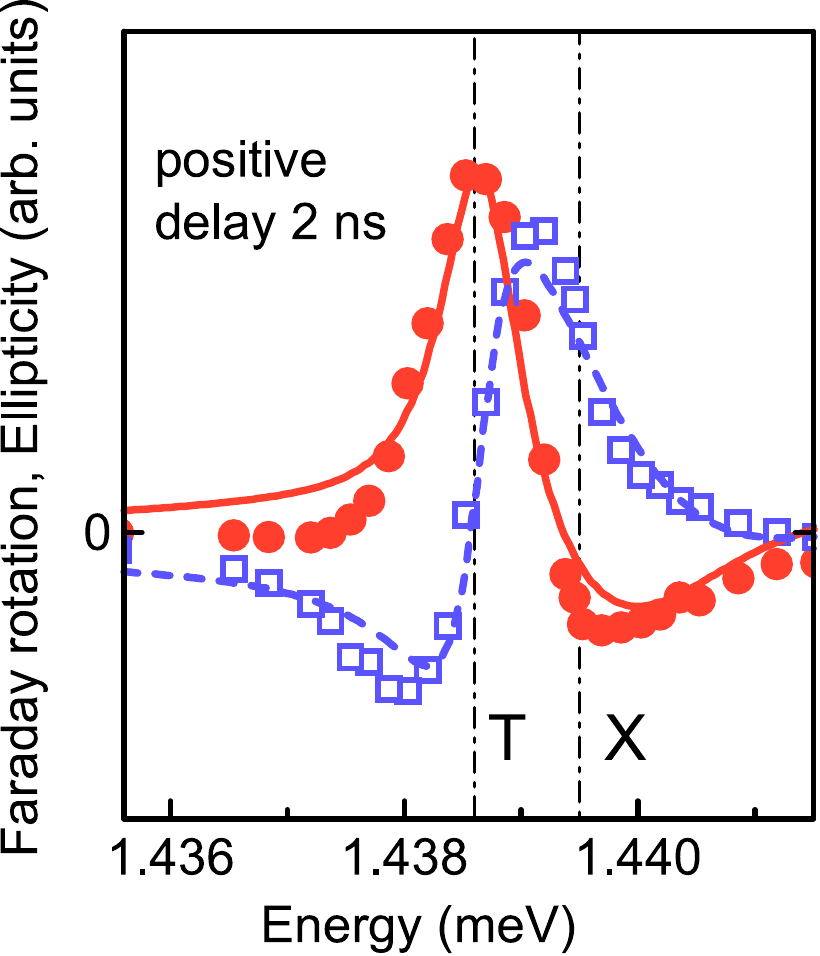}
\caption{Dependence of Faraday rotation (squares) and ellipticity
  (circles) signals on the probe pulse frequency in 
  In$_{0.09}$Ga$_{0.91}$As/GaAs quantum well structure. Quantum well
  width is 8 nm, electron density in the well is  $N\lesssim
  10^{10}$~cm$^{-2}$. Measurements were carried out in the magnetic
  field $B = 0.5$~T at the temperature  $T =
  1.6$~K. Curves are the fits according to Eqs.~\eqref{EFT} and
  \eqref{EFX}. Vertical lines show the positions of trion (T) and
  exciton (X) resonances. Data are reproduced from
  Ref.~\cite{fokina-2010}.}\label{fig:freq}  
\end{figure}

Comparison of Eqs.~\eqref{EFT} and \eqref{EFX} shows that the
ellipticity signs at exciton and trion resonances are
opposite. Indeed, an excess of electrons with the spin projection
$+1/2$ ($N_{+1/2}>N_{-1/2}$) leads to a stronger absorption of
$\sigma^+$ polarized photons at the trion resonance and to a weaker
absorption at the exciton one. Such an ellipticity sign change was
experimentally observed in  
InGaAs/GaAs quantum well
structure~\cite{fokina-2010}. Figure~\ref{fig:freq} represents the
dependence of Faraday rotation signal on the probe pulse frequency
(squares are the experiment, dashed line is the theory) and that of
ellipticity signal (circles are the experiment, solid line is the
theory). The signals were measured at the long enough positive delays
(about 2~ns), which markedly exceed the lifetimes of photoexcited
excitons and trions, therefore these signals correspond to the
resident carrier spin polarization. It is seen that the ellipticity
signals at the trion and exciton resonances have opposite signs and
the whole spectral dependence is rather well described by the
macroscopic theory outlined above. The experiment and calculation
details as well as the fitting parameters values are presented in
Ref.~\cite{fokina-2010}.

We note that a sign change of the spin signals is also observed if
one uses resonances provided by the electron-light-hole complexes
instead of those with the heavy hole~\cite{ast08,Chen20101803}. Such a
sign change is related with the change of the selection rules for
optical interband transitions.

\subsubsection{Quantum wells with high density electron gas}\label{subsec:macro:detect:high}
 
In structures containing the electron gas of high density, where
$Na_B^2\gg 1$, or at high temperatures, the excitons and trions are
unstable, and the light absorption is accompanied by free
electron-hole pairs formation. In this case Eq.~\eqref{rpm} is
unapplicable. The main mechanism of spin Faraday, Kerr 
and ellipticity signals formation in these systems is the
polarization-dependent blocking of optical transitions caused by the
filling of electron spin states. We make use of the general expression
for the doped quantum well reflection coefficient (see e.g.,
Ref.~\cite{PhysRevB.76.045320}), which in the case of small  $|r_\pm|$
value and for negligibly small Coulomb interaction takes form
\begin{equation}
\label{rqw}
r_\pm = \mathrm i Q
\int \frac{\mathrm d \bm k}{(2\pi)^2} \frac{1-f_{\mp 1/2}(\bm k)}{E_g+\frac{\hbar^2k^2}{2\mu}-\hbar\omega-\mathrm i \hbar \Gamma_{\rm eh}}.
\end{equation}
Here  $Q>0$ is a constant introduced in Eq.~(10) of
Ref.~\cite{PhysRevB.76.045320},  $f_{\pm 1/2}(\bm k)$ are the
distribution functions of the carriers with spin components $\pm 1/2$,
$\mu=m_em_h/(m_e+m_h)$ is the electron-hole reduced mass ($m_e$ is the
electron effective mass,  $m_h$ is that of the hole), $E_g$ is the
effective band gap found with allowance for electron and hole size
quantization, $\Gamma_{\rm eh}$ is the nonradiative damping of
electron-hole pair. Assuming electron spin polarization to be small,
one can obtain (c.f.~\cite{zhu07})
\begin{equation}
\label{EFdens}
\mathcal E + \mathrm i \mathcal F \propto 2\mathrm i Q
\int \frac{\mathrm d \bm k}{(2\pi)^2} \frac{s_{z}(\bm k)}{E_g+\frac{\hbar^2k^2}{2\mu}-\hbar\omega-\mathrm i \hbar \Gamma_{\rm eh}},
\end{equation}
where $s_z(\bm k) = [f_{1/2}(\bm k) -
f_{-1/2}(\bm k)]/2$ is the distribution function of
the electron spin $z$ component.
It follows from Eq.~\eqref{EFdens}  that for degenerate electrons where the Fermi energy $E_F$, temperature
$T$, expressed in the energy units, and nonradiative damping satisfy
the conditions
$T \ll   \hbar \Gamma_{\rm eh} \ll E_F$
\begin{equation}
\label{EFdens1}
\mathcal E + \mathrm i \mathcal F \propto \frac{2\mathrm i Q
S_{z}}{E_0-\hbar\omega-\mathrm i \hbar \Gamma_{\rm eh}},
\end{equation}
where $E_0 = E_g +E_F(1+m_e/m_h)$ is the absorption edge energy, $S_z =
(2\pi)^{-2}\int s_z(\bm k) \mathrm d \bm k$ is the electron spin density.
The main contribution to the spin signals in provided by Fermi level
electrons which gives rise to the resonant character of reflection
coefficients. Moreover, it is the imbalance of electrons with
$s_z=+1/2$ and $-1/2$ which determines the difference of  $\sigma^+$
and $\sigma^-$ probe pulse components interaction efficiencies with
the system. For instance, if $N_{1/2}>N_{-1/2}$, then
$\sigma^+$ component of the probe pulse is absorbed better than 
$\sigma^-$ one. As a result, the spectral dependence of the spin
Faraday and ellipticity signals in dense electron gas is analogous to
that observed at a trion resonance.

Let us also analyze the contribution to the Faraday and ellipticity
signals made by the shifts of electron energy levels in the
spin-polarized electron gas due to exchange interaction (Hartree-Fock
effect). The relative shift of optical transition energies in
$\sigma^+$ and $\sigma^-$ polarizations can be recast as~\cite{PhysRevB.60.4826}
\[
2\sum_{\bm k'} V_{\bm k'-\bm k} s_z(\bm k'),
\]
where $V_{\bm k}$ is the Fourier transform of the Coulomb
 potential of interaction between charge carriers.
 Hartree-Fock
 contribution to the ellipticity and Faraday rotation spin signals
 writes in the form
\begin{equation}
\label{rpm:HF}
\mathcal E + \mathrm i \mathcal F \propto 
\end{equation}
\[
-2{\mathrm i Q}\int \frac{\mathrm d \bm k}{(2\pi)^2}
\frac{1-f(\bm k)}{(E_g+\frac{\hbar^2k^2}{2\mu}-\hbar\omega-\mathrm i
  \hbar \Gamma_{\rm eh})^2} \sum_{\bm k'} V_{\bm k- \bm k'} s_{z}(\bm k').
\]
Calculation shows that for degenerate electrons and low degree of spin
polarization
\begin{equation}
\label{EFHF}
\mathcal E + \mathrm i \mathcal F \propto -\frac{2 \mathrm  i Q S_z }{E_0-\hbar\omega-\mathrm i \hbar \Gamma_{\rm eh}}   \frac{F(r_s)}{1+m_e/m_h}.
\end{equation}
Here the gas parameter $r_s=\sqrt{2}m_ee^2/(\ae\hbar^2k_F)\sim
1/(Na_B^2)$ in introduced,
$k_F$ in the wave vector of electron at the Fermi surface, $\ae$ is
the static dielectric constant. Function 
$F$ is defined as follows~\cite{PhysRevB.60.4826}:
\[
F(r_s) = \frac{r_s}{\pi\sqrt{|2-r_s^2|}} \left\{
\begin{array}{cc}
\Arch(\sqrt{2}/r_s), & r_s\leqslant\sqrt{2} \\
\arccos(\sqrt{2}/r_s), & r_s>\sqrt{2}
\end{array}
\right..
\]
It is important to note that in quantum well structures the
contributions caused by the optical transitions blocking and the
exchange interaction differ by the common factor and sign only. In a
high-density electron gas $r_s \ll 1$, and Hartree-Fock contribution,
Eq.~\eqref{EFHF}, is small as compared with the contribution from the
transitions blocking described by Eq.~\eqref{EFdens1}. In real
structures, however, 
$r_s$ can be on the order of unity and both effects can make
comparable contributions to the spin Kerr, Faraday and ellipticity
signals. Experimental investigations of many-body effects in the
pump-probe method and, in particular, the renormalizations caused by
the exchange interaction of electrons were carried out in
Ref.~\cite{Barate2010}.

\subsubsection{Dynamics of electron and magnetic ion spins in 
CdMnTe quantum wells}\label{subsec:macro:detect:Mn}

\begin{figure}[hptb]
\includegraphics[width=0.45\textwidth]{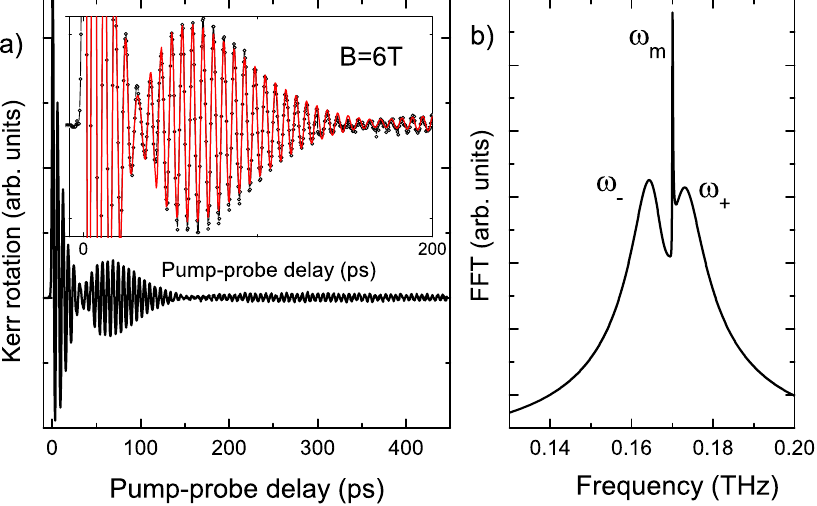}
\caption{(а) The dependence of Kerr rotation signal on time delay
  between pump and probe pulses in the structure with the diluted
  magnetic quantum well CdMnTe. Magnetic ion fraction is $x=0.002$,
  electron density $N=7\times
  10^{10}$~cm$^{-2}$, well width is 10~nm. (b) Fourier
  spectrum of spin beats. Magnetic field $B=6$~T is applied in the
  well plane. Data 
  are reproduced from Ref.~\cite{PhysRevB.78.081305}.}\label{fig:mn}
\end{figure}

In the presence of several spin subsystems: electron, hole and
nuclear, the
spin Faraday and Kerr rotation signals, as well ellipticity bear
information of spin dynamics in all subsystems. The case where one of
the subsystems does not directly interact with light but gives rise to
the spin signals via interaction with electrons is of special
interest. Possible realizations of this situation are the Faraday rotation
induced by spin polarized nuclei considered theoretically for bulk
semiconductors in Ref.~\cite{artemova85}
(see also Refs.~\cite{J.M.Kikkawa01212000,PhysRevLett.99.056804}),
and the Kerr rotation induced by the Manganese ions magnetization in
diluted magnetic structures CdMnTe,
ZnMnSe~\cite{PhysRevLett.55.1128,PhysRevB.50.7689,PhysRevB.56.7574,PhysRevB.78.081305}.  

Experimental dependence of the Kerr rotation signal as function of the
temporal delay between the pump and probe pulses obtained in
Ref.~\cite{PhysRevB.78.081305} in diluted magnetic CdMnTe quantum
well structure are presented in Fig.~\ref{fig:mn}(a). Spin signal
shows a complex temporal dependence whose Fourier analysis presented in
Fig.~\ref{fig:mn}(b) evidences the presence of three spin precession
frequencies. According to the results of
Ref.~\cite{PhysRevB.78.081305} two of these frequencies
$\omega_+$ and $\omega_-$ correspond to the mixed (collective) spin
precession modes of electrons and magnetic ions. In fact, the exchange
interaction between the resident electrons and  $d$-electrons of Mn
ions results in the coherent energy transfer between them and to the
spin beats at a difference frequency $\omega_+-\omega_-$, which are
clearly visible in Fig.~\ref{fig:mn}(a). Spin signals from the mixed
modes decay during the electron spin relaxation time being on the
order of 100~ps. These contributions to the Kerr signal are
proportional, according to Eqs.~ \eqref{EFdens1}, \eqref{EFHF}, to the
spin polarization of electrons in the corresponding spin precession
mode.

Third narrow peak in the spin beats Fourier spectrum at a frequency
$\omega_m$ corresponds to those Mn spin precession modes which are not
coupled with conduction electrons. The modes in question are excited
by the pump pulses and contribute to the Kerr rotation signal owing to
the exchange interaction with holes. Indeed, spin polarized
photocreated hole generates an effective magnetic field oriented along
$z$ axis, which results in the momentary tilt of Mn spins from the
external magnetic field direction and ``triggers'' their
precession~\cite{CrookerPRL96}. The polarization of magnetic ions, on
the other hand, results in the change of optical transition energies
in circular polarizations $\sigma^+$ and
$\sigma^-$, being proportional to  $\pm A_h M_z$, where $M_z$ is the
magnetic ions magnetization and $A_h$ is the effective interaction
constant of the exchange interaction between the Mn and the hole
averaged over the distribution of magnetic ions and hole
wavefunction~\cite{PhysRevB.50.7689}. It can be shown, that the
dependence of the spin signals on the probe pulse frequency in
sufficiently dense electron gas is determined by the expressions
analogous to formulae \eqref{K} and \eqref{EFHF}
\begin{equation}
\label{EFKHF}
\mathcal E + \mathrm i \mathcal F \propto \frac{\mathrm iQ A_h
  M_z}{E_0-\hbar\omega - 
  \mathrm i \hbar\Gamma_{\rm eh}}, \quad \mathcal K = \Im\{\mathrm
e^{2\mathrm i q L} (\mathcal E + \mathrm i \mathcal F)\}.
\end{equation}
The damping time of this contribution to the spin signal is
considerably larger than that of collective modes since the spin
relaxation of Mn ions, which are uncoupled from electrons, is slow.

We note in conclusion of this Subsection that the spin precession of
magnetic ions is directly visible from the pump-probe signals, see long
living signal in Fig.~\ref{fig:mn}(a), because $g$-factor of Mn is
close to 2, and the coupling constants of electron and holes with
$d$-electrons of Mn ions are relatively large. Another situation is 
realized for the nuclei: in the pump-probe experiments the nuclear
spin precession is not seen, while dynamic nuclear polarization
is observed as the Overhauser shift of electron spin precession
frequency~\cite{Li2010450}. 

\section{Microscopic description}\label{sec:micro}

The models of spin coherence excitation and detection set out in
Sec.~\ref{sec:macro} describe successfully qualitative specifics of
spin signals obtained in the pump-probe method in quantum well and
quantum dot array structures. In this Section the microscopic theory
of resident charge carriers spin orientation by short pulses in
quantum dots is put forward. This theory is based on a model of a two
level system. We show that optical pulses do more than merely excite
spin polarization of electrons: the pulses can modify the existing
spin. We also review experimental advances on spin coherence
control. In addition, the microscopic description of spin coherence
detection within the framework of developed two level model is
given here. The stated theory is also applicable with certain
restrictions for the quantum well structures where electrons are
localized, e.g., at interface fluctuations.

\subsection{Two level model for the description of trion resonance
  excitation}\label{subsec:micro:twolevel} 

Consider a planar array of singly charged quantum dots grown from the
zinc blende lattice material along the axis
$z\parallel [001]$. Quantum dot states can be conveniently described
by a four component wave function~\cite{yugova09}
\begin{equation}
 \label{wave}
\Psi = \left[
\psi_{1/2},
\psi_{-1/2},
\psi_{3/2},
\psi_{-3/2}
\right]\:,
\end{equation}
where subscripts $\pm 1/2$ refer to the resident electron spin states
and subscripts $\pm 3/2$ to the photocreated trion states. Electron
spin components are expressed as quantum mechanical averages of the
spin operator $\hat{\bm s} =
(\sigma_x,\sigma_y,\sigma_z)/2$, where  $\sigma_i$ ($i=x,y,z$) are the
Pauli matrices, in the form:
\[
S_z=\left(|\psi_{1/2}|^2-|\psi_{-1/2}|^2\right)/2,\]
\begin{equation}
S_x=\Re(\psi_{1/2}\psi_{-1/2}^*),\quad
~S_y=-\Im(\psi_{1/2}\psi_{-1/2}^*)\:. \label{eq:12} 
\end{equation}
We neglect all other excited states of the system (e.g., triplet trion
states) in the quantum dot description by means of the wave
function Eq.~\eqref{wave}. We assume that the pump pulse duration $\tau_p$
is long enough compared to the field oscillations period at a carrier
(optical) frequency of electromagnetic wave, which we denote as
$\omega_{ \mbox{}_{\rm P}}$, that is $\tau_p \gg 2\pi/\omega_{
  \mbox{}_{\rm P}}$, but it is short enough compared with the electron
spin precession period in the quantum dot subject to an external
magnetic field, $\tau_p \ll 2\pi/\Omega$, and compared to the trion
lifetime in the quantum dot, $\tau_p \ll \tau_r^T$. These relations
between the time scales are typical for pump-probe experiments.

Since  $\sigma^+$ polarized light pulse induces the transition from
the quantum dot state corresponding to the resident electron with the
spin projection $s_z=+1/2$ to the trion state with the hole spin
projection $+3/2$, and $\sigma^-$ polarized pulse couples the states
$-1/2$ and $-3/2$, it is sufficient to limit ourselves with a pair of
states: $[\psi_{1/2},\psi_{3/2}]$ or $[\psi_{-1/2},\psi_{-3/2}]$,
i.e. by a two level model, in order to describe the interaction of the
pump pulse of given circular polarization. Equations describing the
quantum dot wave function under the optical pulse action can be
recast as
\begin{eqnarray}
 \label{system:1}
&&\mathrm i \hbar \dot{\psi}_{3/2} =\hbar \omega_0^{\mathrm T}
\psi_{3/2} + V_+(t) \psi_{1/2} \:, 
\\ &&\mathrm i \hbar \dot{\psi}_{1/2}= V_+^*(t) \psi_{3/2}\:,
\nonumber
\\  \label{system:2}
&&\mathrm i \hbar \dot{\psi}_{-3/2} = \hbar \omega_0^{\mathrm T}
\psi_{-3/2} + V_-(t) \psi_{-1/2} 
\:,\\
&&\mathrm i \hbar \dot{\psi}_{-1/2} = V_-^*(t) \psi_{-3/2}\:.\nonumber
\end{eqnarray}
Here $\dot{\psi} \equiv \mathrm d \psi/\mathrm d t$, and time
dependent matrix elements  
\begin{equation}
\label{VPM}
V_\pm (t) = -\int \mathsf d(\bm
r)E_{\sigma^\pm}(\bm r,t)\mathrm d\bm r
\end{equation}
describe the interaction of the circularly polarized components of the
incident field $E_{\sigma^\pm} =
(E_x \mp \mathrm i E_y)/\sqrt{2} \propto e^{-\mathrm i \omega_{\rm P}
  t}$ with the quantum dot. Function $\mathsf d(\bm r)$ in
Eq.~\eqref{VPM} is an
effective dipole moment of the transition defined in
Ref.~\cite{yugova09}. We stress again that the absence of
contributions responsible for the electron spin precession in Eqs.~(\ref{system:1}),
(\ref{system:2})  is related with the small duration of the pump
pulse. Under the condition $\tau_p\ll \tau_r^{T}$ the quantum mechanical
description is still valid, and for 
$\tau_p \gtrsim 1/\Omega$ one has to allow for the influence of an
external field in Eqs.~\eqref{system:1},
\eqref{system:2}, Ref.~\cite{bayer_long}. However, if $\tau_p \gtrsim
\tau_r^{T}$, then such a pump pulse can be considered as a
quasistationary one (see also Ref.~\cite{PhysRevB.83.033301}).

Electron spin dynamics in  the framework of the given system of
equations was discussed in Ref.~\cite{greilich06} in the case of the
``rectangular'' pulse whose carrier frequency coincides with the trion
resonance frequency,
$\omega_{ \mbox{}_{\rm P}}=\omega_0^{\rm T}$. The solutions of
Eqs.~\eqref{system:1},  
\eqref{system:2} for arbitrary pulse shape and its carrier frequency
detuning from the trion resonance were analyzed in
Ref.~\cite{yugova09}, a discussion of the effects related with the
triplet trion excitation is given in Ref.~\cite{zhukov10}.

Let us consider in more detail the interaction of the quantum dot with
the $\sigma^+$ polarized pulse. Under the experimental conditions,
Refs.~\cite{greilich06,A.Greilich07212006,carter:167403}, the pumping
is carried out by a train of the pulses following with a repetition
period $T_R\sim 10$~ns. Usually, this period exceeds by far the trion
lifetime, $T_R \gg \tau_r^T$, hence the quantum dot is in the ground
state at the moment of the next pulse arrival:
$\psi_{3/2} = \psi_{-3/2}=0$. However, time $T_R$ is, as a rule,
smaller than the relaxation time of the localized
electron spin~\cite{dzhioev97:eng,Kikkawa98,A.Greilich07212006}, hence by the
pulse arrival electron can be spin polarized. It follows from
Eqs.~\eqref{system:2}, that  $\psi_{-1/2}(t)=\mbox{const}$, while
system~\eqref{system:1} can be rewritten as a single equation:
\begin{equation} \label{single}
\ddot{\psi}_{1/2} - \left( \mathrm i \omega'+ \frac{\dot{f}(t)}{f(t)}
\right) \dot{\psi}_{1/2} + f^2(t) \psi_{1/2} =0\:.
\end{equation}
Here $\omega' = \omega_{ \mbox{}_{\rm P}} - \omega_0^{\mathrm T}$ is
the detuning between the pump carrier frequency and the trion
resonance frequency and 
$f(t)$ is the pump pulse smooth envelope defined as
\[
f(t) = -\frac{\mathrm e^{\mathrm i \omega_{\mbox{}_{\rm P}}
    t}}{\hbar}\int \mathsf d(\bm r) 
E_{\sigma_+}(\bm r,t)\mathrm d^3 r\:.
\]
It follows from Eq.~\eqref{single}, that the values of
$\psi_{1/2}(-\infty)$, i.e. before the pump pulse arrival, and the
values 
$\psi_{1/2}(\infty)$ (after the pump pulse arrival) are connected
linearly. This linear relation can be in general written as
\begin{equation} \label{an}
\psi_{1/2}(\infty) = Q \mathrm e^{\mathrm i \Phi}\psi_{1/2}(-\infty)\:,
\end{equation}
where real coefficient $Q$ satisfies the condition $0 \leqslant
Q \leqslant 1$, and phase  $\Phi$ can be chosen in the interval from  $-
\pi$ to $\pi$. These parameters are determined by the pulse shape, its
duration and the detuning from the resonance frequency. Making use of
Eqs.~(\ref{eq:12}) and  (\ref{an}) we relate the electron spin values
before the pump pulse arrival, $\bm S^- = (S_x^-,S_y^-,S_z^-)$, and
right after the pump pulse arrival, $\bm S^+ = (S_x^+,S_y^+,S_z^+)$ as:
\begin{subequations} \label{pm}
\begin{eqnarray}
S_z^+ &=&   \frac{Q^2-1}{4} +\frac{Q^2+1}{2}S_z^-\:, \label{szpm}\\
S_x^+ &=& Q\cos{\Phi} S_x^- +Q\sin{\Phi} S_y^-\:,\label{sxpm} \\
S_y^+ &=& Q\cos{\Phi} S_y^- - Q\sin{\Phi} S_x^-\:.\label{sypm}
\end{eqnarray}
\end{subequations}
System of equations~\eqref{pm} describes the electron spin orientation
and transformation by a short optical pulse in the quantum dot. One
can check that the trion spin polarization defined as $J_z
= (|\psi_{3/2}(\infty)|^2 - |\psi_{-3/2}(\infty)|^2)/2$ right after
the pulse arrival equals to
\begin{equation} \label{Jz}
J_z = S_z^- - S_z^+\:.
\end{equation}
The transformation of electron spin under the action of left
circularly polarized, $\sigma^-$, pump pulse is described by the
analogous set of equations. In such a case, in the first term of
Eq.~(\ref{szpm}) one has to change the sign, and in Eqs.~\eqref{sxpm},
\eqref{sypm} $\Phi$ should be replaced by $- \Phi$.

Expressions~\eqref{pm}, \eqref{Jz} represent a quantum mechanical
generalization of the initial conditions Eq.~\eqref{init:class} for the
coupled electron and trion spin dynamics
Equations~\eqref{system:kin}. It is seen from Eqs.~\eqref{pm} that
$\sigma^+$ pulse changes spin $z$ component by 
$S_z^+-S_z^-=(Q^2-1)(1 + 2 S_z^-)/4$. The pulse also leads to the
rotation of electron spin in the structure plane at  $\Phi\neq 0$ [see
Eqs.~\eqref{sxpm} and \eqref{sypm}]. We come to the question of spin
control later, now we discuss the dependence of the resident electron spin
orientation efficiency on the pump pulse power.

\begin{figure}[hptb]
\includegraphics[width=0.45\textwidth]{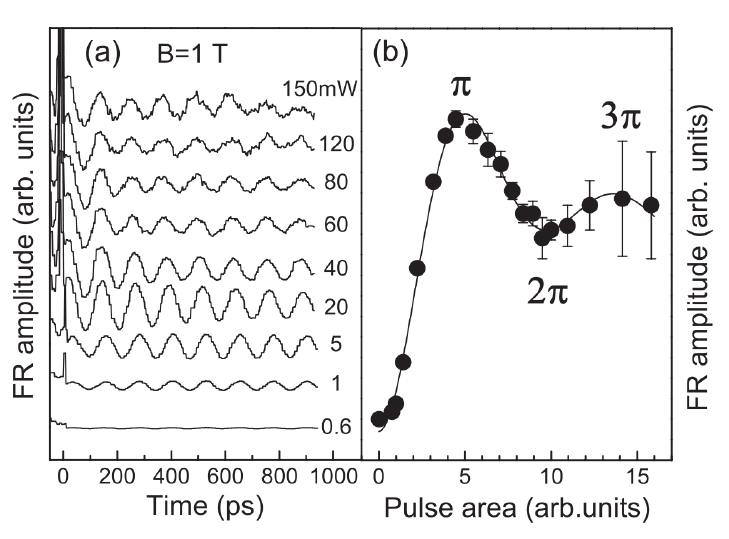}
\caption{(a) Faraday signal dependence on the delay between the pump
  and probe pulses measured in the  InGaAs/GaAs $n$-type quantum dot
  array (20 layers with dot density per layer of  $10^{10}$~cm$^{-2}$,
  structure is doped in a such a way that there is one electron per
  dot on average) at  $B=1$~T. Different curves correspond to the
  different pump pulse powers. (b) Spin beats amplitude dependence on
  the pump pulse area obtained by the scaling of the amplitudes power
  dependence  in panel (a). Data are reproduced from
  Ref.~\cite{greilich06}.}\label{fig:rabi} 
\end{figure}

To begin with, we consider a pulse whose carrier frequency is in
resonance with a trion transition. In these conditions~\cite{greilich06}
\begin{equation}
 \psi_{1/2}(t) = \psi_{1/2}(-\infty)
\cos{\left[\int_{-\infty}^t f(t')\mathrm dt'\right]},
\end{equation}
therefore quantities $Q$ and $\Phi$ in Eq.~\eqref{an} equal to
\begin{equation}
\label{PhiQ}
\Phi\equiv 0, \quad Q
= \cos{(\Theta/2)}, 
\end{equation}
where
\begin{equation} \Theta = 2\int_{-\infty}^\infty
f(t')\mathrm dt' 
\end{equation}
is an effective pulse area. It follows from Eqs.~\eqref{PhiQ} and
\eqref{pm}, that electron spin excited by a single pulse depends
periodically on the pulse area $\Theta$, i.e. on the field amplitude
in the pulse. The power dependence of the electron spin should also
have an oscillatory character typical for two level systems (Rabi
effect)~\cite{ll3_eng}: strong enough pulse does not merely transfer the
system from the ground to the excited state, but can also transfer it
from the excited state to the ground one. It is demonstrated in
Fig.~\ref{fig:rabi}, where the spin beats signals and their amplitudes
measured in the InGaAs/GaAs quantum dot arrays are shown as function
of the pump pulse power.

\begin{figure}[hptb]
\includegraphics[width=0.75\linewidth]{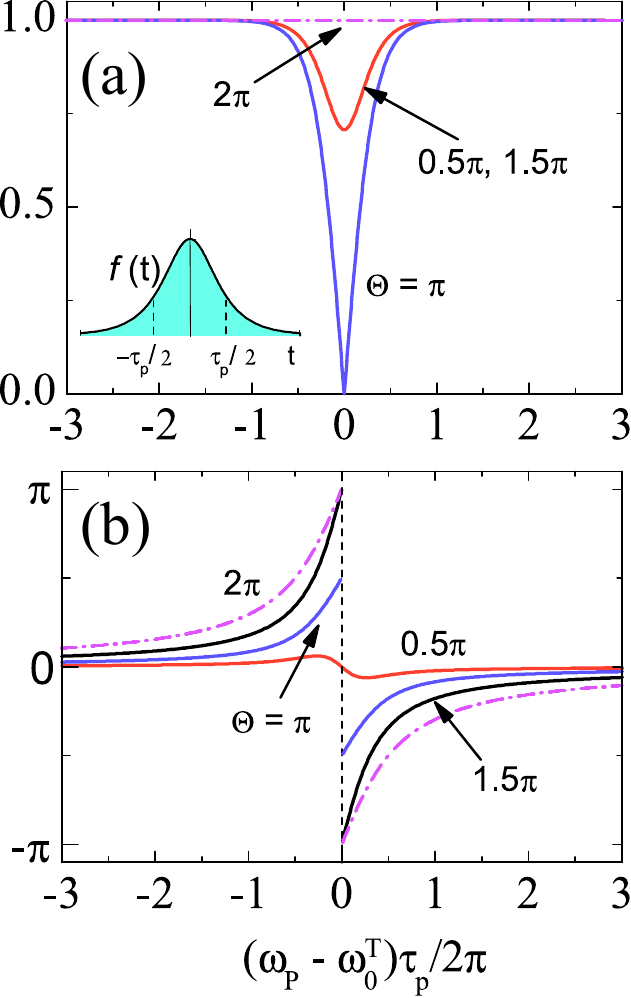}
\caption{Dependence of the parameters $Q$ (a) and  $\Phi$  (b), see
  Eqs.~\eqref{an}, \eqref{pm}, on the detuning $y=(\omega_{
    \mbox{}_{\rm P}}-\omega_0^{\mathrm T})\tau_p/2\pi$ for the Rosen \&
  Zener pulse~\eqref{pulse}, calculated for different pulse areas
  $\Theta = \pi/2, \pi, 
  3\pi/2, 2\pi$.  Inset illustrates the pulse shape. Data are
  reproduced from Ref.~\cite{yugova09}.}
\label{fig:q_phi}
\end{figure}

To conclude this Subsection we discuss briefly the dependence of
quantities   $Q$ and 
$\Phi$ on the detuned pulse parameters. The detailed analysis is given
in Ref.~\cite{yugova09}, here we just dwell on the case of the Rosen
\& Zener pulse~\cite{PhysRev.40.502}:
\begin{equation}\label{pulse}
f(t) =  \frac{\mu}{\cosh{(\pi t/\tau_p)}},
\end{equation}
where the coefficient  $\mu$ characterizes the amplitude of the field
in the maximum of the pulse. The effective pulse area $\Theta$ is  $2
\mu \tau_p$. Solution of Eq.~\eqref{single} for this pulse can be
expressed following Ref.~{\cite{PhysRev.40.502}}, via hypergeometric function
 \begin{equation}\label{sol}
 \psi_{1/2}(t) = \psi_{1/2}(-\infty) \times
\end{equation}
\[
\hF\left[\frac{\Theta}{2\pi},-\frac{\Theta}{2\pi};
 \frac{1}{2}- \mathrm i y;\frac{1}{2}\tanh{\left(\frac{\pi
       t}{\tau_p}\right)} +\frac{1}{2}  \right], 
\]
where the dimensionless detuning from the resonance is $y=
\omega'\tau_p/(2\pi)$. Explicit expressions for $Q$ and $\Phi$ are
given in Ref.~\cite{yugova09}, their dependence on the detuning for
different areas $\Theta$ is shown in Fig.~\ref{fig:q_phi}. It is seen
from the Figure that for  relatively large detunings, 
$|y| \gg 1$, value of 
$Q$ is close to $1$, while 
$\Phi$ tends to $0$. Therefore, detuned pulses do not affect the
quantum dot state. At  $\Theta =
\pi$, function $Q(y)$ has a sharp dip at $y=0$. Such a pulse makes in
plane spin components
$S_{x}^+$ and $S_{y}^+$ zero, and results in the most efficient spin $z$
component generation.

\subsection{Electron spin control by short optical pulses}\label{subsec:micro:control}

As follows from Eqs.~\eqref{pm}, the circularly polarized pulse does
not only generate spin polarization in the quantum dot, but transforms also
the spin, which is already present in the system. For example, pulses with
$Q=0$ erase completely the electron spin components in the structure
plane, it results in the electron spin alignment along  $z$ axis. On
the contrary, detuned pulses with $Q=1$ accomplish the spin rotation
in the structure plane by the angle $\Phi$. Physically, the spin
rotation by the circularly polarized pulse can be interpreted as an
inverse Faraday effect~\cite{PhysRev.143.574}: circularly polarized
pulse induces the splitting of electron spin sublevels with spin
projections $\pm
1/2$ onto $z$ axis. This splitting is equivalent to an effective
magnetic field directed along $z$ axis and results in the spin
components rotation in the $(xy)$ plane. 

In that way, there is a possibility to control electron spins by short
optical pulses~\cite{economou:205415}. Corresponding experiments are
carried out in the three pulse method: first pulse orients resident
electrons by spin, second one serves to control spins and the third
pulse is used for the spin polarization detection. Experimentally spin
rotation by polarized pulses was demonstrated in
Refs.~\cite{phelps:237402} and~\cite{Carter:07} for  CdTe and GaAs
quantum well structures, respectively, and for GaAs quantum dot
structures~\cite{Bonadeo20111998,J.Berezovsky04182008, 
  Greilich2009,Kim2011}. A detailed analysis of the spin polarization
control mechanisms is presented in
Refs.~\cite{economou:205415,Takagahara:10}, while the experimental
achievements are reviewed in Ref.~\cite{Ramsay}. The processes of
optical magnetization control in magnetic media have also attracted a
considerable interest in recent
years~\cite{PhysRevLett.98.047403,PhysRevLett.103.117201}. 

\begin{figure}[hptb]
\includegraphics[width=0.75\linewidth]{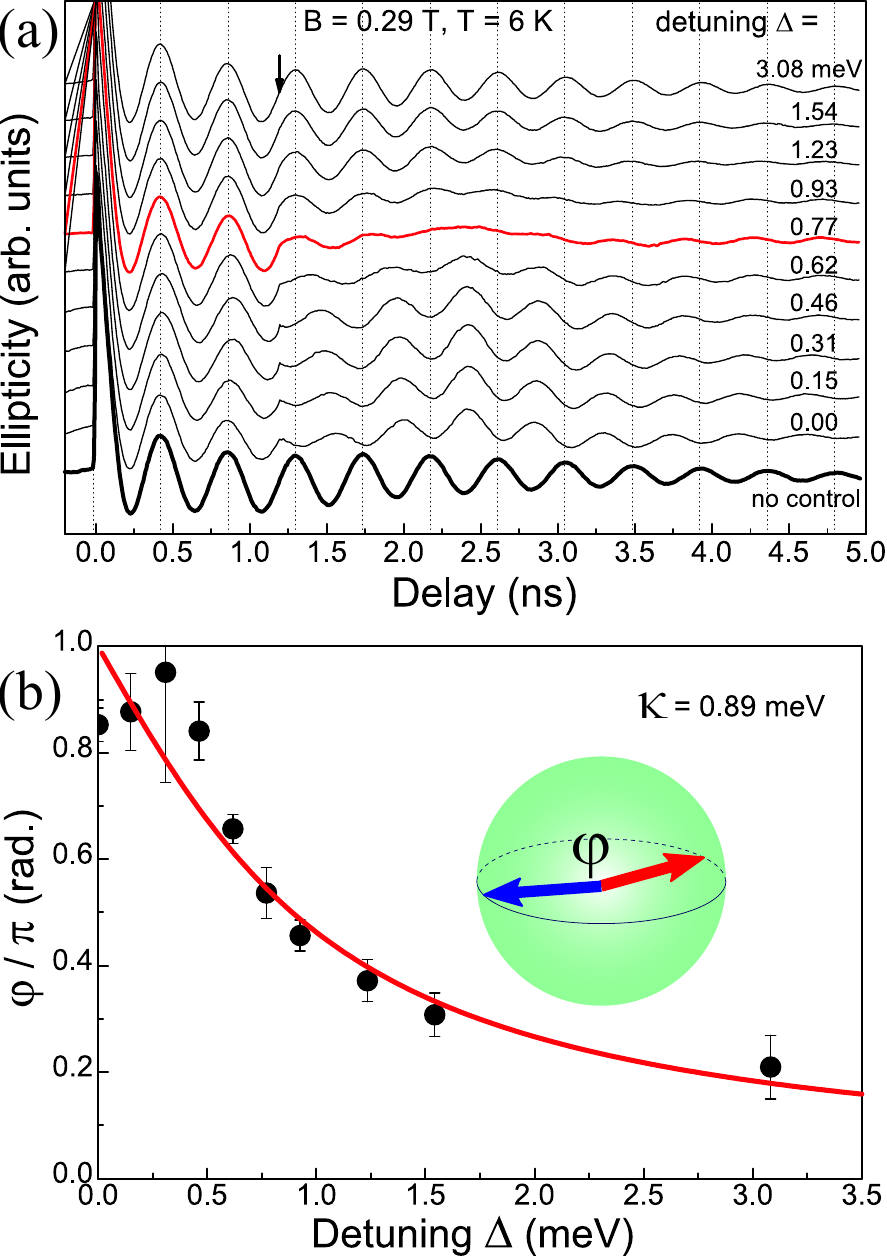}
\caption{(a) Spin ellipticity signals as functions of the temporal
  delay between the pump and probe pulses. Arrow marks the time moment
  of the control pulse arrival (it corresponds to the electron spin in
  the structure plane). Different curves correspond to different
  detunings between the pump pulse and control pulse (indicated above
  curves). (b) Spin rotation angle in the $(xy)$ plane under the
  circularly polarized control pulse action as a function of
  detuning. Points are obtained on the basis of curves shown in panel
  (a), solid curve is the theory~\cite{economou:205415}. Parameter
  $\kappa = \hbar/\tau_p$ is the spectral width of the control
  pulse. An inset shows schematically the spin rotation in the $(xy)$
  plane. The measurements were carried out on the structure with 20
  InGaAs/GaAs quantum dot layers, dot density per layer is
  $10^{10}$~cm$^{-2}$, the structure is doped in a such a way that
  there is one electron per dot on average. Data are reproduced
  from~\cite{Greilich2009}.} 
\label{fig:rotation}
\end{figure}

Spin ellipticity signals measured on InGaAs/GaAs quantum dot array in
Ref.~\cite{Greilich2009} by means of the three pulse method:
pump-control-probe, as functions of time delay between the pump and probe
pulses are shown in Fig.~\ref{fig:rotation}(a). The time moment of the
circularly polarized control pulse arrival is marked by the arrow, it
corresponds to the situation where the electron spin lies in $(xy)$
plane, and different curves correspond to different spectral detunings
between the control pulse and pump pulse. Control pulse power is
adjusted in such a way that its area is close to 
$2\pi$. It is clearly seen that the spin beats amplitude changes after
the control pulse arrival. It corresponds to the spin rotation in the
$(xy)$ plane around $z$ axis, as shown schematically in the inset to
Fig.~\ref{fig:rotation}(b). Figure~\ref{fig:rotation}(b) presents the
dependence of the spin
rotation angle in the structure plane, measured in the units of $\pi$,
on the detuning between the carrier frequency of the pulse and the
trion resonance frequency in quantum dots. Points show the experimental
data obtained from the spin beats analysis presented in
Fug,~\ref{fig:rotation}(a), the 
solid curve is the theoretical calculation in the framework of the
model developed in Ref.~\cite{economou:205415}. Qualitatively the spin
rotation angle dependence is in the agreement with the dependence of  $\Phi$
on the detuning presented in Fig.~\ref{fig:q_phi}(b).

\begin{figure}[hptb]
\includegraphics[width=0.75\linewidth]{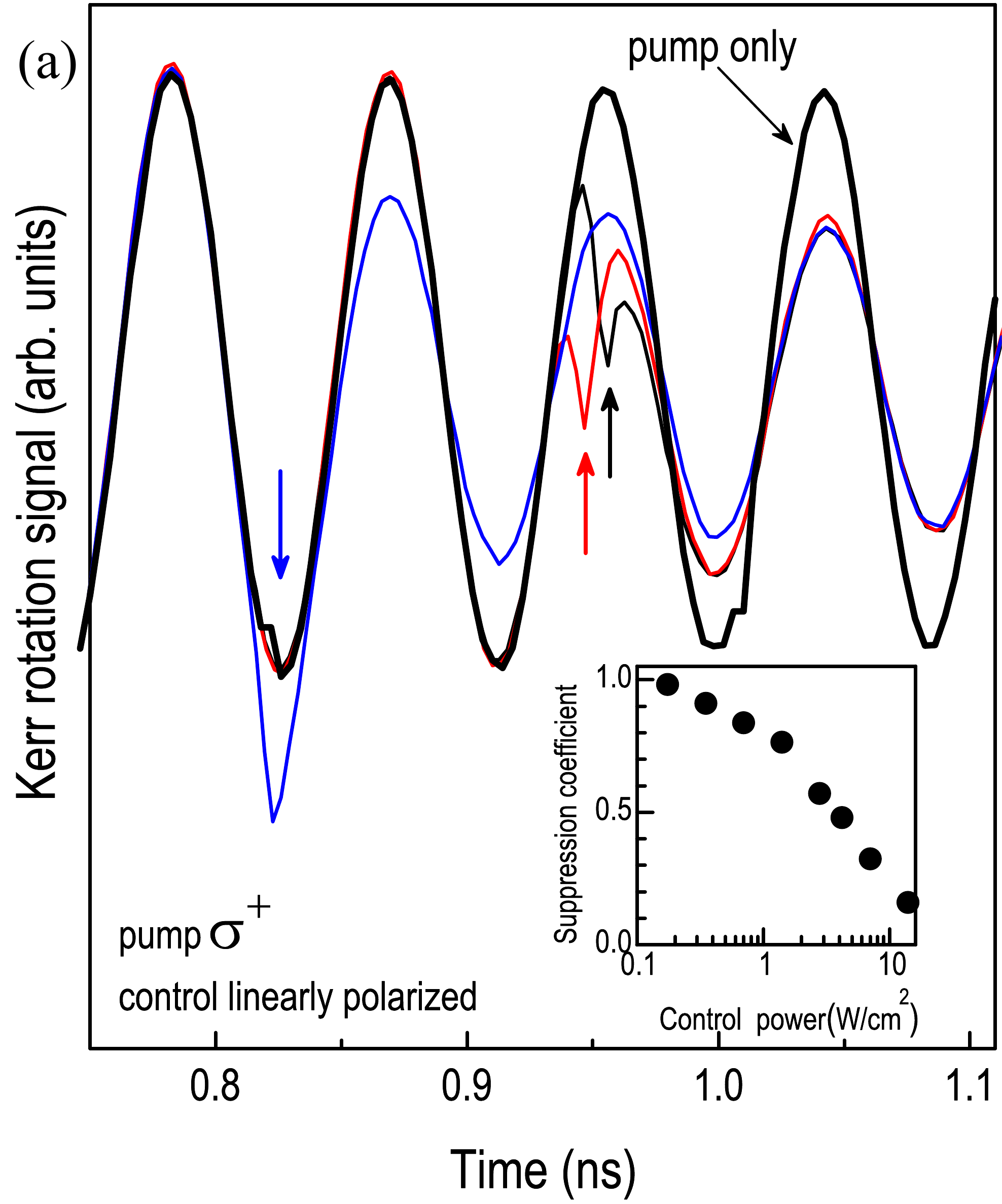} \\
\includegraphics[width=0.75\linewidth]{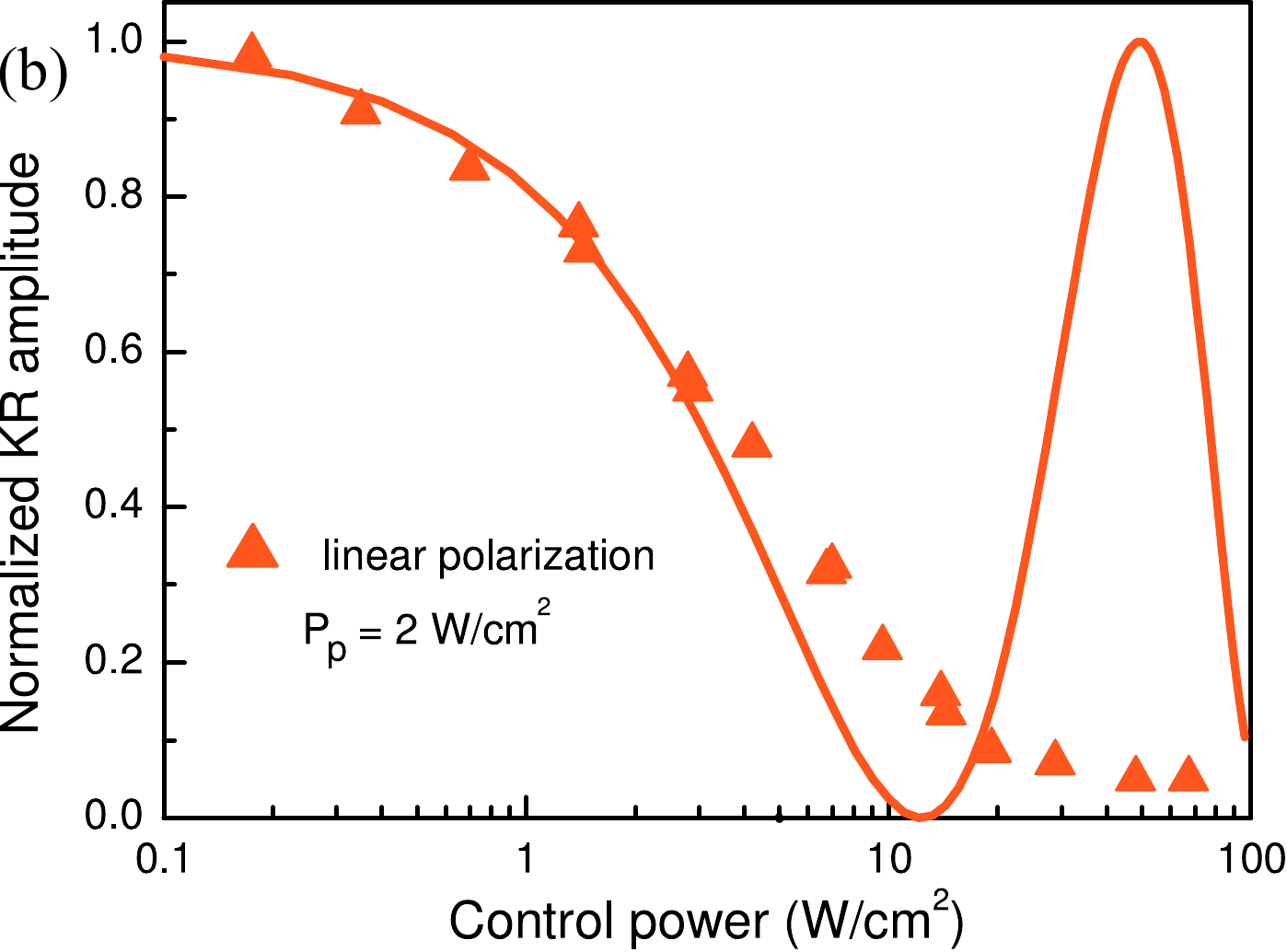}
\caption{(a) Temporal dependence of the Kerr rotation signal obtained
  in the three pulse pump-control-probe technique. Arrows mark the moments of
  the control pulse arrival.  Thick curve is the signal in the
  absence of the control pulse. The power of the pump and control
  pulses is 
    $2.2$~W/cm$^2$. Inset shows the Kerr signal amplitude as function
    of the control pulse power.
  (b) Kerr rotation signal amplitude as function of the control pulse
  power. Solid line is the theory, triangles are the experiment. Pump
  pulse power is 2 W/cm$^2$. Measurements were performed on the five
  CdTe/Cd$_{0.78}$Mg$_{0.22}$Te quantum well structure, well width is
  20~nm. Electron density in each well is $N=2\times
10^{10}$~cm$^{-2}$. Data are reproduced from Ref.~\cite{zhukov10}.}
\label{fig:suppression}
\end{figure}

It is quite unexpected that for the experimental configuration in
question a linearly polarized control pulse makes any effect. Indeed,
in accordance with the electron spin coherence generation model
described in Sec.~\ref{subsec:macro:orient}, the spin coherence
formation requires the spin-dependent photogeneration of
trions. Therefore, linearly polarized  pump pulse does not lead to a
spin coherence generation and to rise of the spin signals in the
pump-probe method. However, experiments carried out in
Ref.~\cite{zhukov10}  at
CdTe/Cd$_{0.78}$Mg$_{0.22}$Te quantum well structures demonstrate that
the linearly polarized control pulse can significantly suppress the
spin beats. It is shown in Fig.~\ref{fig:suppression}(a), where the
Kerr rotation signals measured in the absence of the control pulse
(thick solid curve) and the signals obtained for different time
moments of the control pulse arrival are presented. The dependence of
the spin Kerr signal on the linearly polarized control pulse power is
shown in the inset to Fig.~\ref{fig:suppression}(a) and by triangles
in Fig.~\ref{fig:suppression}(b). It is seen from the Figure that the
signal amplitude drops practically to zero for the control pulse
powers on the order of $10$~W/cm$^2$.

Qualitatively, the spin polarization suppression effect has a simple
explanation. Let  $N_{+1/2}$ be the number of ``spin-up'' electrons
and  
$N_{-1/2}$ be the number of ``spin-down'' electrons by the moment of
the control pulse arrival. We represent the linearly polarized pulse as
a superposition of two circularly polarized ones. Since the trion transition
oscillator strength is proportional to the number of electrons with a
given spin projection, then, due to the  $\sigma^+$ component of the
control pulse, $W N_{+1/2}$ trions are formed, while 
$\sigma^-$ one results in the  $WN_{-1/2}$ trion formation, where $W$ is a
constant proportional to the pulse power. In the simplest model where
the hole-in-trion spin relaxation goes faster than the trion
recombination, all electrons returned from the trions are
depolarized. Therefore,  $z$ component of electron spin changes by 
\begin{equation}
\label{depol}
\Delta S_z = -(WN_{+1/2} - WN_{-1/2})/2 = -W S_z^{(b)},
\end{equation}
where $S_z^{(b)}$ denotes the spin  $z$ component before the control
pulse arrival. Equation~\eqref{depol} describes the total spin
suppression in the system. 

Quantitative description of this effect can be carried out within the
framework of the equation system~\eqref{system:1},
\eqref{system:2}. Calculation shows that the spin after the control
pulse arrival, $\bm S^{(a)}$, and spin before the pulse arrival, $\bm
S^{(b)}$, are linked by a simple relation~\cite{zhukov10}
\begin{equation}
\label{depol:lin}
\bm S^{(a)} = Q_l^2 \bm S^{(b)}.
\end{equation}
Here $Q_l$ is a constant defined according to Eq.~\eqref{an} for
circular components of the control pulse. It is important to note that
the suppression of spin is independent of the control pulse time
moment arrival in agreement with experimental data shown
in Fig.~\ref{fig:suppression}(a).  

Experimental (triangles) and theoretical (line) dependence of the spin
Kerr signal are shown in Fig.~\ref{fig:suppression}(b). The details of
the experimental data fitting are given in Ref.~\cite{zhukov10}. There
is a good agreement of experiment and theory for
moderate control pulse powers. Oscillatory character of the
theoretical dependence of the spin coherence suppression efficiency is
related with the Rabi effect, which was briefly discussed above at the
analysis of the spin coherence excitation. The absence of the
oscillations in the experiment is related with the fact that the two
level model can provide incorrect description of the trion excitation
processes in quantum well structures~\cite{zhukov10}.

\subsection{Microscopic description of probing
  processes}\label{subsec:micro:detect}  

The two level model which we put forward above to describe  electron
spin polarization excitation
and control can be easily extended to
describe the spin probing of electrons and trions.

We decompose a probe pulse whose electric field oscillates along $x$
axis into a superposition of two circularly polarized ones. In the
first order in the probe pulse amplitude the corrections to the
quantum dot wave function can be written as
\begin{eqnarray}  \label{sol1}
\delta \psi_{\pm 3/2} &=& \psi_{\pm 1/2} \int_{-\infty}^t
\frac{V(t')}{\mathrm i\hbar} 
\mathrm e^{-\mathrm i \omega_0^{\mathrm T} (t -t')}  \mathrm dt'\:,\nonumber \\
\delta\psi_{\pm 1/2} &=& \psi_{\pm 3/2}  \int_{-\infty}^t
\frac{V^*(t')}{\mathrm i\hbar} \mathrm e^{\mathrm i \omega_0^{\mathrm
    T} (t -t')} \mathrm dt'\:, 
\end{eqnarray}
%where
\begin{equation}
 V(t) = -\frac{1}{\sqrt{2}} \int \mathsf d(\bm r) E_x^{\rm pr}(\bm r,
 t) \mathrm d^3 r\:. 
\end{equation}
It is assumed that the electric field in a probe pulse has the form $E_{x}^{\rm
  pr}(\bm r, t) \propto \mathrm e^{-\mathrm i \omega_{\rm pr} t}$,
where 
$\omega_{\rm pr}$ is the probe pulse carrier frequency. Before the
probe pulse arrival, the quantum dot state is described by the wave
function Eq.~\eqref{wave}. In this case, generally, both electron and
trion states have nonzero occupations: $n_e
= |\psi_{1/2}|^2 + |\psi_{-1/2}|^2$ and  $n_{tr} =
|\psi_{3/2}|^2 + |\psi_{-3/2}|^2$, respectively, the spin polarization
of the electron and trion can also be present: $S_z =
\left(|\psi_{1/2}|^2-|\psi_{-1/2}|^2\right)/2 \ne 0$ and $J_z =
\left(|\psi_{3/2}|^2-|\psi_{-3/2}|^2\right)/2 \ne 0$.
It can be shown that the probe pulse induced components of dielectric
polarization in the quantum dot have the form~\cite{yugova09}
\begin{eqnarray}
&&\delta  P_{x}^{QD}(\bm r,t) = - \frac{n_e - n_{tr}}{2 \mathrm i
  \hbar} \mathsf d^*(\bm r) \times  \nonumber \\
&& \int
\mathrm d^3 r' \int\limits_{-\infty}^t  \mathrm dt'\mathrm e^{\mathrm i \omega_0^{\mathrm T}
(t' -t)} \mathsf d(\bm r') E_x^{\rm pr}(\bm r',t') + {\rm c.c}\:,\label{px1}\\
&&\delta  P_{y}^{QD}(\bm r,t) = -\frac{S_z - J_z}{\hbar}  \mathsf
d^*(\bm r) \nonumber \\
&& \int \mathrm d^3r'
\int\limits_{-\infty}^t \mathrm dt'\mathrm e^{\mathrm i \omega_0^{\mathrm T} (t' -t)} \mathsf d(\bm r')
E_x^{\rm pr}(\bm r',t')  + {\rm c.c.}\:. \nonumber
%\label{py1}
\end{eqnarray}
It follows from Eqs.~{\eqref{px1}} that the induced polarization in
the quantum dot has two components. One of those, $\delta
P_{x}^{QD}$, is parallel to the polarization plane of the probe pulse,
and its value is proportional to the difference of the electron and
trion states populations, $n_e - n_{tr}$. Another component, $\delta
P_y^{QD}$, is orthogonal to the probe pulse polarization plane, and
its magnitude is determined by the difference of electron and trion
spin projections onto the $z$ axis, $S_z - J_z$. It is the component which
is responsible for the probe pulse polarization plane rotation as well
as the appearance of its ellipticity.

Solution of Maxwell equations for the quantum dot array, whose
dielectric polarization is described by Eqs.~\eqref{px1}, allows us to
determine the magnitudes of the Faraday rotation and ellipticity
signals. In the case of a planar array where the typical distances
between the dots are small as compared with the wavelength we
have~\cite{yugova09} 
\begin{equation}
 \label{farad:fin}
\mathcal E + \mathrm i \mathcal F = 
\end{equation}
\[\frac{3\pi}{q^2\tau_r^T}
N_{QD}^{2{D}} (J_z - S_z) 
\int_{-\infty}^\infty \mathrm d t \int_{-\infty}^t \mathrm dt'
\mathrm e^{\mathrm i \omega_0^{\mathrm T}(t'-t)} E_{0,x}^{{\rm
    pr}*}(t) E_{0,x}^{\rm pr}(t')\:, 
\]
where $N_{QD}^{2D}$ is the two-dimensional density of the dots in the array,
$q=\omega_{\rm pr} \sqrt{\varepsilon_b}/c$ is the wave vector of light
in the system ($\varepsilon_b$ is the background dielectric constant
assumed to be the same both for dots and matrix),  $\tau_r^T$
is the trion radiative lifetime:
\begin{equation}
 \frac{1}{\tau_r^{T}} = \frac{4}{3}\frac{q^3}{\varepsilon_b \hbar}
 \left|\int \mathrm d\bm r \, \mathsf d(\bm r)\right|^2\:.
\end{equation}
Kerr rotation signal is determined by the interference of the probe
pulse reflected from the structure cap layer and from the quantum dot
array. It is described by a standard expression [cf. Eq.~\eqref{K}]
\begin{equation}
 \label{kerr:fin}
\mathcal K = r_{01}t_{01}t_{10} [\cos{(2qL)} \mathcal F + \sin{(2qL)}
\mathcal E]\:, 
\end{equation}
where $r_{01}$ is the reflection coefficient at the air/cap layer
boundary, $t_{01}$ and $t_{10}$ are the transmission coefficients of
this boundary inside and outside, respectively. The effect of the cap
layer on the Faraday and ellipticity effects can be taken into account
by a factor $t_{01}t_{10}$ in the right hand side of Eq.~\eqref{farad:fin}.

\begin{figure}[hptb]
\includegraphics[width=0.75\linewidth]{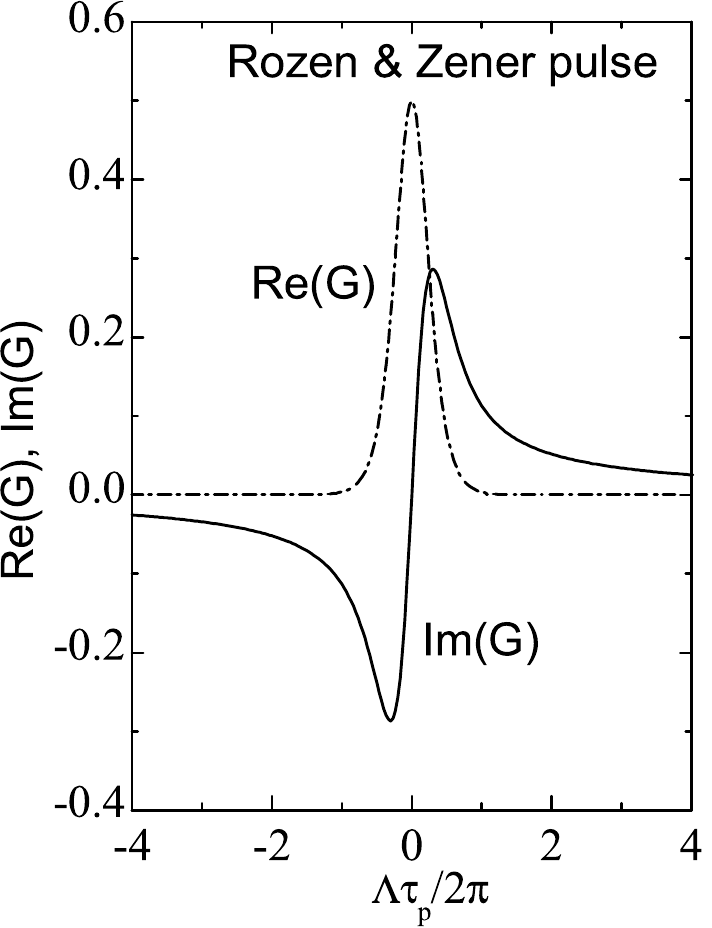}
\caption{Dependence of function $G$ on the detuning 
$\Lambda=\omega_{\rm pr}-\omega_0^{\mathrm T}$. Data are reproduced
from Ref.~\cite{yugova09}.}
\label{fig:probe}
\end{figure}

As it follows from Eqs.~\eqref{farad:fin} and \eqref{kerr:fin}, the
spin signals amplitudes are determined by the difference of the spin $z$
components of trion and electron in a quantum dot. To analyze the
frequency dependence of the Faraday rotation and ellipticity signals,
we represent the probe pulse field as $E_0^{\rm pr}(t) = E^{(0)} s(t)
e^{-\mathrm i\omega_{\rm pr} t}$, where  $s(t)$ is the probe pulse envelope
function. One can check that
\begin{equation}
 \label{freq}
\mathcal F \propto \Im{G(\omega_{\rm pr} - \omega_0^{\mathrm T})}\:,
\quad \mathcal E \propto \Re{G(\omega_{\rm pr} - \omega_0^{\mathrm T})}\:,
\end{equation}
where
\begin{equation}
\label{GL}
 G(\Lambda) = \int_{-\infty}^{\infty} \mathrm dt \int_{-\infty}^t \mathrm dt' s(t) s(t')
 \mathrm e^{\mathrm i \Lambda(t-t')}, \end{equation}
and $\Lambda= \omega_{\rm pr} - \omega_0^{\mathrm T}$. In the
particular case of the Rosen \& Zener pulse, where  $s(t) =
1/\cosh(\pi t/\tau_p)$, we obtain
\begin{equation}\label{G}
G(\Lambda) = 
\cfrac{\tau_p^2}{\pi^2}\zeta{\left(2,\frac{1}{2}-\frac{\mathrm i
\Lambda\tau_p}{2\pi}\right)},
\end{equation}
where $\zeta(a,b)= \sum_{k=0}^\infty (k+b)^{-a}$ is the generalized
Riemann 
$\zeta$-function. 

Figure~\ref{fig:probe} shows real and imaginary parts of the function
$G$ calculated for the Rosen \& Zener pulse. Qualitatively, the
behavior of Faraday rotation and ellipticity as functions of detuning
between the trion resonance and the probe pulse carrier frequency are
analogous to those obtained in Sec.~\ref{subsec:macro:detect}, see
Fig.~\ref{fig:freq:theor}. It is worth to note, that the maximum
sensitivity of Faraday rotation  and ellipticity signals corresponds
to different detunings: Faraday rotation signal takes its maximum
value for detuned pulses $|\Lambda| \tau_p \approx
1$, while ellipticity takes its maximum for the resonant ones. We
shall ascertain below that this leads to different dependence of the
spin signals on the pump-probe pulses delay in inhomogeneous quantum
dot ensembles.

\section{Temporal dependence of the spin Faraday, Kerr and 
induced   ellipticity signals}\label{sec:time}

We discussed in Secs.~\ref{sec:macro} and \ref{sec:micro} the
mechanisms of the spin signals formation in the pump-probe method, the
dependence of the signal amplitudes on the pump power and on spectral
positions of pump and probe pulses, as well as the
possibilities to control the spin polarization by means of optical
pulses. Below we focus on the analysis of the Faraday rotation and
ellipticity signals dependence on the time delay between pump and
probe pulses.

\begin{figure}[htb]
\includegraphics[width=0.7\linewidth]{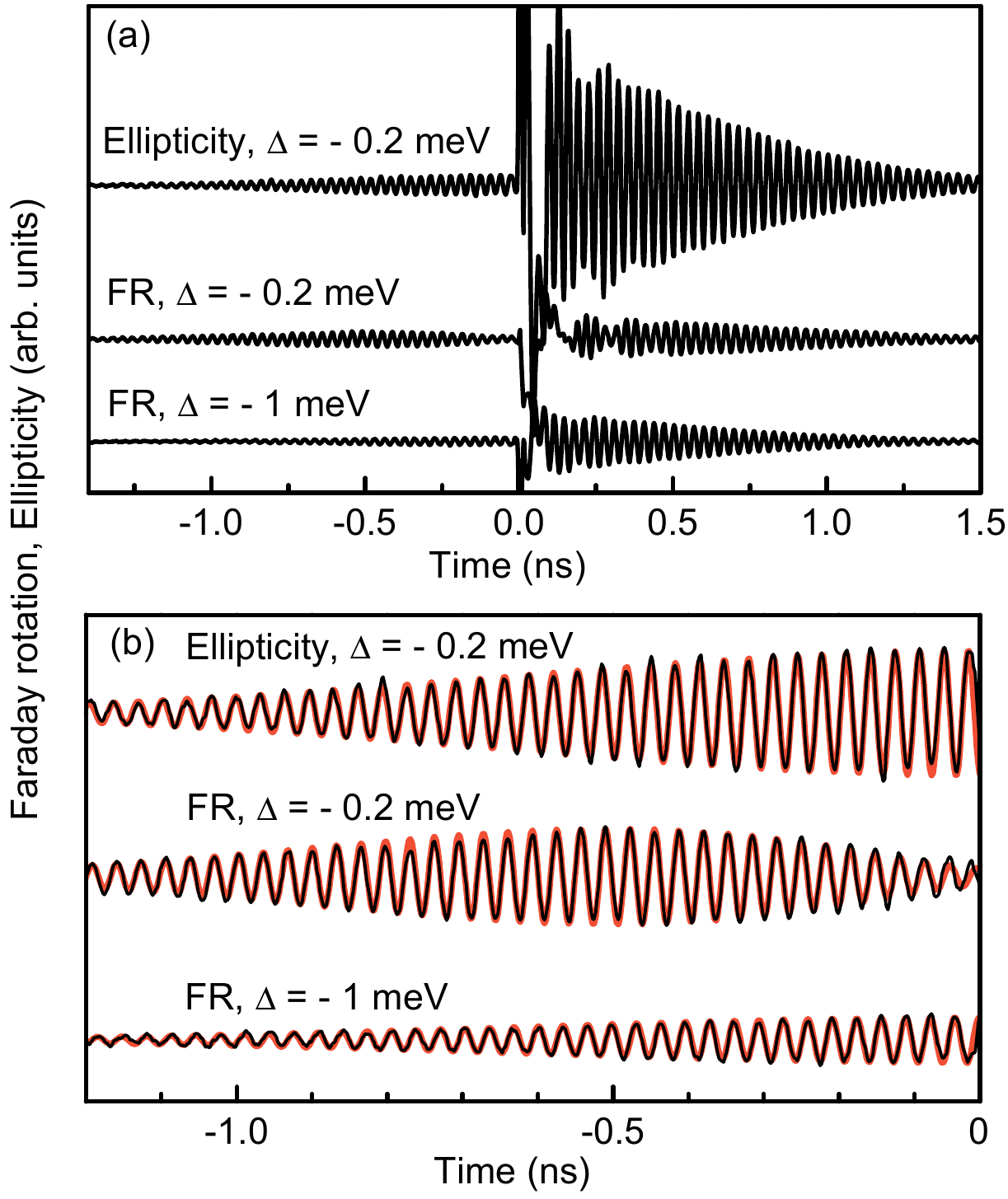}\\
\includegraphics[width=0.7\linewidth]{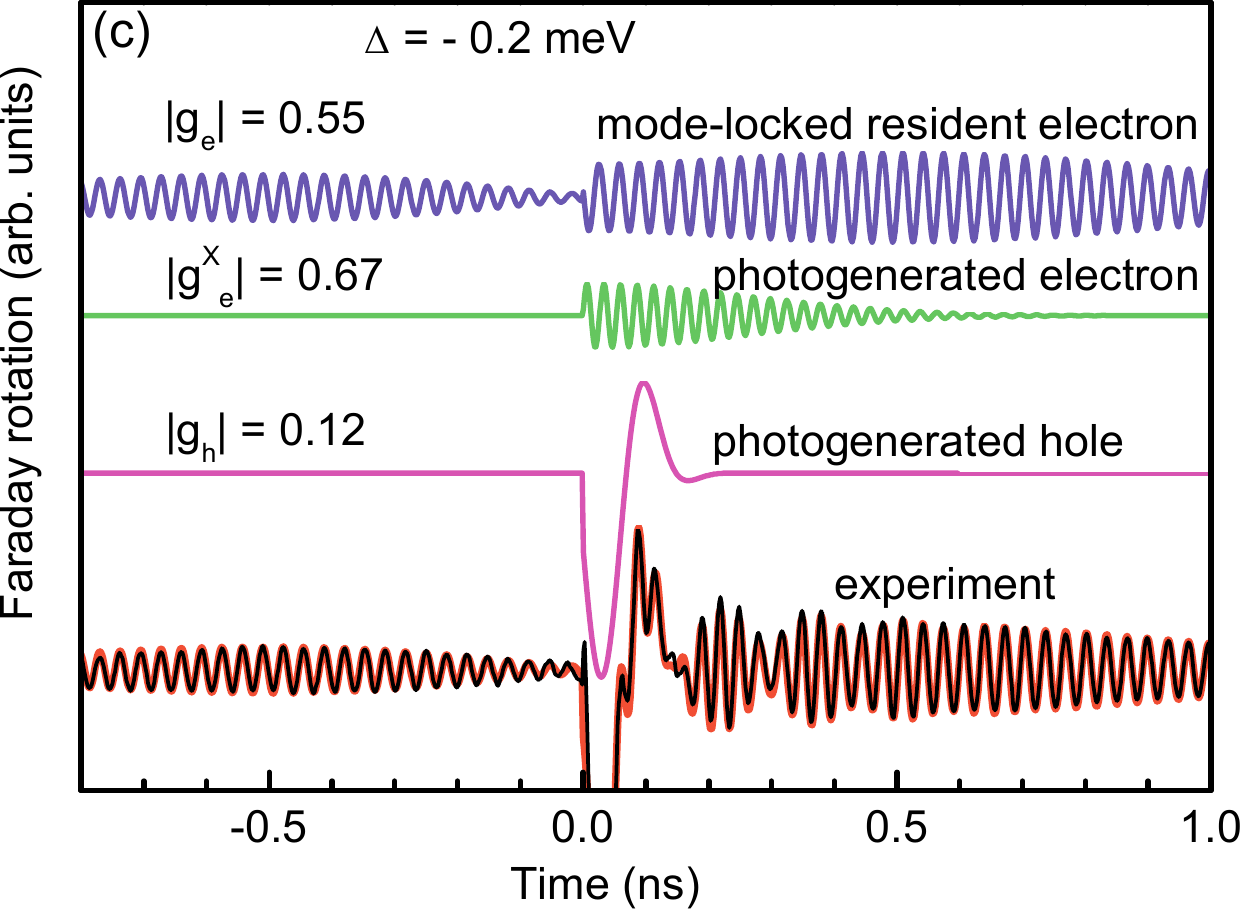}
\caption{(a) Faraday rotation (FR) and ellipticity signals as
  functions of temporal delay between pump and probe pulses. Top two
  curves are the ellipticity and Faraday rotation signals at (almost)
  spectrally degenerate pump and probe lasers ($\hbar\omega_{\rm P} -
  \hbar \omega_{\rm pr} 
  =\Delta=-0.2$~meV), bottom curve shows the Faraday rotation signal
  for a detuned probe pulse
($\Delta=-1.0$~meV). (b) Corresponding signals at negative
delays. Thin curves are experimental data, thick ones are the fitting.
(c) Faraday rotation signal (bottom curve) at almost degenerate pump
and probe pulses. The fitting is superimposed over the experimental
curve. The curves presented above (from top to bottom): signal related
with the long living spin polarization in charged quantum dots, signal
related with electron-in-exciton spin precession in neutral
quantum dots, and signal related with hole spin precession (both in
neutral quantum dots and in trions). Measurements are carried out on
the structure consisting of 20 layers of InGaAs/GaAs quantum dot
layers with the dot density in each layer  $10^{10}$~cm$^{-2}$, the
structure is doped in such a way, that there is one electron per dot
on average. Temperature $T=6$~K, magnetic field $B=4$~T. Data are
reproduced from Ref.~\cite{glazov2010a}.} \label{fig:exp1}
\end{figure}

Figure~\ref{fig:exp1}(a) shows typical spin signals of Faraday
rotation and ellipticity obtained in Ref.~\cite{glazov2010a} on the
$n$-type InGaAs quantum dot array. The measurements were carried out
in the so-called ``two color'' (nondegenerate) pump-probe technique,
where the pump and probe pulses are generated by different lasers,
hence their carrier frequencies can be tuned independently. The pulses
themselves are synchronized with high precision (about 10~fs). Pumping
and probing is carried out by a periodic sequence of the pulses
following with a repetition period $T_R=13.2$~ns.

Key experimental observations are the following:
\begin{enumerate}
\item Signals at positive delays have a complex character
  corresponding to the superposition of oscillations with different
  frequencies. Analysis of the oscillation frequencies and decay times
  (Fig.~\ref{fig:exp1}(c)) allows one to establish that the observed
  signal is a superposition of the resident electron spin signal, as
  well as photocreated (in neutral dots) electron and hole
  signals~\cite{glazov2010a,yu07}. 
\item The remarkable signals take place also at negative delays, that
  is when the probe pulse arrives before the next pump pulse. Effects
  of spin signals generation at negative delays and spin accumulation
  at the excitation of quantum wells and quantum dots by a periodic
  sequence of the pulses are described below in
  Secs.~\ref{subsec:RSA}, \ref{subsec:NIFF}. 
\item Under the conditions of spectral degeneracy of pump and probe
  pulses, the amplitude of the Faraday rotation signal induced by
  resident electrons behaves nonmonotonically as a function of delay:
  at small delays the signal amplitude builds up, afterwards it
  decays. It is clearly seen in Fig.~\ref{fig:exp1}(b), middle curve,
  there the area of negative delays is shown. Ellipticity signal
  demonstrates an expected behavior in this case: damped oscillations.
  Section~\ref{subsec:emerge:FR} is devoted to the buildup of the
  Faraday rotation. 
\end{enumerate}

\subsection{Resonant spin amplification and spin precession
  mode-locking}\label{subsec:RSA} 

As we have already noted above, the spin signals at negative
delays arise due to the fact, that electron spin does not fully relax
during the pulse repetition period. Depending on the relation between the
electron spin precession period and the pump pulse repetition period,
the spin polarization in the system can either accumulate or get
suppressed. 

\begin{figure}[hptb]
\includegraphics[width=0.65\linewidth]{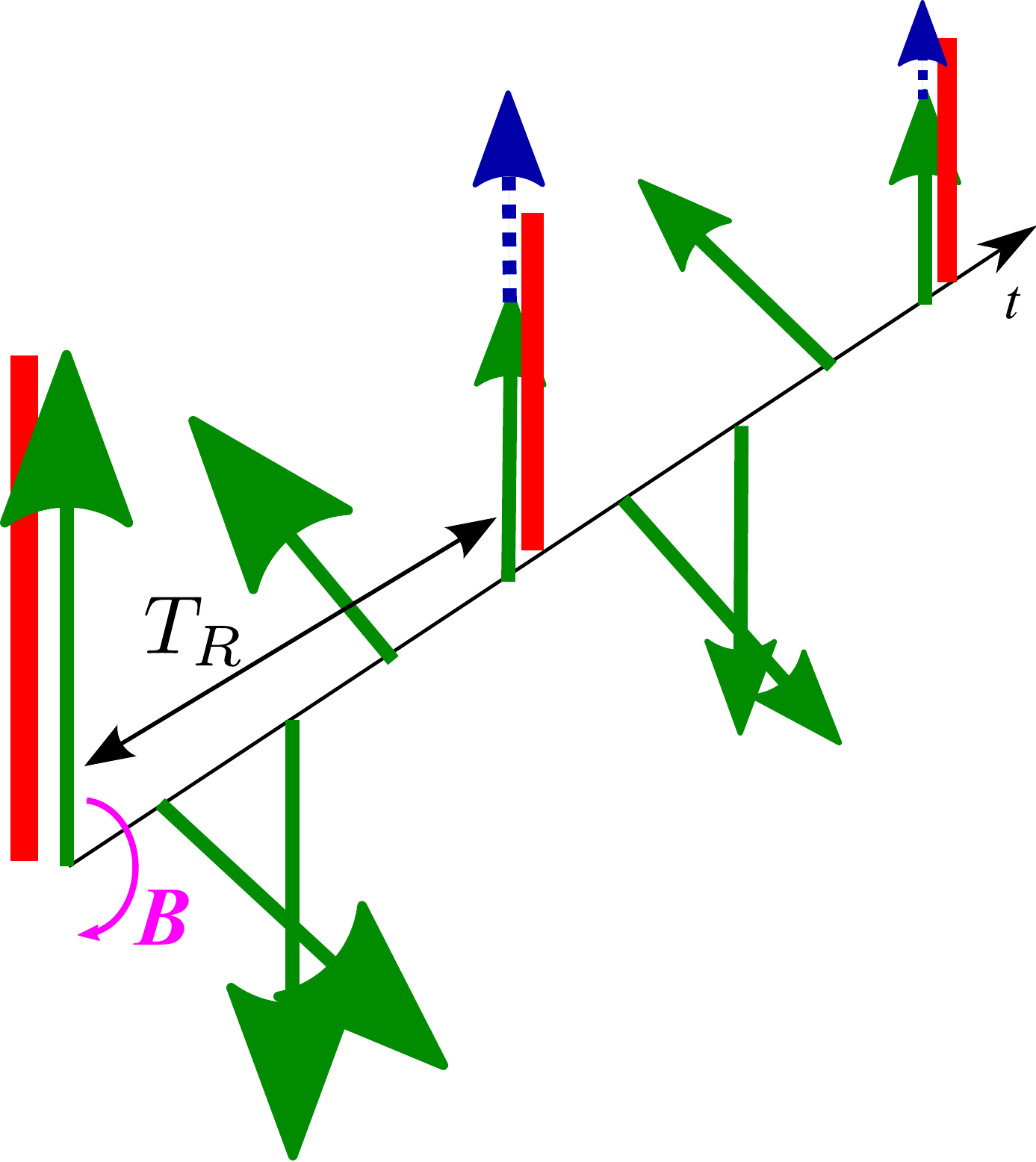}\\
\includegraphics[width=0.65\linewidth]{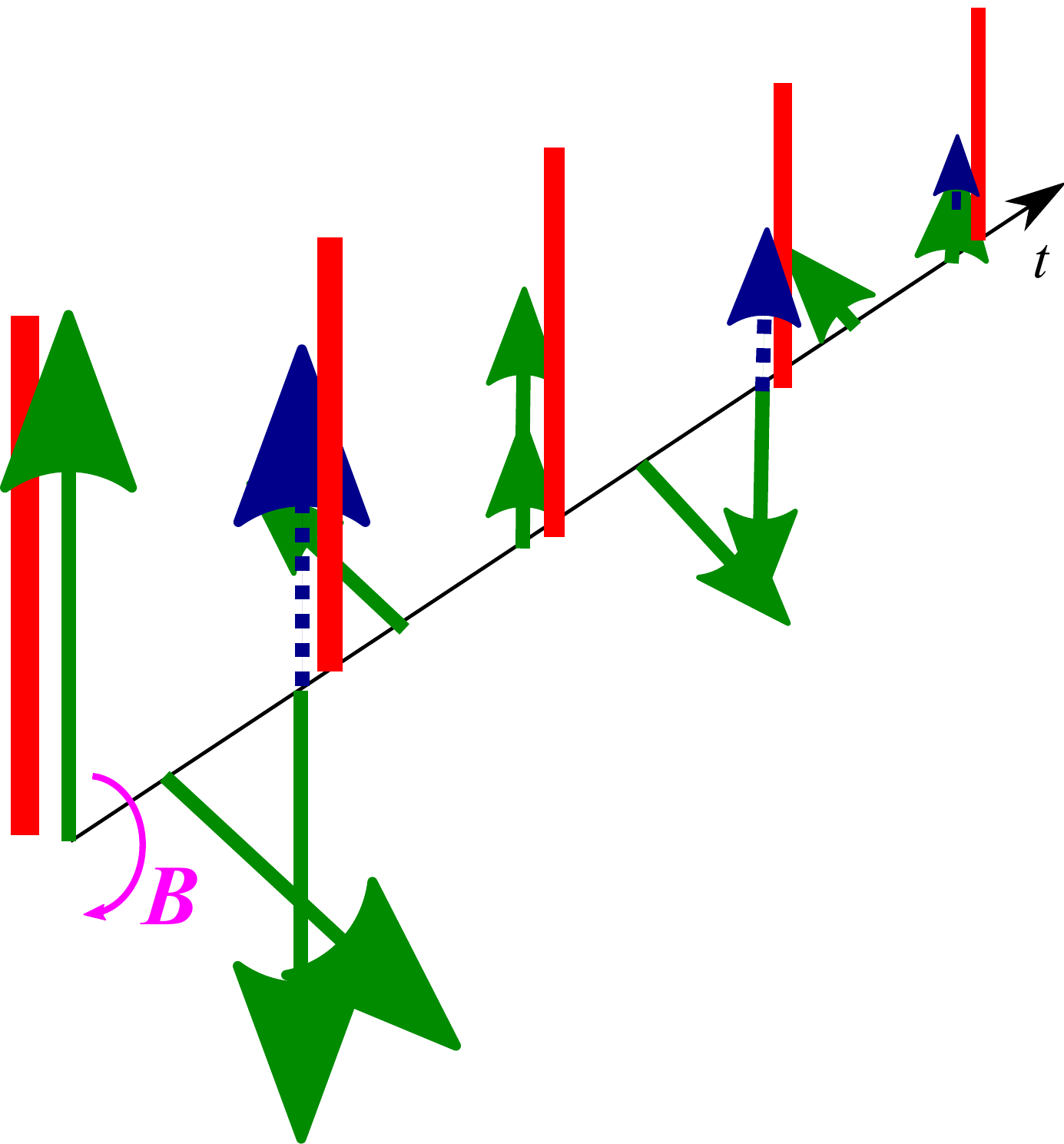}
\caption{Schematic illustration of the resonant spin
  amplification. Pump pulse arrival time moments are marked. Solid
  arrows show the electron spin orientation at different time
  moments. Dotted arrows shown the spin polarization generated by the
  pump pulses. Top panel shows the case of equal spin precession and
  repetition periods, $T_R = T_L$, in the bottom panel pulses arrive
  twice more often: $T_R=T_L/2$.} \label{fig:rsa}
\end{figure}

Indeed, as it is shown in Fig.~\ref{fig:rsa}, if the pump pulse
repetition period, $T_R$, is a multiple of the electron spin
precession period in the external field $T_L = 2\pi/\Omega$,
\begin{equation}
\label{PSC}
T_R = NT_L = \frac{2\pi N}{\Omega}, \quad N =1, 2, \ldots \ ,
\end{equation}
then the next pump pulse adds the spin in phase with the precessing
one. In this case, the spin polarization in the system is enhanced as
compared with that formed by a single pulse. This effect is known as
resonant spin amplification. If condition~\eqref{PSC} is not
fulfilled, the phase synchronization fails and spin polarization is
suppressed.

In experiments the spin dynamics of electron ensemble is studied,
as a rule. Optical excitation of the quantum dot array or the quantum
well results in the spin polarization of the charge carriers with
different energies being spread within the pump pulse spectral width $\sim
\hbar/\tau_p$. Electron $g$-factor values are different, so the
electron spin precession frequencies are different. An additional
contribution to the spread of spin precession frequencies is given by
the hyperfine interaction of electron spins with the spins of lattice
nuclei. Spin beats damping is characterized by the following parameters:
$T_2\equiv \tau_s$ being the transverse to the external field electron
spin components relaxation time, $T_2^* = T_2T_{\rm
  inh}/(T_2+T_{\rm inh})$ being the dephasing time of electron spin in
the ensemble contributed both by the relaxation processes and the
spread of Larmor precession frequencies, the latter one is
characterized by the time $T_{\rm inh} \sim (\Delta \Omega)^{-1}$,
where $\Delta\Omega$ is the spin precession frequency spread. The
specifics of spin dynamics under the excitation by a periodic train of
the pulses is determined by the repetition period of the pulses $T_R$
and the spin beats decay times. Obviously, if $T_2 \ll T_R$, the
effects of spin polarization accumulation are unessential, since spin
relaxes before the next pulse arrival. Below we assume that  $T_2
\gtrsim T_R$ and analyze two important cases: (i) of weak dephasing
related with the spin precession frequency spread $T_{\rm inh} \gg
T_R$, where the resonant spin amplification is realized, and (ii) the
regime of strong dephasing, $T_{\rm inh} < T_R$, where the spin
precession mode-locking becomes possible. 

\subsubsection{Resonant spin amplification}\label{subsubsec:RSA}

We start with a situation where the spread of the spin precession
frequencies is not relevant and the spin beats damping is determined
by the spin relaxation processes. We assume for simplicity that the
average electron spin is small (both generated by a single pump pulse
and accumulated by a pump pulse train). Therefore, we obtain from
Eq.~\eqref{system:kin} for the periodic 
sequence consisting of large enough number of pulses:
\begin{equation}
\label{szrsa}
S_z^{\rm tot}(\Delta t) = \sum_{n=1}^\infty S_z(0) \mathrm e^{-(\Delta
  t+ nT_R)/T_2} \cos{[\omega(\Delta t+ nT_R)]}.
\end{equation}
Here $S_z(0)$ is the electron spin created by a single pulse,  $\Delta
t$ is the delay between the probe pulse and the nearest next pump
pulse, it can take any negative value in the interval $\Delta t \in
(-T_R, 0]$. It is assumed in derivation of Eq.~\eqref{szrsa} that
$\tau_r^T \gg \tau_s^T$. 

Calculation~\cite{Kikkawa98,beschoten} shows that
\begin{equation}\label{lit}
S^{\rm tot}_z(\Delta t) = \frac{S_z(0)}{2}\ {\rm e}^{ - ( T_{R} + \Delta
t)/T_2}\times
\end{equation}
\[ \frac{ \cos{ ({\Omega} \Delta t) } - {\rm e}^{T_{R}/T_2}  \cos{ [ {\Omega} ( T_{R} + \Delta t) ] } }{
\cos{ ({\Omega} T_{R} )} - \cosh{(T_{R}/T_2)} }\:.
\]
It follows from Eq.~\eqref{lit} that the electron spin dependence on
the Larmor precession frequency $\Omega$ (and, correspondingly, on the
magnetic field) at a fixed delay $\Delta t$ consists of a sequence of
maxima corresponding to the condition ${\Omega T_{R}} \approx 2\pi
N$, where $N$ is an integer. In the vicinity of the maximum where
$|\Omega T_{R} - 2\pi N| \ll 
1$ and $|\Delta t/T_{R}|\ll 1$, expression \eqref{lit} can be
rewritten in the form of Lorentz function
\begin{equation}
 \label{lorentz}
S^{\rm tot}_z(0,\Omega T_R) = S_z(0) \frac{1 - \mathrm e^{-\frac{T_{R}}{T_2}}}
{(\Omega T_{R} - 2 \pi N)^2 + 2 \left[\cosh(T_{R}/T_2) - 1 \right]}.
\end{equation}
In this approximation the peak width is determined by the quantity
\begin{equation}
 \label{width}
\Delta = \sqrt{2\left[\cosh(T_{R}/T_2) - 1 \right]},
\end{equation}
which, in the limit of long spin relaxation times, $T_{R}/T_2 \ll
1$, passes to  $\Delta \approx T_{R}/ T_2$, and the width is the
smaller the longer spin relaxation time $T_2$.

\begin{figure}[hptb]
\includegraphics[width=0.95\linewidth]{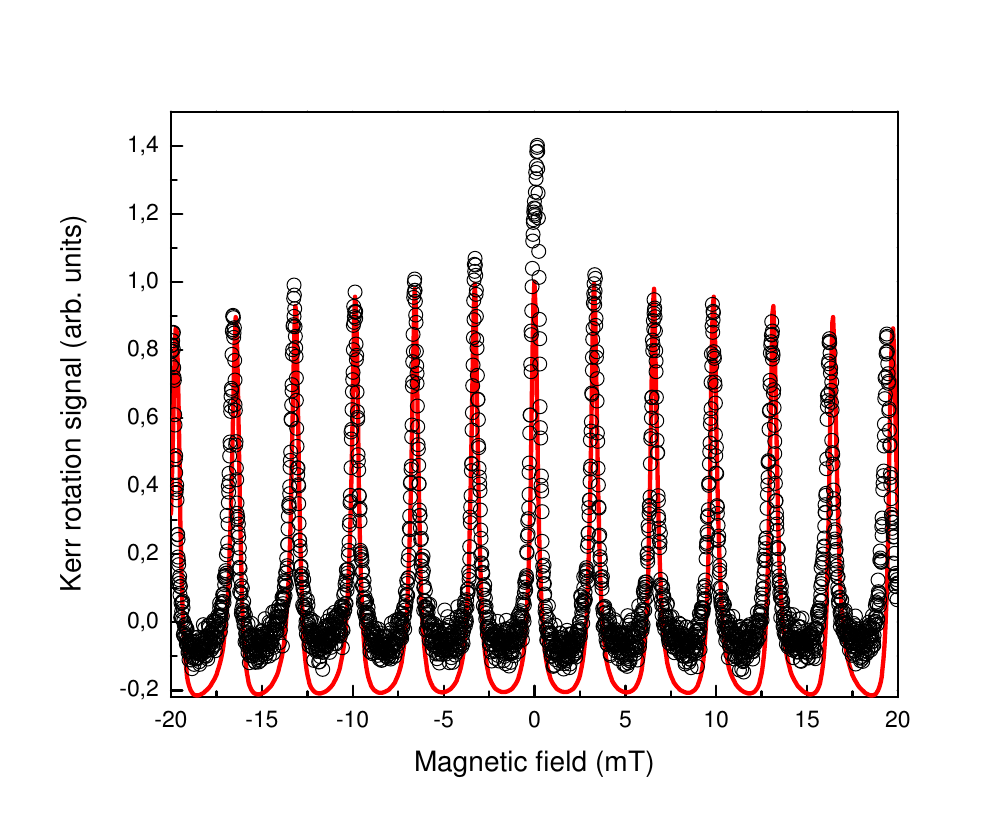}
\caption{Kerr rotation signal dependence on the external magnetic
  field. Points are the experimental data obtained at the 
  CdTe/Cd$_{0.78}$Mg$_{0.22}$Te quantum well structure (five wells of
  20~nm, electron density in the well $N\approx 10^{10}$~cm$^{-2}$)
  for small negative delay $\Delta t = -80$~ps, pulse repetition
  period is 
  $T_R = 12.5$~ns (data are reproduced from Ref.~\cite{ast08}). Solid
  curve is the fit of experimental data, transverse spin relaxation time $T_2 = 30$~ns, mean
  $g$-factor $g=1.64$, spread of 
  $g$-factor values $\Delta g/g=0.4\%$.} \label{fig:rsa:TE}
\end{figure}

Figure~\ref{fig:rsa:TE} presents the resonant spin amplification
spectrum: the dependence of the Kerr rotation signal, $S_z^{\rm
  tot}$, on the magnetic field obtained at 
CdTe/Cd$_{0.78}$Mg$_{0.22}$Te quantum well structure at small negative
delay~\cite{ast08}. Peaks correspond to the commensurability condition
of the spin precession period and the pulse repetition
period, Eq.~\eqref{PSC}. Note, that the zero field peak is larger than
adjacent ones, seemingly, it is related with the influence of nuclear
effects on the electron spin. Peak height decreases monotonously with
its number and the peaks themselves become slightly broadened. It is
caused by the $g$-factor spread in the localized electron ensemble,
which gives rise to an additional spin dephasing. The experimental
data fitting with allowance for the $g$-factor values spread carried
out by using theoretical formulae obtained in Ref.~\cite{glazov08a}, see
also~\cite{ast08} and shown by a solid line in Fig.~\ref{fig:rsa:TE},
describes experiment rather well. The comparison of the theoretical
calculation and the experimental data allows us to establish the main
parameters of the resident electrons spin kinetics~\cite{ast08}: mean
$g$-factor value, $g=1.64$, its spread  $\Delta g/g=0.4\%$ (this spread was
modelled by a Gaussian distribution in our theory), transverse spin
relaxation time  $T_2 = 30$~ns. These parameters obtained by means of
resonant spin amplification technique agree well with the values
extracted from the Hanle effect and temporal dependence of the spin
signals on the same structure.

Classical expression for the resonant spin amplification spectrum
Eq.~\eqref{lit} can be easily extended to allow for the arbitrary spin
relaxation anisotropy~\cite{glazov08a} inherent to semiconductor quantum
well structures. As compared with the isotropic spin relaxation case,
the maxima at $B\ne 0$ do not change, while the zero field maximum can
be either suppressed (if spin $z$ component relaxation time is shorter
than that of the in plane spin components) or enhanced (if $z$ spin
component turns out to be longer living one). Spin relaxation
anisotropy was discovered by Hanle effect measurements in
Ref.~\cite{averkiev06} and by Kerr signal spin beats in magnetic
field~\cite{larionov:033302,larionov11:eng}. The experimental studies of
the spin relaxation anisotropy in the resonant spin amplification
technique are impeded since the peak at $B=0$ is strongly affected by
the hyperfine interaction of electron and nuclear spins.

An additional specifics of the resonant spin amplification spectra can
be brought about by the hole-in-trion (in $n$-type
structures) and by the electron-in-tron (in $p$-type structures)
spin dynamics. If trion spin relaxation is suppressed, then the
considerable spin polarization of resident charge carriers appears in
magnetic field only, as discussed above in
Sec.~\ref{subsec:macro:orient:tr}. At that, the resonant spin
amplification peak amplitude increases with an increase of the peak
number in moderate magnetic fields and the spectrum envelop has a
smooth bat-like shape. One can manage to extract the spin relaxation
times of unpaired charge carrier in the trion from such
spectra~\cite{sokolova09,fokina-2010,korn_njp}. 

Equation~\eqref{lit} does not account for the spin polarization
saturation at the periodic pumping. The corresponding extension of the
treatment in the framework of the two level model describing electron
spin excitation by short circularly polarized pulses outlined in
Sec.~\ref{subsec:micro:twolevel} was carried out in
Refs.~\cite{yugova09,sokolova09,fokina-2010}. With an increase of the
pumping the peaks become broader and the spin polarization dependence
on magnetic field becomes smoother.

\subsubsection{Spin precession mode-locking}\label{subsubsec:modelock}

Let us turn now to the opposite limiting case where the spread of
electron 
$g$-factors and the random nuclear fields result in fast electron spin
dephasing, i.e.
\begin{equation}
\label{modelock}
T_2^* \approx T_{\rm inh}  < T_R.
\end{equation}
In this situation \emph{prima facie} any remarkable spin signals at
negative delays can not be expected, because electron spin gets
dephased before the next pulse arrival.

\begin{figure}[hptb]
\includegraphics[width=\linewidth]{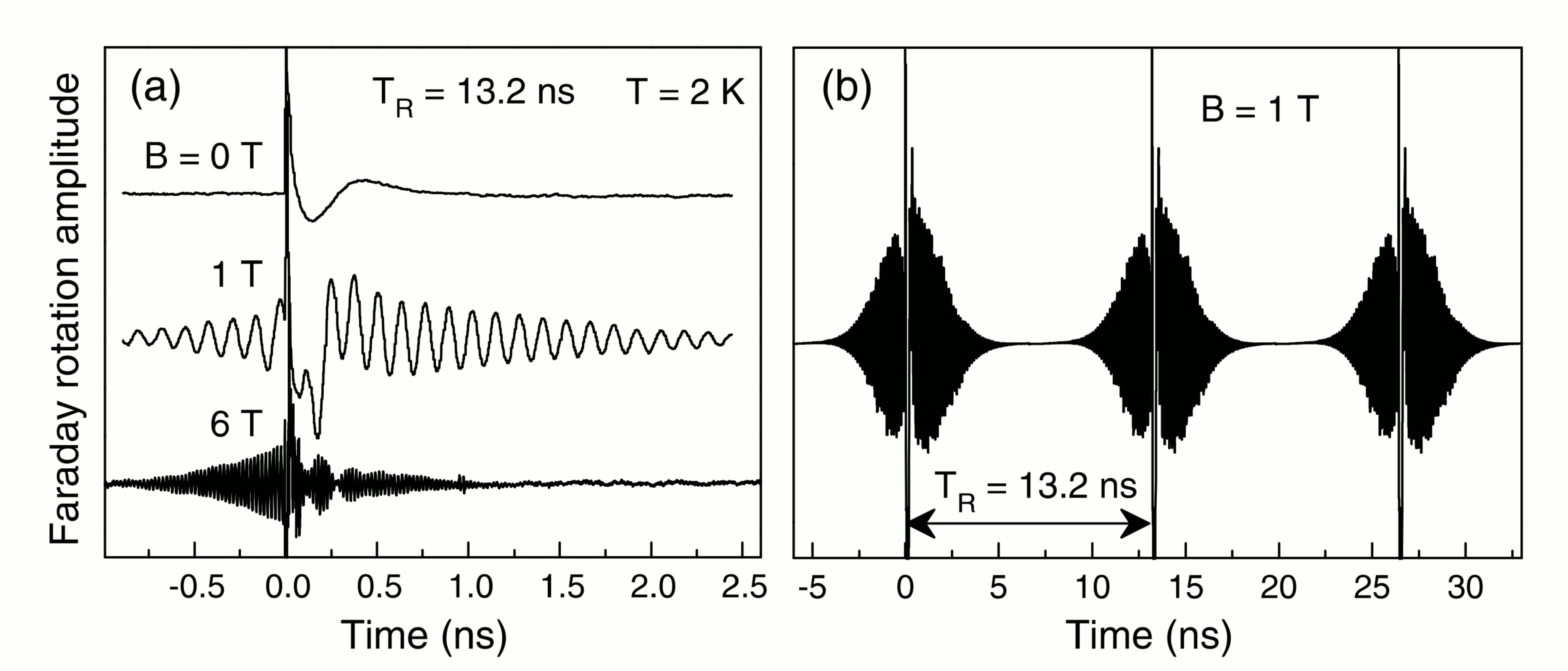}
\caption{(a) Faraday rotation signal obtained in the pump-probe method
  for different values of magnetic field. The measurements were
  carried out on a structure containing 20 layers of InGaAs/InAs
  quantum dots with the dot density per layer of $10^{10}$~cm$^{-2}$,
  the structure is doped in a such a way that there is one electron
  per dot on average. The complex signal shape at positive delays is
  related with the interference of resident electron spin beats and
  those of electron and hole in neutral dots
  (cf. Fig.~\ref{fig:exp1}), signal at negative delays is caused by
  the spin precession mode-locking. (b) Faraday signal measured at a
  larger temporal interval including three pulse repetition
  periods. Panel (a) is reproduced from
  Ref.~\cite{A.Greilich07212006}, panel (b) is from Ref.~\cite{yakovlev_bayer}.} \label{fig:modelock_several} 
\end{figure}

However, as mentioned above, the spin of a given electron is conserved
during long time which exceeds by far the pulse repetition
period. Moreover, if condition Eq.~\eqref{modelock} holds, the wide
spectrum of spin precession frequencies is excited. Hence, there are
the spins in the whole ensemble of the precessing ones for which the
spin precession and the pump pulse repetition frequencies are
synchronized. Obviously, the spins of these electrons will always be in
phase at the time moments $t=0, T_R, 2T_R,
\ldots$, i.e. when the next pump pulse arrives. The spins of remaining
charge carriers in these time moments have random precession phases,
and they do not contribute to the observed signal. Therefore, if
condition Eq.~\eqref{modelock} is satisfied, the spin signal decays
during the time scale on the order of $T_2^*$, and afterwards emerges
by the next pump pulse arrival during approximately the same
time. This effect was observed in the pump-probe experiments on InGaAs
quantum dot arrays~\cite{A.Greilich07212006} and was named
\emph{spin precession mode-locking}. Typical behavior of the Faraday
rotation signal as function of the time delay between the pump and probe pulses,
measured on the InGaAs/GaAs quantum dot sample, are presented in
Fig.~\ref{fig:modelock_several}. 

The spin precession mode-locking phenomenon allows one, in certain
extent, to overcome the electron spin dephasing effects related with
an inhomogeneity of the electron ensemble. About a $10^6$ electron spins
precess with commensurable frequencies under the conditions of
experiment, Ref.~\cite{A.Greilich07212006}. Since the spread of spin
precession frequencies decreases with the magnetic field decrease, at
small fields the situation where one or two spin precession modes are
excited can be achieved, as shown experimentally in
Ref.~\cite{greilich09}. Application of spin precession mode-locking
allowed one to determine experimentally the transverse relaxation time
of electron spin, $T_2$, and to excite the spin echo under the pumping by
a sequence containing pairs of circularly polarized
pulses~\cite{A.Greilich07212006}. 

Evidently, the ratio of the long living (electron) spin signal
amplitudes at
negative and positive delays $A_{\rm neg}/A_{\rm pos}$ must be
determined by the fraction of electrons, whose spins satisfy the
mode-locking condition, Eq.~\eqref{PSC}. Indeed, the contribution to
the signal at negative delays is given by only those electrons, whose spin
precession is synchronous with the pump pulses, while at positive
delays all electrons from the ensemble contribute to the
signal. Analysis shows that this ratio should not exceed $0.2 \ldots
0.3$. It is seen from the experimental data shown in
Fig.~\ref{fig:modelock_several} that it is not the case: $A_{\rm neg}$
is just slightly smaller than $A_{\rm pos}$. It means, that
synchronization condition~\eqref{PSC} is satisfied in almost all the
dots. The reason for this is described in the next Section.

\subsection{Electron spin precession frequency focusing provided by
  interaction with lattice nuclei}\label{subsec:NIFF}

Up to this point we excluded from consideration the nuclear spin
subsystem. Indeed, during the time interval of about ten nanoseconds
which corresponds to the repetition period of the pulses, nuclear
spins may be considered frozen. Owing to hyperfine
interaction, nuclear spin fluctuations contribute
to the spread of electron spin precession frequencies and result in the
electron spin dephasing~\cite{PhysRevB.64.125316, merkulov02,
  PhysRevLett.88.186802, 
  PhysRevB.66.161318}, since electron spin precession frequency
$\Omega_{\rm eff}$ is determined by the total magnetic field,
including both the external field and Overhauser field, acting from
the side of nuclei. For instance, in the box model, where the
hyperfine interaction constant of electron with nuclei $\alpha_{\rm
  hf}$ is the same for all quantum dot nuclei~\cite{Kozlov2007}, 
$$\bm \Omega_{\rm eff} = \bm
\Omega + \alpha_{\rm hf} \bm m,$$ 
where $\bm m = \sum_i \bm I_i$ is the total nuclear spin ($\bm I_i$
are mean values of the nuclear spin vectors, $i$ enumerates nuclei
interacting with the electron). If nuclei are on average unpolarized,
 vectors $\bm m$ in different dots are oriented randomly, and
frequency $\bm \Omega_{\rm 
  eff}$ fluctuates from the dot to the dot. However, in the pump-probe
experiments optical excitation takes place by the long train of
circularly polarized pulses, and nuclear spin polarization $\bm m$ may
change, both due to the interaction with an external field and with
electron spin. In the pump-probe regime, it leads to an unconventional
dynamics of electron and nuclear spins~\cite{A.Greilich09282007}.

\begin{figure}[hptb]
\includegraphics[width=0.85\linewidth]{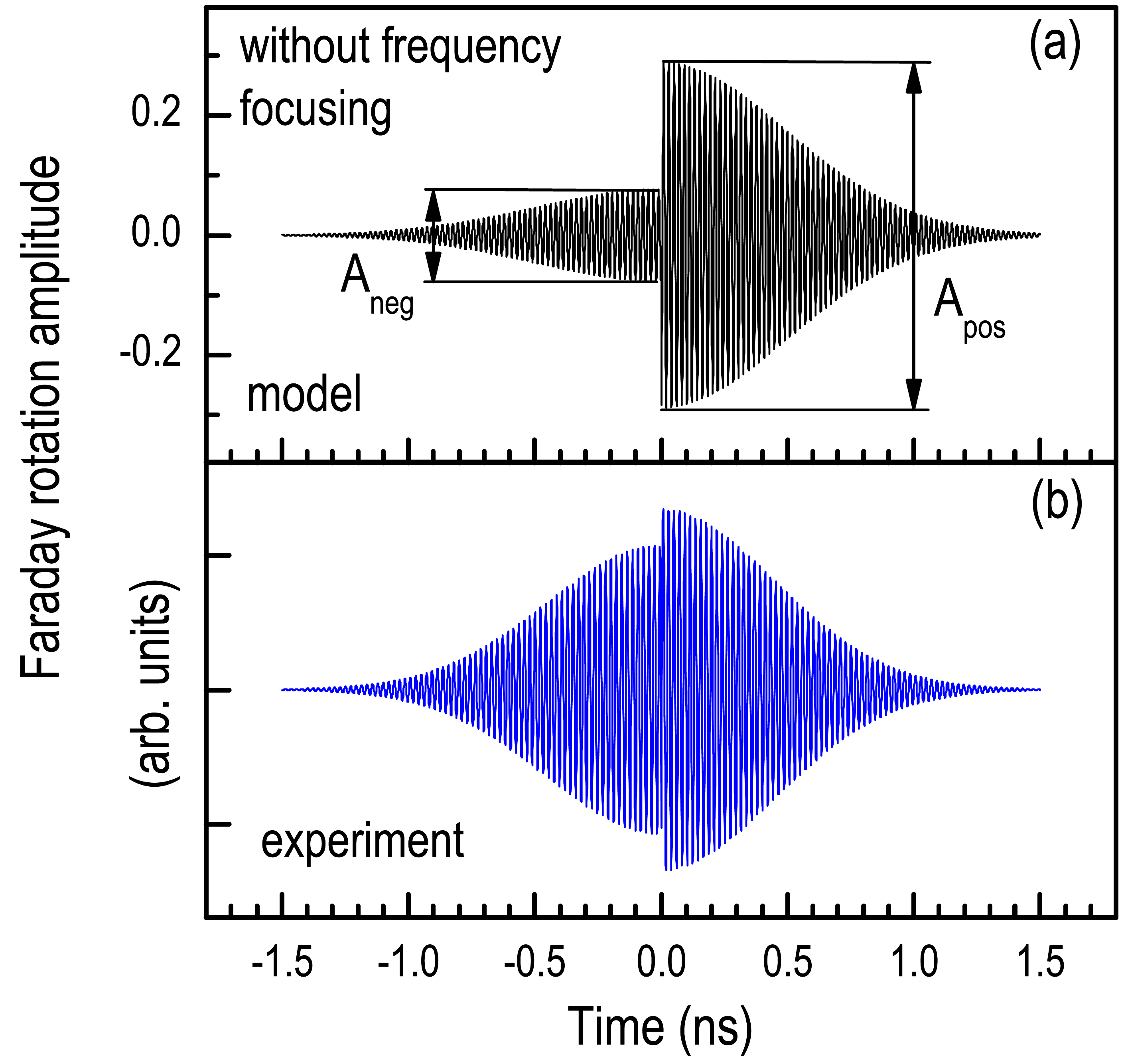}
\caption{(a) Faraday rotation signal from quantum dot array calculated
  for the experimental conditions~\cite{A.Greilich09282007} neglecting
  nuclear effects. (b) Experimentally measured Faraday
  signal. Measurements were carried out on a structure consisting of
  20 InGaAs/GaAs quantum dot layers with the dot density per layer of
  $10^{10}$~cm$^{-2}$, the structure doped in a such a way that there
  is one electron per dot on average. Data are reproduced from
  Ref.~\cite{A.Greilich09282007}.} \label{fig:ff} 
\end{figure}

Figure~\ref{fig:ff} shows calculated (a) and measured (b) spin Faraday
signals in the InGaAs quantum dot
array~\cite{A.Greilich09282007}. Figures shows dramatic difference of
signal amplitudes at negative delays in the experiment and in the
calculation which does not take into account nuclear effects. The
conclusion was drawn in Ref.~\cite{A.Greilich09282007} that it is the
interaction of electrons with nuclear spins being responsible for the
observed effect: in the process of electron spin coherence excitation
by the periodic pulse train nuclear spins orient in such a way, that
the electron spin precession period becomes a divisor of the pump
pulse repetition period.  

There are two theoretical approaches aimed at the description of the frequency
focusing process. In the model suggested in
Ref.~\cite{A.Greilich09282007}, see also Ref.~\cite{carter:167403},
the random nuclear spin flips are considered, which are caused by the
hyperfine interaction. The rate of these processes can be estimated
as~\cite{dp74,A.Greilich09282007}  
\begin{equation}
\label{gamma:efros}
\gamma \sim \frac{\alpha_{\rm hf}^2}{\Omega^2\tau_{\rm c}},
\end{equation}
where  $\tau_{\rm
  c}$ is the electron spin correlation time in the quantum dot. The
key assumption of this approach is that the electron spin correlation
time is governed by the processes of the pump pulse interaction with
the quantum dot, therefore an estimate holds~\cite{A.Greilich09282007}:
\begin{equation}
\tau_{\rm c} \sim \frac{T_R}{W_{\rm tr}}.
\end{equation}
Here $W_{\rm tr}$ is the trion formation probability by a single pump
pulse. In those dots where the phase synchronization condition with
allowance for the nuclear field is
fulfilled: 
\begin{equation}
\label{PSCn}
\Omega_{\rm eff} T_R = 2\pi N
\end{equation}
by the moment of the next pump pulse arrival $|S_z|=1/2$, and the trion is
not formed. Therefore,  $W_{\rm tr} =0$, correlation time $\tau_c \to
\infty$, and nuclear spin flips stop. In those dots where the spin
precession phase synchronization condition is not fulfilled,
nuclear spin flips take place until $\Omega_{\rm eff}$ changes in 
random manner in such a way that the condition Eq.~\eqref{PSCn} is
reached. Similar effects were discussed also in
Ref.~\cite{2010arXiv1006.5144K}. 

Alternative approach to the nuclei induced electron spin precession
frequency focusing is
suggested in Ref.~\cite{yugova11}. The description of the coupled
dynamics of electron and nuclear spins is carried out within a
classical model, where the electron spin, $\bm S$, and nuclear spin
fluctuation, $\bm m$, are treated as classical vectors, which precess
around the external field and around each other. The analysis of the
spin dynamics equations shows that the nuclear spin precession in the
external magnetic field with the frequency $\omega$ results, owing to
the hyperfine interaction, in  additional small oscillations of
electron spin with this very frequency. These electron spin
oscillations, in turn, affect the nuclear spin dynamics, in fact, they
cause nuclear magnetic resonance. As a result, the projection of
nuclear spin onto the external magnetic field, $m_x$, start to vary
slowly, until the electron spin precession frequency $\Omega_{\rm
  eff} \approx \Omega + \alpha_{\rm hf} m_x$ satisfies the
mode-locking condition Eq.~\eqref{PSCn}. This process is no more
random and its rate can be estimated as~\cite{yugova11}
\begin{equation}
\label{tau:nf}
\frac{1}{\tau_{\rm nf}} \sim \frac{\alpha_{\rm hf}^3m}{\omega\Omega^2T_R}.
\end{equation}
The rate of such a process is, by factor $\alpha_{\rm hf} m/\omega
\sim 10$, larger than that of the nuclear spin flips in the model of
random flips, Eq.~\eqref{gamma:efros}. Estimations by
Eq.~\eqref{tau:nf} show that in experimental
conditions~\cite{A.Greilich09282007} the nuclei spin focusing time
ranges from units to tens of seconds, which is in the satisfactory
agreement with the experimental data. For detailed description of the
coupled dynamics of electron and nuclear spins under the pump-probe
conditions further experimental studies are needed, in particular, the
analysis of the focusing time as function of the magnetic field and
pump pulse power.

\subsection{Buildup of Faraday rotation signal}\label{subsec:emerge:FR}

So far, we established the origins of the spin signals at negative
delays between the pump and probe pulses, shown in
Fig.~\ref{fig:exp1}. Let us turn now to the discussion of the last
bright experimental fact: the amplitude of the Faraday rotation signal,
related with the resident electron (top curve in
Fig.~\ref{fig:exp1}(c)) increases with time before decaying. It is
especially remarkable at negative delays: with an increase of $|\Delta
t|$ spin beats amplitude first increases, and decreases afterwards.
It is evident, that nuclear effects can not determine such a behavior:
firstly, the buildup of Faraday rotation signal goes fast,
approximately during  $0.5$~ns, and, secondly, the ellipticity signal
behavior is quite standard, the oscillation amplitude decays with
time. Therefore, nonmonotonic behavior of the Faraday signal amplitude
can be related only with the specifics of this signal spectral sensitivity.

In order to describe this effect qualitatively and quantitatively, we
note that electron $g$-factor depends on the localization energy of
this charge carrier. Indeed, the renormalization of the $g$-factor in
direct band semiconductors is determined mostly by the admixture of
the valence band states to the conduction band
states~\cite{PhysRev.114.90,ivchenko_kiselev92:eng,ivchenko05a}. Since
trion excitation energy is related with the electron localization
energy, the resident carrier  $g$-factor in a quantum dot is coupled
with the optical transition frequency $\omega_0^{\mathrm T}$. This
dependence can be quite accurately described by a linear function~\cite{A.Greilich07212006,PhysRevB.75.245302}:
\begin{equation} \label{g:omega0}
|g_e(\omega_0^{\mathrm T})| = a \hbar\omega_0^{\mathrm T} + c\:,
\end{equation}
where  $a$ and $c$ are some parameters dependent on the quantum dot
ensemble material.

The pump pulse excites out of a broad distribution of quantum dots
over energy an ensemble, whose spectral width amounts to $\hbar/\tau_p \sim
1$~meV for pulses with duration $\tau_p \sim 1$~ps. We introduce a
function $S_z^+(\omega_0^{\mathrm T}, \Omega,\omega_{\rm P})$,  which
describes the magnitude of the electron spin $z$ component in the
quantum dot with the resonant frequency $\omega_0^{\mathrm T}$ and
spin precession frequency $
\Omega$ right after the pump pulse arrival with the carrier frequency
$\omega_{\rm P}$.  Note, that the Larmor frequency in a quantum dot is
determined, in general, not only the $g$-factor
value Eq.~\eqref{g:omega0}, but also the nuclear spin polarization
fluctuation. Spin ellipticity $\mathcal
E(t)$ and Faraday rotation $\mathcal F(t)$ signals detected by a probe
pulse with the carrier frequency
$\omega_{\rm pr}$ are given (up to a common factor) in accordance with
Eq.~\eqref{freq} by the expression~\cite{yugova09,glazov2010a}:
\begin{multline}
\label{signals}
\mathcal E(t) + \mathrm i \mathcal F(t) = \int \mathrm d\omega_0^{\mathrm
  T} \mathrm d\Omega p(\omega_0^{\mathrm
  T},\Omega) G(\omega_{\rm pr} - \omega_0^{\mathrm T}) \times \\
S_z^{+}(\omega_0^{\mathrm T},\Omega, \omega_{\rm P})
\cos{[\Omega t + \varphi]} \exp(- t/\tau_s) . 
\end{multline}
Here the delay between the pump and probe pulse $t>0$, function
$p(\omega_0^{\mathrm T},\Omega)$ is the joint distribution of optical
and Larmor frequencies in quantum dots (in the absence of nuclear
fluctuations $p(\omega_0^{\mathrm T},\Omega) = \delta[\Omega -
g(\omega_0^{\mathrm T})\mu_B 
B/\hbar]$), and
$G(\Lambda)$ describes spectral sensitivity of the signals, see
Eq.~\eqref{GL}. Last two factors in Eq.~\eqref{signals} describe the
single spin
dynamics in the quantum dot, 
$\tau_s$ is the spin relaxation time and 
$\varphi=\varphi(\omega_0^{\mathrm T},\Omega,\omega_{\rm P})$ is the
initial phase of the spin precession. Functions  
$S_z^{+}(\omega_0^{\mathrm T},\Omega, \omega_{\rm P})$ and
$\varphi(\omega_0^{\mathrm T},\Omega,\omega_{\rm P})$ can be found
from the general solution of spin dynamics equations, given in
Ref.~\cite{yugova09}. 

The detailed analysis and modeling of electron spin dynamics in
quantum dots described by Eq.~\eqref{signals} is carried out in
Ref.~\cite{glazov2010a}. Here we consider a simplest model, which allows
us to obtain qualitative description of the situation. Let us take
that $G(\Lambda)$ has the form
\begin{equation}
\label{Glambda}
G(\Lambda) =  (1+ 2 \mathrm i \Lambda \tau_p)\exp{[-(\Lambda\tau_p)^2]}.
\end{equation}
At moderate values of detunings, $\Lambda=\omega_{\rm pr} -
\omega_0^{\mathrm T}$, between the probe carrier frequency and
resonance frequency of the quantum dot,
$\Lambda\tau_p\lesssim 1$, function $G(\Lambda)$ given by
Eq.~\eqref{Glambda} has a shape similar to that for Rosen \& Zener
pulse [see Eq.~\eqref{G}]. At 
$\Lambda\tau_p \gg 1$ the imaginary part of  $G$ (that is Faraday
rotation signal sensitivity) decays faster than the exact function,
which behaves as $\Im{G(\Lambda)} \sim 1/(\Lambda\tau_p)$. It results
only
in quantitative difference of the Faraday signal behavior calculated
in this model and obtained in the exact
calculation~\cite{glazov2010a}. We assume further, that nuclear
effects are absent and spin precession frequency is stringently linked
with the quantum dot resonance frequency $\Omega(\omega_0^{\mathrm T})
= g(\omega_0^{\mathrm T}) \mu_B B /\hbar$. 
Moreover, we choose  $S_z^+$ function in the form
\begin{equation}
\label{sz0}
S_z^{+}(\omega_0^{\mathrm T}, \omega_{\rm P}) = S_0
\exp{[-(\omega_0^{\mathrm T} - \omega_{\rm P})^2\tau_p^2]}, 
\end{equation}
where $S_0$ is some constant which depends on the pump pulse area and
we set $\varphi\equiv 0$, $\tau_s \to \infty$. The effects related
with the spin precession 
mode-locking are discussed below.

Integration in Eq.~\eqref{signals} gives
\begin{subequations}
\label{ell1far1}
\begin{equation}
\label{ell1}
\mathcal E(t) = \sqrt{\frac{\pi}{2\tau_p^2}}
\exp{\left[\frac{-\Delta^2\tau_p^2/(2\hbar^2)-(\Omega't)^2}{8\tau_p^2}\right]}
\cos{\left(\tilde{\Omega}_0  t\right)} , 
\end{equation}
\begin{multline}
\label{far1}
\mathcal F(t) =
\frac{1}{2}
\sqrt{\frac{\pi}{2\tau_p^2}}
 \exp{\left[
 \frac{-\Delta^2\tau_p^2/(2\hbar^2)-(\Omega't)^2}{8\tau_p^2}
 \right]}\times \\
\left[\frac{2\Delta\tau_p}{\hbar}
\cos{\left(\tilde{\Omega}_0 t\right)}
+\frac{\Omega' t}{\tau_p}
\sin{\left(\tilde{\Omega}_0 t\right)}
\right].
\end{multline}
\end{subequations}
Here the following notations are introduced: $\Omega'=\mathrm
d\Omega/\mathrm d\omega_0^{\mathrm T}$,
$\Delta/\hbar = \omega_{\rm P} - \omega_{\rm pr}$ is the pump and
probe pulses detuning, $\tilde{\Omega}_0 = \Omega_0 +
\Omega'\Delta/(2\hbar)$ is the observed spin precession frequency, and
$\hbar\Omega_0=g_e(\omega_{ \rm pr})\mu_B B$.

It is seen from Eqs.~\eqref{ell1far1} that the temporal dependence of
Faraday rotation and ellipticity signals can be qualitatively
different. The ellipticity signal amplitude simply decays with time,
this decay rate is determined by the spread of Larmor frequencies of
``excited'' electrons. Faraday signal has two contributions [see
Eq.~\eqref{far1}]: first one is similar to the ellipticity, but its
amplitude depends sharply on the detuning, $\sim \Delta \tau_p$, white
the amplitude of the second contribution  ($\propto \sin{\tilde
  \Omega_0 t}$) contains the linear in time factor. For degenerate
pump and probe pulses ($\Delta =0$) Faraday signal, described by the
second term in brackets of Eq.~\eqref{far1}, first grows and
afterwards decay, in agreement with experimental data, presented in
Fig.~\ref{fig:exp1}. 

\begin{figure}[hptb]
\includegraphics[width=\linewidth]{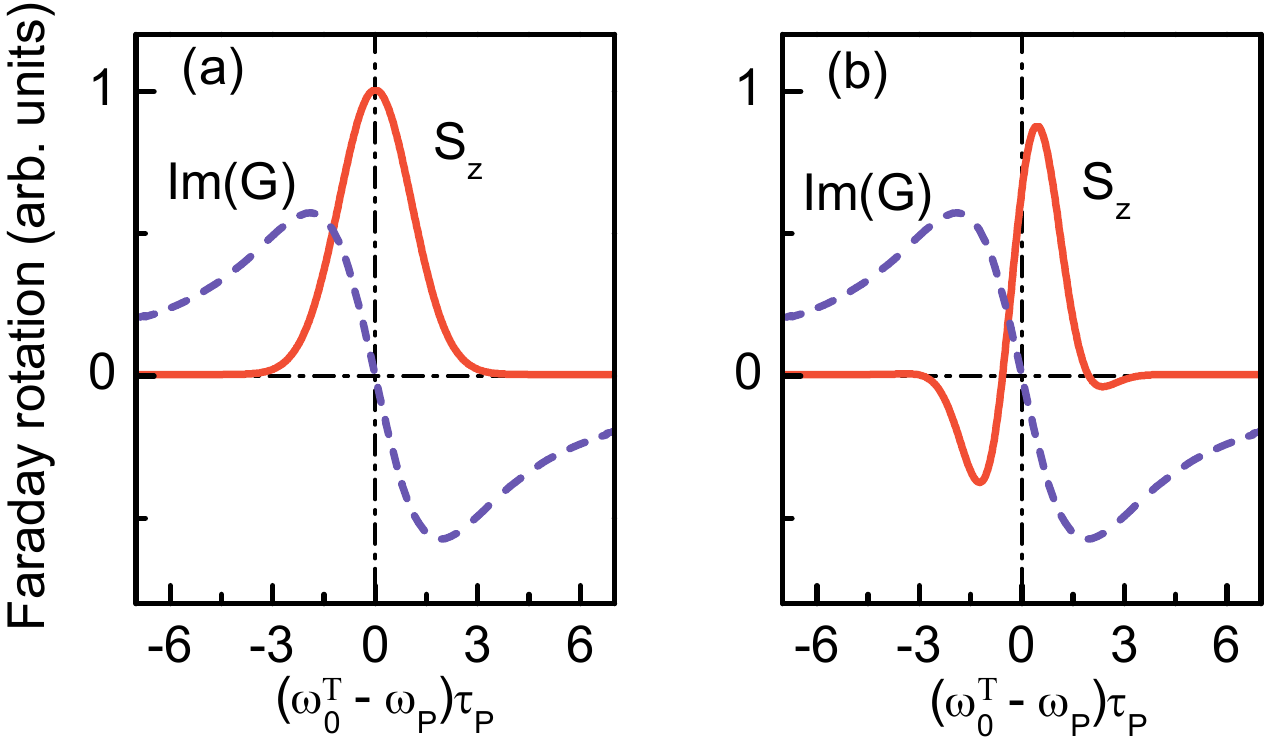}
\caption{Schematic illustration of Faraday rotation signal formation
  for spectrally degenerate pump and probe pulses, 
  $\omega_{\rm pr} = \omega_{\rm P}$. Panel (a) corresponds to the
  zero delay between 
  the pump and probe pulses, panel (b) corresponds to the positive delay,
  $t>0$. Solid curve shows the distribution of the spin $z$ component,
  while dashed one shows the spectral sensitivity of Faraday signal, $\Im
  G(\omega_0^{\rm T}
  - \omega_{\rm pr})$. Data are reproduced from
  Ref.~\cite{glazov2010a}.} \label{fig:detect_farad}  
\end{figure}

Such a temporal behavior of spin signals is related with the different
spectral sensitivity of the Faraday and ellipticity signals, and it is
a direct consequence of the correlation between the Larmor frequency
and optical transition frequency in a quantum
dot, Eq.~\eqref{g:omega0}. Figure~\ref{fig:detect_farad} illustrates the
spin Faraday signal formation under the conditions of degenerate pump
and probe pulses. At $t=0$ the spin distribution is a symmetric
function of  $\omega_0^{\mathrm T}-\omega_{\rm P}$ and it makes no
contribution to the Faraday signal, since it is determined by the
convolution of  $S_z^+$ and odd function $\Im{G(\Lambda)}$, as shown
in Fig.~\ref{fig:detect_farad}(a). The spin distribution becomes
asymmetric with a course of time, since [for $a>0$, $c>0$
in Eq.~\eqref{g:omega0}] spins in quantum dots with larger optical
transition energies, $\omega_0^{\mathrm T}$, precess faster than those
in dots with smaller transition energies. Therefore, with an increase
of the pump and probe pulses temporal separation, the spin distribution
function becomes asymmetric with respect to the carrier frequency of
the optical pulse, as shown in
Fig.~\ref{fig:detect_farad}(b). Therefore, Faraday rotation signal
becomes nonzero at $t>0$. At large enough delays electron spin
dephases and Faraday rotation decays.

\begin{figure}[htb]
\includegraphics[width=\linewidth]{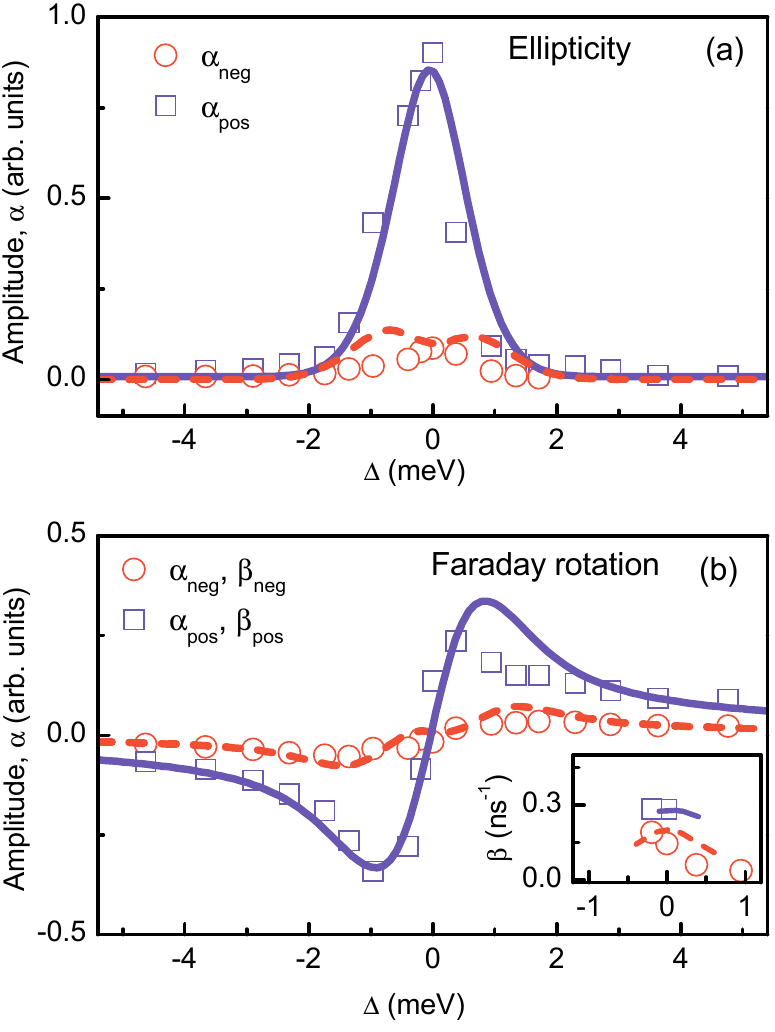}
\caption{Amplitudes of ellipticity  (a) and Faraday rotation (b)
  signals as functions of the detuning between the pump and probe
  pulses. Circles show amplitudes of the decaying contributions to
  spin signals at negative delays, $\alpha_{\rm neg}$, squares show
  the amplitudes for positive ones, $\alpha_{\rm pos}$. In inset to panel (b)
  the amplitudes of the growing with time component of the Faraday
  rotation signal are presented: $\beta_{\rm neg}$ (circles) at
  negative delays and $\beta_{\rm pos}$ (squares) at positive
  ones. Solid lines are the modeling results. Measurements were
  carried out on the structure consisting of 20 InGaAs/GaAs quantum dot
  layers with dot density per layer being $10^{10}$~cm$^{-2}$, the
  structure is doped in such a way, that there is one electron per
  dot on average. Data are reproduced from Ref.~\cite{glazov2010a}.}
\label{fig:amp1}
\end{figure}

Note, that the detuning between the probe and pump pulses already
introduces asymmetry into the spin distribution with respect to the
probe pulse carrier frequency, $\omega_{\rm pr}$, and results in the
Faraday signal formation even at $t=0$. Therefore, at  $\Delta \ne
0$ the decaying with time component of the spin Faraday signal
appears, which is described by the first term in brackets of
Eq.~\eqref{far1}. The ellipticity spectral sensitivity,
$\Re{G(\Lambda)}$, is even, therefore ellipticity signal reflects
ensemble average $z$ spin component. Induced ellipticity decays as a
function of time due to the spread of the Larmor frequencies. It is in
the agreement with the experimental data shown in Fig.~\ref{fig:exp1}.

The model suggested qualitatively describes also the differences between the
Faraday rotation and ellipticity at negative delays. In the spin
precession mode-locking regime, considered in
Sec.~\ref{subsubsec:modelock}, the distribution
function of spin  $z$ component,
$S_z^+(\omega_0^{\mathrm T},\Omega, \omega_{\rm P})$, has sharp maxima
for those quantum dots, where $\Omega(\omega_0^{\mathrm T})T_R=2\pi
N$. If one allows for the synchronized modes only, the spin signals
become even functions of the pump-probe delay, $t$. It implies, that
the Faraday signal at the zero detuning and 
$\Delta t<0$ first builds-up and afterwards decays with an increase of
$|\Delta t|$, see Fig.~\ref{fig:exp1}(b).  Presence of other spin
precession frequencies gives rise to the additional contribution to
the signal, which decays at $t>0$ and which is absent at negative
delays. Note, that the focusing of electron spin precession frequency
induced by the electron interaction with the lattice nuclei
breaks the correlation between optical transition and spin precession
frequencies and weakens the growing part of Faraday signal.

The simple model outlined here is not free of drawbacks: due to
simplified form of the function Eq.~\eqref{Glambda}, it does not
describe the amplitudes of Faraday signal at large detunings. For the
same reasons, the spectral dependence of $g$-factor, extracted
from the Faraday rotation experiment is different from the predicted
by Eq.~\eqref{far1}. The complete description of experimental data
shown in Fig.~\ref{fig:exp1}, was carried out in
Ref.~\cite{glazov2010a}. Microscopic calculation by using exact function
$G(\Lambda)$ and functions $S_z^+(\omega_0^{\mathrm T},\Omega, \omega_{\rm P})$,
$\varphi(\omega_0^{\mathrm T}, \Omega,\omega_{\rm P})$, obtained with
allowance for the spin precession mode-locking, are in good agreement
with experiment. Comparison of the theory and experiment is presented
in Fig.~\ref{fig:amp1}. Figure shows spectral dependence of the
amplitudes of induced ellipticity [panel (a)] and Faraday rotation
[panel (b)]. Circles and squares are the experimental data obtained at
negative and positive delays, respectively, by means of experimental
data fitting using the formula: 
\[
\mathcal S \propto \left[\alpha\cos{\Omega t} + \beta t\sin{\Omega t}
\right] \exp{\left[-\frac{t^2}{{(T_2^*)}^2}\right]}.
\]
Curves in Fig.~\ref{fig:amp1} are the result of the
calculation. Figure shows amplitudes, $\alpha$, of the Faraday and
ellipticity signals decaying components. The inset to panel (b) shows
amplitudes of the growing in time contribution to the Faraday signal,
$\beta$. A good agreement of spectral dependence of signal amplitudes
is seen from the Figure.

Thus, spin Faraday and ellipticity signals in inhomogeneous arrays of
quantum dots are formed by different resident electron ensembles. As a
result, their behavior as function of the pump and probe pulses delay
can be qualitatively different. It is most brightly manifested in the
buildup of the Faraday rotation signal as a function of time, being a
result of the link between the electron $g$-factor and optical transition
energy in quantum dot.

\section{Conclusion}\label{sec:conclusions}
 
In this review, the electron, exciton and trion spin dynamics in semiconductor
nanostructures observed by the pump-probe method is examined. The main
physical mechanisms of resident electron spin 
orientation at excitation of excitons and trions by circularly
polarized light pulses are established, the principles of charge
carriers spin coherence detection by linearly polarized optical pulses
are considered. The possibilities to control spins by light pulses of
different polarization are analyzed.

Two approaches to describe electron spins interaction with optical
pulses are suggested. One of those is macroscopic, it is based on the
consideration of electron and electron-hole complexes ensemble. In the
other, microscopic, approach, the
single electron interaction with the optical pulse is treated within a two
level model.

We have demonstrated the electron spin dynamics specifics in quantum
well and quantum dot structures under excitation of spins by a
periodic train of pulses. The spin precession mode-locking and the
focusing of electron spin precession frequencies by means of charge
carrier interaction with lattice nuclei are examined. It is shown that
the different spin signals in the pump-probe method: Faraday and Kerr
rotation, as well as induced ellipticity are sensitive to the spin
dynamics in different electron ensembles. It may result in the
different temporal behavior of spin Faraday and ellipticity signals,
this fact should be taken into account at extraction of electron and
hole spin dynamics parameters from experimental data.

The applications of spin Faraday and Kerr signals for studies of
nuclear spin dynamics in semi\-con\-duct\-ors and semiconductor
nanostructures as well as for studies of spin-dependent processes in
semiconductor microcavities may serve as prospective directions for
the development of the pump-probe method.

Author is grateful to M. Bayer, M.R. Vladimirova, E.L. Ivchenko,
T. Korn, Al.L. Efros, I.A. Yugova, and D.R. Yakovlev for help and
useful discussions. This work was partially supported by the grants of
RFBR, Dynasty Foundation -- ICFPM, and EU projects: SpinOptronics and
POLAPHEN.

% %\bibliographystyle{gost71u_mg}
% %\bibliographystyle{apsrev4-1long_mg}
% \bibliographystyle{misha}
% \bibliography{/home/misha/Work/Coherent/Bibliography/all}
% %\bibliography{/media/1302-316D/Work/Coherent/Bibliography/all}

\begin{thebibliography}{100}
\providecommand{\selectlanguage}[1]{\relax}
%%
%% This is file `babelbst.tex',
%% generated with the docstrip utility.
%%
%% The original source files were:
%%
%% tktl.dtx  (with options: `babelbst')
%% 
%% This is a generated file.
%% 
%% Copyright (C) 2002 Mikael Puolakka.
%% 
%% This file may be distributed and/or modified under the
%% conditions of the LaTeX Project Public License, either
%% version 1.2 of this license or (at your option) any later
%% version. The latest version of this license is in
%% 
%%    http://www.latex-project.org/lppl.txt
%% 
%% and version 1.2 or later is part of all distributions of
%% LaTeX version 1999/12/01 or later.
%% 
%% \CharacterTable
%%  {Upper-case    \A\B\C\D\E\F\G\H\I\J\K\L\M\N\O\P\Q\R\S\T\U\V\W\X\Y\Z
%%   Lower-case    \a\b\c\d\e\f\g\h\i\j\k\l\m\n\o\p\q\r\s\t\u\v\w\x\y\z
%%   Digits        \0\1\2\3\4\5\6\7\8\9
%%   Exclamation   \!     Double quote  \"     Hash (number) \#
%%   Dollar        \$     Percent       \%     Ampersand     \&
%%   Acute accent  \'     Left paren    \(     Right paren   \)
%%   Asterisk      \*     Plus          \+     Comma         \,
%%   Minus         \-     Point         \.     Solidus       \/
%%   Colon         \:     Semicolon     \;     Less than     \<
%%   Equals        \=     Greater than  \>     Question mark \?
%%   Commercial at \@     Left bracket  \[     Backslash     \\
%%   Right bracket \]     Circumflex    \^     Underscore    \_
%%   Grave accent  \`     Left brace    \{     Vertical bar  \|
%%   Right brace   \}     Tilde         \~}
%%
%% This is file `babelbst.tex',
%% generated with the docstrip utility.
%%
%% The original source files were:
%%
%% merlin.mbs  (with options: `bblbst')
%%
%% Copyright 1994-1999 Patrick W Daly
 % ===============================================================
 % IMPORTANT NOTICE:
 % This bibliographic style (bst) file has been generated from one or
 % more master bibliographic style (mbs) files, listed above.
 %
 % This generated file can be redistributed and/or modified under the terms
 % of the LaTeX Project Public License Distributed from CTAN
 % archives in directory macros/latex/base/lppl.txt; either
 % version 1 of the License, or any later version.
 % ===============================================================
 % Name and version information of the main mbs file:
 % \ProvidesFile{merlin.mbs}[1999/05/28 3.89 (PWD)]
 % This is babelbst.tex for English.
 % It should serve as a model for other languages.
 % Alternatively, store it under a different name (e.g. englbst.tex)
 % and then \input it with a command in babelbst.tex.
%%
%% This is file `englbst.tex',
%% generated with the docstrip utility.
%%
%% The original source files were:
%%
%% tktl.dtx  (with options: `englbst')
%% 
%% This is a generated file.
%% 
%% Copyright (C) 2002 Mikael Puolakka.
%% 
%% This file may be distributed and/or modified under the
%% conditions of the LaTeX Project Public License, either
%% version 1.2 of this license or (at your option) any later
%% version. The latest version of this license is in
%% 
%%    http://www.latex-project.org/lppl.txt
%% 
%% and version 1.2 or later is part of all distributions of
%% LaTeX version 1999/12/01 or later.
%% 
%% \CharacterTable
%%  {Upper-case    \A\B\C\D\E\F\G\H\I\J\K\L\M\N\O\P\Q\R\S\T\U\V\W\X\Y\Z
%%   Lower-case    \a\b\c\d\e\f\g\h\i\j\k\l\m\n\o\p\q\r\s\t\u\v\w\x\y\z
%%   Digits        \0\1\2\3\4\5\6\7\8\9
%%   Exclamation   \!     Double quote  \"     Hash (number) \#
%%   Dollar        \$     Percent       \%     Ampersand     \&
%%   Acute accent  \'     Left paren    \(     Right paren   \)
%%   Asterisk      \*     Plus          \+     Comma         \,
%%   Minus         \-     Point         \.     Solidus       \/
%%   Colon         \:     Semicolon     \;     Less than     \<
%%   Equals        \=     Greater than  \>     Question mark \?
%%   Commercial at \@     Left bracket  \[     Backslash     \\
%%   Right bracket \]     Circumflex    \^     Underscore    \_
%%   Grave accent  \`     Left brace    \{     Vertical bar  \|
%%   Right brace   \}     Tilde         \~}
%%
%% This is file `englbst.tex',
%% generated with the docstrip utility.
%%
%% The original source files were:
%%
%% merlin.mbs  (with options: `bblbst')
%%
%% Copyright 1994-1999 Patrick W Daly
 % ===============================================================
 % IMPORTANT NOTICE:
 % This bibliographic style (bst) file has been generated from one or
 % more master bibliographic style (mbs) files, listed above.
 %
 % This generated file can be redistributed and/or modified under the terms
 % of the LaTeX Project Public License Distributed from CTAN
 % archives in directory macros/latex/base/lppl.txt; either
 % version 1 of the License, or any later version.
 % ===============================================================
 % Name and version information of the main mbs file:
 % \ProvidesFile{merlin.mbs}[1999/05/28 3.89 (PWD)]
 % This is babelbst.tex for English.
 % It should serve as a model for other languages.
 % Alternatively, store it under a different name (e.g. englbst.tex)
 % and then \input it with a command in babelbst.tex.
\def\bbland{and}
\def\bbletal{et~al} % Piste sijoitetaan tarvittaessa perддn!
\def\bbleditors{editors}        \def\bbleds{eds.}
\def\bbleditor{editor}          \def\bbled{ed.}
\def\bbledby{edited by}
\def\bbledition{edition}        \def\bbledn{edn.}
\def\bblvolume{volume}          \def\bblvol{vol.}
\def\bblof{of}
\def\bblnumber{number}          \def\bblno{no.}
\def\bblin{in}
\def\bblpages{pages}            \def\bblpp{pp.}
\def\bblpage{page}              \def\bblp{p.}
\def\bblchapter{chapter}        \def\bblchap{chap.}
\def\bbltechreport{Technical Report}
\def\bbltechrep{Tech. Rep.}
\def\bblmthesis{Master's thesis}
\def\bblphdthesis{Ph.D. thesis}
\def\bblfirst{first}            \def\bblfirsto{1st}
\def\bblsecond{second}          \def\bblsecondo{2nd}
\def\bblthird{third}            \def\bblthirdo{3rd}
\def\bblfourth{fourth}          \def\bblfourtho{4th}
\def\bblfifth{fifth}            \def\bblfiftho{5th}
\def\bblst{st}  \def\bblnd{nd}  \def\bblrd{rd}
\def\bblth{th}
\def\bbljan{January}  \def\bblfeb{February}  \def\bblmar{March}
\def\bblapr{April}    \def\bblmay{May}       \def\bbljun{June}
\def\bbljul{July}     \def\bblaug{August}    \def\bblsep{September}
\def\bbloct{October}  \def\bblnov{November}  \def\bbldec{December}
\def\bblins{in} % Oma lisдys: "in" series.
\def\bblinj{in} % Oma lisдys: "in" journal.
%%
%%
%% End of file `babelbst.tex'.
%% 
%%
%% End of file `englbst.tex'.

%%
%% This is file `ruslbst.tex',
%% generated with the docstrip utility.
%%
%% The original source files were:
%%
%% tktl.dtx  (with options: `englbst')
%% 
%% This is a generated file.
%% 
%% Copyright (C) 2002 Mikael Puolakka.
%% 
%% This file may be distributed and/or modified under the
%% conditions of the LaTeX Project Public License, either
%% version 1.2 of this license or (at your option) any later
%% version. The latest version of this license is in
%% 
%%    http://www.latex-project.org/lppl.txt
%% 
%% and version 1.2 or later is part of all distributions of
%% LaTeX version 1999/12/01 or later.
%% 
%% \CharacterTable
%%  {Upper-case    \A\B\C\D\E\F\G\H\I\J\K\L\M\N\O\P\Q\R\S\T\U\V\W\X\Y\Z
%%   Lower-case    \a\b\c\d\e\f\g\h\i\j\k\l\m\n\o\p\q\r\s\t\u\v\w\x\y\z
%%   Digits        \0\1\2\3\4\5\6\7\8\9
%%   Exclamation   \!     Double quote  \"     Hash (number) \#
%%   Dollar        \$     Percent       \%     Ampersand     \&
%%   Acute accent  \'     Left paren    \(     Right paren   \)
%%   Asterisk      \*     Plus          \+     Comma         \,
%%   Minus         \-     Point         \.     Solidus       \/
%%   Colon         \:     Semicolon     \;     Less than     \<
%%   Equals        \=     Greater than  \>     Question mark \?
%%   Commercial at \@     Left bracket  \[     Backslash     \\
%%   Right bracket \]     Circumflex    \^     Underscore    \_
%%   Grave accent  \`     Left brace    \{     Vertical bar  \|
%%   Right brace   \}     Tilde         \~}
%%
%% This is file `englbst.tex',
%% generated with the docstrip utility.
%%
%% The original source files were:
%%
%% merlin.mbs  (with options: `bblbst')
%%
%% Copyright 1994-1999 Patrick W Daly
 % ===============================================================
 % IMPORTANT NOTICE:
 % This bibliographic style (bst) file has been generated from one or
 % more master bibliographic style (mbs) files, listed above.
 %
 % This generated file can be redistributed and/or modified under the terms
 % of the LaTeX Project Public License Distributed from CTAN
 % archives in directory macros/latex/base/lppl.txt; either
 % version 1 of the License, or any later version.
 % ===============================================================
 % Name and version information of the main mbs file:
 % \ProvidesFile{merlin.mbs}[1999/05/28 3.89 (PWD)]
 % This is babelbst.tex for English.
 % It should serve as a model for other languages.
 % Alternatively, store it under a different name (e.g. englbst.tex)
 % and then \input it with a command in babelbst.tex.
\def\bbland{и}
\def\bbletal{и др.} 
\def\bbleditors{ред.}        \def\bbleds{ред.}
\def\bbleditor{ред.}          \def\bbled{ред.}
\def\bbledby{п. ред.}
\def\bbledition{вып.}        \def\bbledn{вып.}
\def\bblvolume{том}          \def\bblvol{том}
\def\bblof{of}
\def\bblnumber{номер}          \def\bblno{н.}
\def\bblin{в}
\def\bblpages{стр.}            \def\bblpp{стр.}
\def\bblpage{стр.}              \def\bblp{стр.}
\def\bblchapter{глава}        \def\bblchap{глава.}
\def\bbltechreport{Technical Report}
\def\bbltechrep{Tech. Rep.}
\def\bblmthesis{Master's thesis}
\def\bblphdthesis{Ph.D. thesis}
\def\bblfirst{первый}            \def\bblfirsto{1ый}
\def\bblsecond{второй}          \def\bblsecondo{2ой}
\def\bblthird{третий}            \def\bblthirdo{3ий}
\def\bblfourth{четверый}          \def\bblfourtho{4ый}
\def\bblfifth{пятый}            \def\bblfiftho{5ый}
\def\bblst{st}  \def\bblnd{nd}  \def\bblrd{rd}
\def\bblth{th}
\def\bbljan{Январь}  \def\bblfeb{Февраль}  \def\bblmar{Март}
\def\bblapr{Апрель}    \def\bblmay{Май}       \def\bbljun{Июнь}
\def\bbljul{Июль}     \def\bblaug{Август}    \def\bblsep{Сентябрь}
\def\bbloct{Октябрь}  \def\bblnov{Ноябрь}  \def\bbldec{Декабрь}
\def\bblins{в} 
\def\bblinj{в} 
%%
%%
%% End of file `babelbst.tex'.
%% 
%%
%% End of file `ruslbst.tex'.

%%
%%
%% End of file `babelbst.tex'.
%% 
%%
%% End of file `babelbst.tex'.

\newcommand{\Capitalize}[1]{\uppercase{#1}}
\newcommand{\capitalize}[1]{\expandafter\Capitalize#1}

\bibitem{lampel68}
G.~Lampel.
\newblock \emph{Nuclear Dynamic Polarization by Optical Electronic Saturation
  and Optical Pumping in Semiconductors}.
\newblock Phys. Rev. Lett. \textbf{20}, 491 (1968).

\bibitem{optor:eng}
{\selectlanguage{english}\emph{Optical orientation}}/ Eds. F. Meier
and B. P. Zakharchenya, North-Holland, Amsterdam  (1984).

\bibitem{PhysRevLett.55.1128}
D.~D. Awschalom, J.~M. Halbout, S.~von Molnar, T.~Siegrist, F.~Holtzberg.
\newblock \emph{Dynamic Spin Organization in Dilute Magnetic Systems}.
\newblock Phys. Rev. Lett. \textbf{55}, 1128 (1985).

\bibitem{Zheludev1994823}
N.~I. Zheludev, M.~A. Brummell, R.~T. Harley, A.~Malinowski, S.~V. Popov, D.~E.
  Ashenford, B.~Lunn.
\newblock \emph{Giant specular inverse $\mbox{F}$araday effect in
  $\mbox{Cd}_{0.6}\mbox{Mn}_{0.4}\mbox{Te}$}.
\newblock Solid State Communications \textbf{89}, 823  (1994).

\bibitem{Kikkawa98}
J.~M. Kikkawa, D.~D. Awschalom.
\newblock \emph{Resonant Spin Amplification in $n$-Type $\mbox{GaAs}$}.
\newblock Phys. Rev. Lett. \textbf{80}, 4313 (1998).

\bibitem{Kikkawa29081997}
J.~M. Kikkawa, I.~P. Smorchkova, N.~Samarth, D.~D. Awschalom.
\newblock \emph{Room-Temperature Spin Memory in Two-Dimensional Electron
  Gases}.
\newblock Science \textbf{277}, 1284 (1997).

\bibitem{A.Greilich07212006}
A.~Greilich, D.~R. Yakovlev, A.~Shabaev, A.~L. Efros, I.~A. Yugova, R.~Oulton,
  V.~Stavarache, D.~Reuter, A.~Wieck, M.~Bayer.
\newblock \emph{{Mode Locking of Electron Spin Coherences in Singly Charged
  Quantum Dots}}.
\newblock Science \textbf{313}, 341 (2006).

\bibitem{brand02}
M.~A. Brand, A.~Malinowski, O.~Z. Karimov, P.~A. Marsden, R.~T. Harley, A.~J.
  Shields, D.~Sanvitto, D.~A. Ritchie, M.~Y. Simmons.
\newblock \emph{Precession and Motional Slowing of Spin Evolution in a High
  Mobility Two-Dimensional Electron Gas}.
\newblock Phys. Rev. Lett. \textbf{89}, 236601 (2002).

\bibitem{leyland06}
W.~J.~H. Leyland, G.~H. John, R.~T. Harley, M.~M. Glazov, E.~L. Ivchenko, D.~A.
  Ritchie, I.~Farrer, A.~J. Shields, M.~Henini.
\newblock \emph{Enhanced spin-relaxation time due to electron-electron
  scattering in semiconductor quantum wells}.
\newblock Phys. Rev. B \textbf{75}, 165309 (2007).

\bibitem{PhysRevB.80.241314}
M.~Griesbeck, M.~M. Glazov, T.~Korn, E.~Y. Sherman, D.~Waller, C.~Reichl,
  D.~Schuh, W.~Wegscheider, C.~Sch\"uller.
\newblock \emph{Cyclotron effect on coherent spin precession of two-dimensional
  electrons}.
\newblock Phys. Rev. B \textbf{80}, 241314 (2009).

\bibitem{Kikkawa1999}
J.~M. Kikkawa, D.~D. Awschalom.
\newblock \emph{Lateral drag of spin coherence in gallium arsenide}.
\newblock Nature \textbf{397}, 139 (1999).

\bibitem{kato04}
Y.~K. Kato, R.~C. Myers, A.~C. Gossard, D.~D. Awschalom.
\newblock \emph{Observation of the Spin $\mbox{H}$all Effect in
  Semiconductors}.
\newblock Science \textbf{306}, 1910 (2004).

\bibitem{PhysRevLett.94.236601}
S.~A. Crooker, D.~L. Smith.
\newblock \emph{Imaging Spin Flows in Semiconductors Subject to Electric,
  Magnetic, and Strain Fields}.
\newblock Phys. Rev. Lett. \textbf{94}, 236601 (2005).

\bibitem{awschalom_book}
D.~Awschalom, D.~Loss, N.~Samarth (eds.).
\newblock \emph{Semiconductor spintronics and quantum computation} (Springer:
  Berlin, New York, 2002).

\bibitem{dyakonov_book}
M.~I. Dyakonov (ed.).
\newblock \emph{Spin Physics in Semiconductors} (Springer-Verlag: Berlin,
  Heidelberg, 2008).

\bibitem{ssc:optor}
Y.~Kusraev, G.~Landwehr (eds.).
\newblock \emph{Semiconductor Science and Technology, Special Issue: Optical
  Orientation}, vol.~23 (IOP Publishing, 2008).

\bibitem{aronovivchenko:eng}
A.~G. Aronov, E.~L. Ivchenko.
\newblock \emph{Dichroism and optical anisotropy of media with oriented spins
  of free electrons}.
\newblock Sov. Phys. Solid State \textbf{15}, 160 (1973).

\bibitem{zhukov10}
E.~A. Zhukov, D.~R. Yakovlev, M.~M. Glazov, L.~Fokina, G.~Karczewski,
  T.~Wojtowicz, J.~Kossut, M.~Bayer.
\newblock \emph{Optical control of electron spin coherence in
  $\mbox{CdTe/(Cd,Mg)Te}$ quantum wells}.
\newblock Phys. Rev. B \textbf{81}, 235320 (2010).

\bibitem{PhysRevB.77.165309}
S.~O'Leary, H.~Wang.
\newblock \emph{Manipulating nonlinear optical responses from spin-polarized
  electrons in a two-dimensional electron gas via exciton injection}.
\newblock Phys. Rev. B \textbf{77}, 165309 (2008).

\bibitem{phelps:237402}
C.~Phelps, T.~Sweeney, R.~T. Cox, H.~Wang.
\newblock \emph{Ultrafast Coherent Electron Spin Flip in a Modulation-Doped
  $\mbox{CdTe}$ Quantum Well}.
\newblock Phys. Rev. Lett. \textbf{102}, 237402 (2009).

\bibitem{Greilich2009}
A.~Greilich, S.~E. Economou, S.~Spatzek, D.~R. Yakovlev, D.~Reuter, A.~D.
  Wieck, T.~L. Reinecke, M.~Bayer.
\newblock \emph{Ultrafast optical rotations of electron spins in quantum dots}.
\newblock Nature Physics \textbf{5}, 262 (2009).

\bibitem{yakovlev_bayer}
D.~Yakovlev, M.~Bayer.
\newblock \emph{Coherent spin
  dynamics of carriers}, in Spin physics in semiconductors
ed. M. Dyakonov, p. 135 (Springer, 2008).

\bibitem{Chen20101803}
Z.~Chen, S.~G. Carter, R.~Bratschitsch, S.~T. Cundiff.
\newblock \emph{Optical excitation and control of electron spins in
  semiconductor quantum wells}.
\newblock Physica E \textbf{42}, 1803  (2010).

\bibitem{mikkelsen07}
M.~H. Mikkelsen, J.~Berezovsky, N.~G. Stoltz, L.~A. Coldren, D.~D. Awschalom.
\newblock \emph{Optically detected coherent spin dynamics of a single electron
  in a quantum dot}.
\newblock Nature Physics \textbf{3}, 770 (2007).

\bibitem{atature07}
M.~Atature, J.~Dreiser, A.~Badolato, A.~Imamoglu.
\newblock \emph{Observation of $\mbox{F}$araday rotation from a single confined
  spin}.
\newblock Nature Physics \textbf{3}, 101 (2007).

\bibitem{dzhioev97:eng}
R.~I. Dzhioev, B.~P. Zakharchenya, V.~L. Korenev, M.~N. Stepanova.
\newblock \emph{Spin diffusion of optically oriented electrons and photon
  entrainment in n-gallium arsenide}.
\newblock Physics of the Solid State \textbf{39}, 1765 (1997).

\bibitem{fokina-2010}
L.~V. Fokina, I.~A. Yugova, D.~R. Yakovlev, M.~M. Glazov, I.~A. Akimov,
  A.~Greilich, D.~Reuter, A.~D. Wieck, M.~Bayer.
\newblock \emph{Spin dynamics of electrons and holes in $\mbox{InGaAs/GaAs}$
  quantum wells at millikelvin temperatures}.
\newblock Phys. Rev. B \textbf{81}, 195304 (2010).

\bibitem{shabaev:201305}
A.~Shabaev, A.~L. Efros, D.~Gammon, I.~A. Merkulov.
\newblock \emph{Optical readout and initialization of an electron spin in a
  single quantum dot}.
\newblock Phys. Rev. B \textbf{68}, 201305 (2003).

\bibitem{PhysRevB.68.235316}
J.~Tribollet, F.~Bernardot, M.~Menant, G.~Karczewski, C.~Testelin, M.~Chamarro.
\newblock \emph{Interplay of spin dynamics of trions and two-dimensional
  electron gas in a $n$-doped $\mbox{CdTe}$ single quantum well}.
\newblock Phys. Rev. B \textbf{68}, 235316 (2003).

\bibitem{dp_orientation:eng}
M.~I. Dyakonov, V.~I. Perel.
\newblock \emph{Theory of optical spin orientation of electrons and nuclei in
  semiconductors}. in Optical orientation, eds. F. Meier and
B. P. Zakharchenya, p. 11 (1984).

\bibitem{Mar99}
X.~Marie, T.~Amand, P.~Le~Jeune, M.~Paillard, P.~Renucci, L.~E. Golub, V.~D.
  Dymnikov, E.~L. Ivchenko.
\newblock \emph{Hole spin quantum beats in quantum-well structures}.
\newblock Phys. Rev. B \textbf{60}, 5811 (1999).

\bibitem{kennedy:045307}
T.~A. Kennedy, A.~Shabaev, M.~Scheibner, A.~L. Efros, A.~S. Bracker, D.~Gammon.
\newblock \emph{Optical initialization and dynamics of spin in a remotely doped
  quantum well}.
\newblock Phys. Rev. B \textbf{73}, 045307 (2006).

\bibitem{PhysRevLett.94.227403}
M.~V.~G. Dutt, J.~Cheng, B.~Li, X.~Xu, X.~Li, P.~R. Berman, D.~G. Steel, A.~S.
  Bracker, D.~Gammon, S.~E. Economou, R.-B. Liu, L.~J. Sham.
\newblock \emph{Stimulated and Spontaneous Optical Generation of Electron Spin
  Coherence in Charged $\mbox{GaAs}$ Quantum Dots}.
\newblock Phys. Rev. Lett. \textbf{94}, 227403 (2005).

\bibitem{zhu07}
E.~A. Zhukov, D.~R. Yakovlev, M.~Bayer, M.~M. Glazov, E.~L. Ivchenko,
  G.~Karczewski, T.~Wojtowicz, J.~Kossut.
\newblock \emph{Spin coherence of a two-dimensional electron gas induced by
  resonant excitation of trions and excitons in $\mbox{CdTe/(Cd,Mg)Te}$ quantum
  wells}.
\newblock Phys. Rev. B \textbf{76}, 205310 (2007).

\bibitem{PhysRevB.75.115320}
Z.~Chen, R.~Bratschitsch, S.~G. Carter, S.~T. Cundiff, D.~R. Yakovlev,
  G.~Karczewski, T.~Wojtowicz, J.~Kossut.
\newblock \emph{Electron spin polarization through interactions between
  excitons, trions, and the two-dimensional electron gas}.
\newblock Phys. Rev. B \textbf{75}, 115320 (2007).

\bibitem{sokolova09}
I.~A. Yugova, A.~A. Sokolova, D.~R. Yakovlev, A.~Greilich, D.~Reuter, A.~D.
  Wieck, M.~Bayer.
\newblock \emph{Long-Term Hole Spin Memory in the Resonantly Amplified Spin
  Coherence of $\mbox{InGaAs/GaAs}$ Quantum Well Electrons}.
\newblock Phys. Rev. Lett. \textbf{102}, 167402 (2009).

\bibitem{ast08}
G.~V. Astakhov, M.~M. Glazov, D.~R. Yakovlev, E.~A. Zhukov, W.~Ossau, L.~W.
  Molenkamp, M.~Bayer.
\newblock \emph{Time-resolved and continuous-wave optical spin pumping of
  semiconductor quantum wells}.
\newblock Semiconductor Science and Technology \textbf{23}, 114001
  (2008).

\bibitem{Korn2010415}
T.~Korn.
\newblock \emph{Time-resolved studies of electron and hole spin dynamics in
  modulation-doped $\mbox{GaAs/AlGaAs}$ quantum wells}.
\newblock Physics Reports \textbf{494}, 415  (2010).

\bibitem{korn_njp}
T.~Korn, M.~Kugler, M.~Griesbeck, R.~Schulz, A.~Wagner, M.~Hirmer, C.~Gerl,
  D.~Schuh, W.~Wegscheider, C.~Schueller.
\newblock \emph{Engineering ultralong spin coherence in two-dimensional hole
  systems at low temperatures}.
\newblock New Journal of Physics \textbf{12}, 043003 (2010).

\bibitem{Machnikowski10}
P.~Machnikowski, T.~Kuhn.
\newblock \emph{Theory of the time-resolved $\mbox{K}$err rotation in ensembles
  of trapped holes in semiconductor nanostructures}.
\newblock Phys. Rev. B \textbf{81}, 115306 (2010).

\bibitem{PhysRevB.81.045322}
B.~Eble, P.~Desfonds, F.~Fras, F.~Bernardot, C.~Testelin, M.~Chamarro,
  A.~Miard, A.~Lema\^\i{}tre.
\newblock \emph{Hole and trion spin dynamics in quantum dots under excitation
  by a train of circularly polarized pulses}.
\newblock Phys. Rev. B \textbf{81}, 045322 (2010).

\bibitem{dzhioev98:eng}
R.~Dzhioev, B.~Zakharchenya, V.~Korenev, P.~Pak, D.~Vinokurov, O.~Kovalenkov,
  I.~Tarasov.
\newblock \emph{Optical orientation of donor-bound excitons in nanosized
  $\mbox{InP/InGaP}$ islands}.
\newblock Physics of the Solid State \textbf{40}, 1587 (1998).

\bibitem{cortez02}
S.~Cortez, O.~Krebs, S.~Laurent, M.~Senes, X.~Marie, P.~Voisin, R.~Ferreira,
  G.~Bastard, J.-M. Gerrard, T.~Amand.
\newblock \emph{Optically Driven Spin Memory in $n$-Doped $\mbox{InAs-GaAs}$
  Quantum Dots}.
\newblock Phys. Rev. Lett. \textbf{89}, 207401 (2002).

\bibitem{Ignatiev05}
M.~Ikezawa, B.~Pal, Y.~Masumoto, I.~V. Ignatiev, S.~Y. Verbin, I.~Y. Gerlovin.
\newblock \emph{Submillisecond electron spin relaxation in $\mbox{InP}$ quantum
  dots}.
\newblock Phys. Rev. B \textbf{72}, 153302 (2005).

\bibitem{laurent06}
S.~Laurent, M.~Senes, O.~Krebs, V.~K. Kalevich, B.~Urbaszek, X.~Marie,
  T.~Amand, P.~Voisin.
\newblock \emph{Negative circular polarization as a general property of
  $n$-doped self-assembled $\mbox{InAs/GaAs}$ quantum dots under nonresonant
  optical excitation}.
\newblock Phys. Rev. B \textbf{73}, 235302 (2006).

\bibitem{springerlink:10.1134/S0030400X09030114}
I.~Ignatiev, S.~Verbin, I.~Gerlovin, R.~Cherbunin, Y.~Masumoto.
\newblock \emph{Negative circular polarization of $\mbox{InP}$ $\mbox{QD}$
  luminescence: Mechanism of formation and main regularities}.
\newblock Optics and Spectroscopy \textbf{106}, 375 (2009).

\bibitem{tarasenko98}
L.~E. Golub, E.~L. Ivchenko, S.~A. Tarasenko.
\newblock \emph{Interaction of free carriers with localized excitons in quantum
  wells}.
\newblock Solid State Commun. \textbf{108}, 799 (1998).

\bibitem{sizov_book}
F. F. Sizov and Yu. I. Ukhanov, \emph{Magnetooptical Faraday and
Voigt Effects in Semiconductors} (in Russian) (Naukova dumka, Kiev,
(1979)).

\bibitem{Kosaka2009}
H.~Kosaka, T.~Inagaki, Y.~Rikitake, H.~Imamura, Y.~Mitsumori, K.~Edamatsu.
\newblock \emph{Spin state tomography of optically injected electrons in a
  semiconductor}.
\newblock Nature \textbf{457}, 702 (2009).

\bibitem{yugova09}
I.~A. Yugova, M.~M. Glazov, E.~L. Ivchenko, A.~L. Efros.
\newblock \emph{Pump-probe Faraday rotation and ellipticity in an ensemble of
  singly charged quantum dots}.
\newblock Phys. Rev. B \textbf{80}, 104436 (2009).

\bibitem{springerlink:10.1134/1.1130188}
E.~Ivchenko, V.~Kochereshko, A.~Platonov, D.~Yakovlev, A.~Waag, W.~Ossau,
  G.~Landwehr.
\newblock \emph{Resonance optical spectroscopy of long-period quantum-well
  structures}.
\newblock Physics of the Solid State \textbf{39}, 1852 (1997).

\bibitem{PhysRevB.74.073407}
H.~Hoffmann, G.~V. Astakhov, T.~Kiessling, W.~Ossau, G.~Karczewski,
  T.~Wojtowicz, J.~Kossut, L.~W. Molenkamp.
\newblock \emph{Optical spin pumping of modulation-doped electrons probed by a
  two-color $\mbox{K}$err rotation technique}.
\newblock Phys. Rev. B \textbf{74}, 073407 (2006).

\bibitem{gourdon:230}
C.~Gourdon, V.~Jeudy, M.~Menant, D.~Roditchev, L.~A. Tu, E.~L. Ivchenko,
  G.~Karczewski.
\newblock \emph{Magneto-optical imaging with diluted magnetic semiconductor
  quantum wells}.
\newblock Applied Physics Letters \textbf{82}, 230 (2003).

\bibitem{astakhov00}
G.~V. Astakhov, V.~P. Kochereshko, D.~R. Yakovlev, W.~Ossau, J.~Nurnberger,
  W.~Faschinger, G.~Landwehr.
\newblock \emph{Oscillator strength of trion states in $\mbox{ZnSe}$-based
  quantum wells}.
\newblock Phys. Rev. B \textbf{62}, 10345 (2000).

\bibitem{astakhov02}
G.~V. Astakhov, V.~P. Kochereshko, D.~R. Yakovlev, W.~Ossau, J.~Nurnberger,
  W.~Faschinger, G.~Landwehr, T.~Wojtowicz, G.~Karczewski, J.~Kossut.
\newblock \emph{Optical method for the determination of carrier density in
  modulation-doped quantum wells}.
\newblock Phys. Rev. B \textbf{65}, 115310 (2002).

\bibitem{PhysRevLett.75.2554}
T.~\"Ostreich, K.~Sch\"onhammer, L.~J. Sham.
\newblock \emph{Theory of Spin Beatings in the $\mbox{F}$araday Rotation of
  Semiconductors}.
\newblock Phys. Rev. Lett. \textbf{75}, 2554 (1995).

\bibitem{wang04}
P.~Palinginis, H.~Wang.
\newblock \emph{Vanishing and Emerging of Absorption Quantum Beats from
  Electron Spin Coherence in $\mbox{GaAs}$ Quantum Wells}.
\newblock Phys. Rev. Lett. \textbf{92}, 037402 (2004).

\bibitem{shen05}
Y.~Shen, A.~M. Goebel, G.~Khitrova, H.~M. Gibbs, H.~Wang.
\newblock \emph{Nearly degenerate time-resolved $\mbox{F}$araday rotation in an
  interacting exciton system}.
\newblock Phys. Rev. B \textbf{72}, 233307 (2005).

\bibitem{PhysRevB.72.245202}
P.~Nemec, Y.~Kerachian, H.~M. van Driel, A.~L. Smirl.
\newblock \emph{Spin-dependent electron many-body effects in $\mbox{GaAs}$}.
\newblock Phys. Rev. B \textbf{72}, 245202 (2005).

\bibitem{PhysRevB.74.125316}
M.~Combescot, O.~Betbeder-Matibet.
\newblock \emph{Faraday rotation in photoexcited semiconductors: A
  composite-exciton many-body effect}.
\newblock Phys. Rev. B \textbf{74}, 125316 (2006).

\bibitem{PhysRevLett.103.056405}
N.~H. Kwong, S.~Schumacher, R.~Binder.
\newblock \emph{Electron-Spin Beat Susceptibility of Excitons in Semiconductor
  Quantum Wells}.
\newblock Phys. Rev. Lett. \textbf{103}, 056405 (2009).

\bibitem{PhysRevB.76.045320}
N.~S. Averkiev, M.~M. Glazov.
\newblock \emph{Light-matter interaction in doped microcavities}.
\newblock Phys. Rev. B \textbf{76}, 045320 (2007).

\bibitem{PhysRevB.60.4826}
G.-H. Chen, M.~E. Raikh.
\newblock \emph{Exchange-induced enhancement of spin-orbit coupling in
  two-dimensional electronic systems}.
\newblock Phys. Rev. B \textbf{60}, 4826 (1999).

\bibitem{Barate2010}
P.~Barate, S.~Cronenberger, M.~Vladimirova, D.~Scalbert, F.~Perez, J.~G\`omez,
  B.~Jusserand, H.~Boukari, D.~Ferrand, H.~Mariette, J.~Cibert, M.~Nawrocki.
\newblock \emph{Collective nature of two-dimensional electron gas spin
  excitations revealed by exchange interaction with magnetic ions}.
\newblock Phys. Rev. B \textbf{82}, 075306 (2010).

\bibitem{PhysRevB.78.081305}
M.~Vladimirova, S.~Cronenberger, P.~Barate, D.~Scalbert, F.~J. Teran, A.~P.
  Dmitriev.
\newblock \emph{Dynamics of the localized spins interacting with
  two-dimensional electron gas: Coexistence of mixed and pure modes}.
\newblock Phys. Rev. B \textbf{78}, 081305 (2008).

\bibitem{artemova85}
E.~S. Artemova, I.~A. Merkulov.
\newblock \emph{Nuclear field and Faraday effect in semiconductors}.
\newblock Sov. Phys. Solid State \textbf{27}, 941 (1985).

\bibitem{J.M.Kikkawa01212000}
J.~M. Kikkawa, D.~D. Awschalom.
\newblock \emph{{All-Optical Magnetic Resonance in Semiconductors}}.
\newblock Science \textbf{287}, 473 (2000).

\bibitem{PhysRevLett.99.056804}
P.~Maletinsky, A.~Badolato, A.~Imamoglu.
\newblock \emph{Dynamics of Quantum Dot Nuclear Spin Polarization Controlled by
  a Single Electron}.
\newblock Phys. Rev. Lett. \textbf{99}, 056804 (2007).

\bibitem{PhysRevB.50.7689}
J.~J. Baumberg, S.~A. Crooker, D.~D. Awschalom, N.~Samarth, H.~Luo, J.~K.
  Furdyna.
\newblock \emph{Ultrafast $\mbox{F}$araday spectroscopy in magnetic
  semiconductor quantum structures}.
\newblock Phys. Rev. B \textbf{50}, 7689 (1994).

\bibitem{PhysRevB.56.7574}
S.~A. Crooker, D.~D. Awschalom, J.~J. Baumberg, F.~Flack, N.~Samarth.
\newblock \emph{Optical spin resonance and transverse spin relaxation in
  magnetic semiconductor quantum wells}.
\newblock Phys. Rev. B \textbf{56}, 7574 (1997).

\bibitem{CrookerPRL96}
S.~A. Crooker, J.~J. Baumberg, F.~Flack, N.~Samarth, D.~D. Awschalom.
\newblock \emph{Terahertz Spin Precession and Coherent Transfer of Angular
  Momenta in Magnetic Quantum Wells}.
\newblock Phys. Rev. Lett. \textbf{77}, 2814 (1996).

\bibitem{Li2010450}
B.~Li, P.~Coles, J.~A. Reimer, P.~Dawson, C.~A. Meriles.
\newblock \emph{Optical pumping of nuclear spin magnetization in GaAs/AlAs
  quantum wells of variable electron density}.
\newblock Solid State Communications \textbf{150}, 450  (2010).

\bibitem{bayer_long}
S.~Spatzek, S.~Varwig, M.~M. Glazov, I.~A. Yugova, A.~Schwan, D.~R. Yakovlev,
  D.~Reuter, A.~D. Wieck, M.~Bayer.
\newblock \emph{Generation and detection of mode-locked spin coherence in
  (In,Ga)As/GaAs quantum dots by laser pulses of long duration}.
\newblock Phys. Rev. B \textbf{84}, 115309 (2011).

\bibitem{PhysRevB.83.033301}
V.~Loo, L.~Lanco, O.~Krebs, P.~Senellart, P.~Voisin.
\newblock \emph{Single-shot initialization of electron spin in a quantum dot
  using a short optical pulse}.
\newblock Phys. Rev. B \textbf{83}, 033301 (2011).

\bibitem{greilich06}
A.~Greilich, R.~Oulton, E.~A. Zhukov, I.~A. Yugova, D.~R. Yakovlev, M.~Bayer,
  A.~Shabaev, A.~L. Efros, I.~A. Merkulov, V.~Stavarache, D.~Reuter, A.~Wieck.
\newblock \emph{Optical Control of Spin Coherence in Singly Charged
  $\mbox{(In,Ga)As/GaAs}$ Quantum Dots}.
\newblock Phys. Rev. Lett. \textbf{96}, 227401 (2006).

\bibitem{carter:167403}
S.~G. Carter, A.~Shabaev, S.~E. Economou, T.~A. Kennedy, A.~S. Bracker, T.~L.
  Reinecke.
\newblock \emph{Directing Nuclear Spin Flips in InAs Quantum Dots Using Detuned
  Optical Pulse Trains}.
\newblock Phys. Rev. Lett. \textbf{102}, 167403 (2009).

\bibitem{ll3_eng}
L.~Landau, E.~Lifshitz.
\newblock {\selectlanguage{english}\emph{Quantum Mechanics: Non-Relativistic
  Theory (vol. 3)}} (Butterworth-Heinemann, Oxford, 1977).

\bibitem{PhysRev.40.502}
N.~Rosen, C.~Zener.
\newblock \emph{Double $\mbox{Stern-Gerlach}$ Experiment and Related Collision
  Phenomena}.
\newblock Phys. Rev. \textbf{40}, 502 (1932).

\bibitem{PhysRev.143.574}
P.~S. Pershan, J.~P. van~der Ziel, L.~D. Malmstrom.
\newblock \emph{Theoretical Discussion of the Inverse $\mbox{F}$araday Effect,
  $\mbox{R}$aman Scattering, and Related Phenomena}.
\newblock Phys. Rev. \textbf{143}, 574 (1966).

\bibitem{economou:205415}
S.~E. Economou, L.~J. Sham, Y.~Wu, D.~G. Steel.
\newblock \emph{Proposal for optical $\mbox{U}$(1) rotations of electron spin
  trapped in a quantum dot}.
\newblock Phys. Rev. B \textbf{74}, 205415 (2006).

\bibitem{Carter:07}
S.~G. Carter, Z.~Chen, S.~T. Cundiff.
\newblock \emph{Ultrafast below-resonance $\mbox{R}$aman rotation of electron
  spins in $\mbox{GaAs}$ quantum wells}.
\newblock Phys. Rev. B \textbf{76}, 201308 (2007).

\bibitem{Bonadeo20111998}
N.~H. Bonadeo, J.~Erland, D.~Gammon, D.~Park, D.~S. Katzer, D.~G. Steel.
\newblock \emph{Coherent Optical Control of the Quantum State of a Single
  Quantum Dot}.
\newblock Science \textbf{282}, 1473 (1998).

\bibitem{J.Berezovsky04182008}
J.~Berezovsky, M.~H. Mikkelsen, N.~G. Stoltz, L.~A. Coldren, D.~D. Awschalom.
\newblock \emph{{Picosecond Coherent Optical Manipulation of a Single Electron
  Spin in a Quantum Dot}}.
\newblock Science \textbf{320}, 349 (2008).

\bibitem{Kim2011}
D.~Kim, S.~G. Carter, A.~Greilich, A.~S. Bracker, D.~Gammon.
\newblock \emph{Ultrafast optical control of entanglement between two
  quantum-dot spins}.
\newblock Nat Phys \textbf{7}, 223 (2011).

\bibitem{Takagahara:10}
T.~Takagahara.
\newblock \emph{Theory of unitary spin rotation and spin-state tomography for a
  single electron and two electrons}.
\newblock J. Opt. Soc. Am. B \textbf{27}, A46 (2010).

\bibitem{Ramsay}
A.~J. Ramsay.
\newblock \emph{A review of the coherent optical control of the exciton and
  spin states of semiconductor quantum dots}.
\newblock Semiconductor Science and Technology \textbf{25}, 103001 (2010).

\bibitem{PhysRevLett.98.047403}
V.~V. Pavlov, R.~V. Pisarev, V.~N. Gridnev, E.~A. Zhukov, D.~R. Yakovlev,
  M.~Bayer.
\newblock \emph{Ultrafast Optical Pumping of Spin and Orbital Polarizations in
  the Antiferromagnetic $\mbox{M}$ott Insulators $\mbox{R}_{2}\mbox{CuO}_{4}$}.
\newblock Phys. Rev. Lett. \textbf{98}, 047403 (2007).

\bibitem{PhysRevLett.103.117201}
K.~Vahaplar, A.~M. Kalashnikova, A.~V. Kimel, D.~Hinzke, U.~Nowak,
  R.~Chantrell, A.~Tsukamoto, A.~Itoh, A.~Kirilyuk, T.~Rasing.
\newblock \emph{Ultrafast Path for Optical Magnetization Reversal via a
  Strongly Nonequilibrium State}.
\newblock Phys. Rev. Lett. \textbf{103}, 117201 (2009).

\bibitem{glazov2010a}
M.~M. Glazov, I.~A. Yugova, S.~Spatzek, A.~Schwan, S.~Varwig, D.~R. Yakovlev,
  D.~Reuter, A.~D. Wieck, M.~Bayer.
\newblock \emph{Effect of pump-probe detuning on the $\mbox{F}$araday rotation
  and ellipticity signals of mode-locked spins in $\mbox{(In,Ga)As/GaAs}$
  quantum dots}.
\newblock Phys. Rev. B \textbf{82}, 155325 (2010).

\bibitem{yu07}
I.~A. Yugova, A.~Greilich, E.~A. Zhukov, D.~R. Yakovlev, M.~Bayer, D.~Reuter,
  A.~D. Wieck.
\newblock \emph{Exciton fine structure in $\mbox{InGaAs/GaAs}$ quantum dots
  revisited by pump-probe Faraday rotation}.
\newblock Phys. Rev. B \textbf{75}, 195325 (2007).

\bibitem{beschoten}
B.~Beschoten.
\newblock \emph{Spin coherence in semiconductors}, in ``Magnetism goes
Nano'', 36th Sping School (2005), Schriften des Forschungzentrum
Julich, Matter and Materials, vol. 26, eds. S. Blugel,
T. Bruckel, and C.M. Schneider(2005). 

\bibitem{glazov08a}
M.~M. Glazov, E.~L. Ivchenko.
\newblock \emph{Resonant Spin Amplification in Nanostructures with Anisotropic
  Spin Relaxation and Spread of the Electronic g-Factor}.
\newblock Semiconductors \textbf{42}, 951 (2008).

\bibitem{averkiev06}
N.~S. Averkiev, L.~E. Golub, A.~S. Gurevich, V.~P. Evtikhiev, V.~P.
  Kochereshko, A.~V. Platonov, A.~S. Shkolnik, Y.~P. Efimov.
\newblock \emph{Spin-relaxation anisotropy in asymmetrical (001)
  $\mbox{Al}_x\mbox{Ga}_{ 1 - x}\mbox{As}$ quantum wells from
  $\mbox{H}$anle-effect measurements: Relative strengths of $\mbox{Rashba}$ and
  $\mbox{Dresselhaus}$ spin-orbit coupling}.
\newblock Phys. Rev. B \textbf{74}, 033305 (2006).

\bibitem{larionov:033302}
A.~V. Larionov, L.~E. Golub.
\newblock \emph{Electric-field control of spin-orbit splittings in
  $\mbox{GaAs/Al}_{x}\mbox{Ga}_{1 - x}\mbox{As}$ coupled quantum wells}.
\newblock Phys. Rev. B \textbf{78}, 033302 (2008).

\bibitem{larionov11:eng}
A.~Larionov, A.~Sekretenko, A.~Il'in.
\newblock \emph{Control of electron spin dynamics in a wide GaAs quantum well
  by a lateral confining potential}.
\newblock JETP Letters \textbf{93}, 269 (2011).

\bibitem{greilich09}
A.~Greilich, S.~Spatzek, I.~A. Yugova, I.~A. Akimov, D.~R. Yakovlev, A.~L.
  Efros, D.~Reuter, A.~D. Wieck, M.~Bayer.
\newblock \emph{Collective single-mode precession of electron spins in an
  ensemble of singly charged $\mbox{(In,Ga)As/GaAs}$ quantum dots}.
\newblock Phys. Rev. B \textbf{79}, 201305 (2009).

\bibitem{PhysRevB.64.125316}
A.~V. Khaetskii, Y.~V. Nazarov.
\newblock \emph{Spin-flip transitions between Zeeman sublevels in semiconductor
  quantum dots}.
\newblock Phys. Rev. B \textbf{64}, 125316 (2001).

\bibitem{merkulov02}
I.~A. Merkulov, A.~L. Efros, M.~Rosen.
\newblock \emph{Electron spin relaxation by nuclei in semiconductor quantum
  dots}.
\newblock Phys. Rev. B \textbf{65}, 205309 (2002).

\bibitem{PhysRevLett.88.186802}
A.~V. Khaetskii, D.~Loss, L.~Glazman.
\newblock \emph{Electron Spin Decoherence in Quantum Dots due to Interaction
  with Nuclei}.
\newblock Phys. Rev. Lett. \textbf{88}, 186802 (2002).

\bibitem{PhysRevB.66.161318}
L.~M. Woods, T.~L. Reinecke, Y.~Lyanda-Geller.
\newblock \emph{Spin relaxation in quantum dots}.
\newblock Phys. Rev. B \textbf{66}, 161318 (2002).

\bibitem{Kozlov2007}
G.~Kozlov.
\newblock \emph{Exactly solvable spin dynamics of an electron coupled to a
  large number of nuclei; the electron-nuclear spin echo in a quantum dot}.
\newblock JETP \textbf{105}, 803
  (2007).

\bibitem{A.Greilich09282007}
A.~Greilich, A.~Shabaev, D.~R. Yakovlev, A.~L. Efros, I.~A. Yugova, D.~Reuter,
  A.~D. Wieck, M.~Bayer.
\newblock \emph{{Nuclei-Induced Frequency Focusing of Electron Spin
  Coherence}}.
\newblock Science \textbf{317}, 1896 (2007).

\bibitem{dp74}
M.~I. {Dyakonov}, V.~I. {Perel'}.
\newblock \emph{{Optical orientation in a system of electrons and lattice
  nuclei in semiconductors. Theory}}.
\newblock Soviet Journal of Experimental and Theoretical Physics \textbf{38},
  177 (1974).

\bibitem{2010arXiv1006.5144K}
V.~L. Korenev.
\newblock \emph{Multiple stable states of a periodically driven electron spin
  in a quantum dot using circularly polarized light}.
\newblock Phys. Rev. B \textbf{83}, 235429 (2011).

\bibitem{yugova11}
M.~M. {Glazov}, I.~A. {Yugova}, A.~L. {Efros}.
\newblock \emph{{Electron spin synchronization induced by optical nuclear
  magnetic resonance feedback}}.
\newblock preprint Arxiv cond-mat:1103.3249  (2011).

\bibitem{PhysRev.114.90}
L.~M. Roth, B.~Lax, S.~Zwerdling.
\newblock \emph{Theory of Optical Magneto-Absorption Effects in
  Semiconductors}.
\newblock Phys. Rev. \textbf{114}, 90 (1959).

\bibitem{ivchenko_kiselev92:eng}
E.~L. Ivchenko, A.~A. Kiselev.
\newblock \emph{Electron g factor of quantum wells and superlattices}.
\newblock Sov. Phys. Semicond. \textbf{26}, 827 (1992).

\bibitem{ivchenko05a}
E.~L. Ivchenko.
\newblock \emph{Optical Spectroscopy of Semiconductor Nanostructures} (Alpha
  Science, Harrow UK, 2005).

\bibitem{PhysRevB.75.245302}
I.~A. Yugova, A.~Greilich, D.~R. Yakovlev, A.~A. Kiselev, M.~Bayer, V.~V.
  Petrov, Y.~K. Dolgikh, D.~Reuter, A.~D. Wieck.
\newblock \emph{Universal behavior of the electron $g$ factor in
  $\mbox{GaAs/Al}_x\mbox{Ga}_{1-x} \mbox{As}$ quantum wells}.
\newblock Phys. Rev. B \textbf{75}, 245302 (2007).

\end{thebibliography}

\end{document}